\documentclass[useAMS,usenatbib]{mn2e}
\usepackage{afterpage}
\usepackage{amsmath}
\usepackage{psfrag} % for nicer fonts in figures

\usepackage{natbib}
\usepackage{graphicx}
\usepackage{txfonts}

\usepackage{relsize}%for a bigger sum / smaller symbols
\usepackage{pbox}
\usepackage[usenames,dvipsnames]{color}
\usepackage{bm}% for  fat greek letters!

\usepackage{soul}% for  corssing out

\newcommand{\Eqref}[1]{Eq.\,(\ref{#1})}
\newcommand{\eqsref}[1]{Eqs.\,(\ref{#1})}

\newcommand{\figref}[1]{Fig.\,\ref{#1}}
\newcommand{\tabref}[1]{Tab.\,\ref{#1}}
\newcommand{\secref}[1]{Sec.\,\ref{#1}}

\newcommand{\inst}[1]{$^{#1}$}

\newcommand{\ober}{\citet{Obergaulinger_et_al_2009} }

\newcommand{\half}{\frac{1}{2}}

\newcommand{\tto}{\text{--}}

\newcommand{\deltt}{\text{\small$\Delta$}}

\newcommand{\MMM}{\mathcal{M}_{r\phi}}

\newcommand{\der}{\mathrm{d}} 
\newcommand{\alf}{Alfv\'{e}n }

\newcommand{\rrr}{\rightarrow }

\newcommand{\divv}{\nabla \cdot}

\newcommand{\bz}{b_{0z}}

\newcommand{\Ree}{R_{\mathrm{e} }}
\newcommand{\Rm}{R_{\mathrm{m} }}
\newcommand{\Pm}{P_{\mathrm{m} }}

\newcommand{\caz}{c_{\mathrm{A} z}}
\newcommand{\cac}{c_{ \mathrm{Ac}}}

\newcommand{\cs}{c_{\mathrm{s}}}

\newcommand{\vek}[1]{\bf{#1}}

\newcommand{\vc}{v_{\mathrm{c}}}

\newcommand{\tildevc}{\tilde{v}_{\mathrm{c}}}

\newcommand{\bc}{b_{\mathrm{c}}}

\newcommand{\bfvc}{{\bf v}_{\mathrm{c}}}
\newcommand{\bfvp}{{\bf v}_{\mathrm{p}}}
\newcommand{\tildebc}{\tilde{b}_{\mathrm{c}}}

\newcommand{\bfbc}{{ \bf b}_{\mathrm{c}}}
\newcommand{\bfbp}{{ \bf b}_{\mathrm{p}}}
\newcommand{\gammap}{\gamma_{\mathrm{p}}}

\newcommand{\phikh}{\phi_{\mathsmaller{\mathrm{KH}}}}

\newcommand{\lambdatm}{\lambda_{\mathsmaller{\mathrm{TM}}}}
\newcommand{\lambdamri}{\lambda_{\mathsmaller{\mathrm{MRI}}}}

\newcommand{\gammamri}{\gamma_{\mathsmaller{\mathrm{MRI}}}}

\newcommand{\gammamrismall}{\gamma_{\mathsmaller{\mathrm{mri}}}}

\newcommand{\kmri}{k_{\mathsmaller{\mathrm{MRI}}}}
\newcommand{\kmrismall}{k_{\mathsmaller{\mathrm{mri}}}}
\newcommand{\phitm}{\phi_{\mathsmaller{\mathrm{TM}}}}
\newcommand{\phip}{\phi_{\mathrm{p}}}
\newcommand{\lambdap}{\lambda_{\mathrm{p}}}

\newcommand{\jcop}{J. Comput. Phys.}

\newcommand{\apj}{ApJ}
\newcommand{\mnras}{MNRAS}

\newcommand{\aap}{A{\&}A}

\newcommand{\prd}{Phys. Rev. D}
\newcommand{\apjl}{ApJL}

\newcommand{\apjs}{ApJS}

\newcommand{\ie}{i.e.~}

\newcommand{\s}{\textrm{s}}

\newcommand{\ms}{\textrm{ms}}
\newcommand{\km}{\textrm{km}}
\newcommand{\cm}{\textrm{cm}}

\newcommand{\gccm}{\textrm{g cm}^{-3}}

\title[MRI in CCSNe]%
{Termination of the magnetorotational instability via parasitic
  instabilities in core-collapse supernovae }

\author[T.~Rembiasz et al.]%
{%%
  T.\,Rembiasz\inst{1,2},
  M.\,Obergaulinger\inst{2},
  P.\,Cerd\'a-Dur\'an\inst{2}, 
  E.\,M{\"u}ller\inst{1}, 
  M.A.\,Aloy\inst{2}  \\
\inst{1} Max-Planck-Institut f{\"u}r Astrophysik, Karl-Schwarzschild-Str.~1, D-85748 Garching, Germany \\
\inst{2} Departamento de Astronom\'{\i}a y Astrof\'{\i}sica,  Universidad de Valencia,  C/ Dr.~Moliner 50, E-46100 Burjassot, Spain  
}

\begin{document}

\date{Accepted 2015 December 9. Received 2015 December 9; in original  2015 July 31}

\pagerange{\pageref{firstpage}--\pageref{lastpage}} \pubyear{2015}

\maketitle

\label{firstpage}

%%%%%%%%%%%%%%%%%%%%%%%%%%%%%%%%%%%%%%%%%%%%%%%%%%%%%%%%%%%%%
%% Abstract
%%%%%%%%%%%%%%%%%%%%%%%%%%%%%%%%%%%%%%%%%%%%%%%%%%%%%%%%%%%%%
\begin{abstract}
  The magnetorotational instability (MRI) can be a powerful mechanism amplifying the magnetic field in core collapse
  supernovae. Whether initially weak magnetic fields can be amplified by this instability to dynamically
  relevant strengths is still a matter of debate. One of the main uncertainties concerns the process
  that terminates the growth of the instability.  Parasitic instabilities of both Kelvin-Helmholtz and tearing-mode type
  have been suggested to play a crucial role in this process, disrupting MRI channel flows and quenching magnetic field
  amplification.  We perform two-dimensional and three-dimensional sheering-disc simulations of a differentially
  rotating proto-neutron star layer in non-ideal magnetohydrodynamics with unprecedented high numerical accuracy, finding 
  that Kelvin-Helmholtz parasitic modes dominate tearing modes in the regime of large hydrodynamic and magnetic Reynolds
  numbers, as encountered close to the surface of proto-neutron stars. They also determine the maximum magnetic field stress achievable
  during the exponential growth of the MRI.  Our results are consistent with the theory of parasitic instabilities based
  on a local stability analysis. To simulate the Kelvin-Helmholtz instabilities properly a very high numerical
  resolution is necessary. Using 9th order spatial reconstruction schemes, we find that at least $8$ grid zones per MRI
  channel are necessary to simulate the growth phase of the MRI and reach an accuracy of $\sim 10\%$ in the growth rate,
  while more than $\sim 60$ zones per channel are required to achieve convergent results for
  the value of the magnetic stress at MRI termination. 
\end{abstract}

%%%%%%%%%%%%%%%%%%%%%%%%%%%%%%%%%%%%%%%%%%%%%%%%%%%%%%%%%%%%%
%%%%%%%%%%%%%%%%%%%%%%%%%%%%%%%%%%%%%%%%%%%%%%%%%%%%%%%%%%%%%

\section{Introduction}
\label{sec:intro}

Originally discovered by \cite{Velikhov__1959__SovPhys__MRI} and
\cite{Chandrasekhar__1960__PNAS__MRI}, the magnetorotational
instability (MRI) was suggested by \citet[][BH91
hereafter]{Balbus_Hawley__1991__ApJ__MRI} to be the physical mechanism
driving the redistribution of angular momentum required for the
accretion process in Keplerian discs orbiting compact objects
\citep[see, e.g.,][for a review]{Balbus_Hawley__1998__RMP__MRI}.

Keplerian discs have a positive radial gradient in angular momentum,
and therefore are linearly (Rayleigh-)stable.  Purely hydrodynamic
perturbations are unlikely to grow to amplitudes at which the
associated stresses can account for efficient angular-momentum
transport.  In the presence of a weak magnetic field, however, a
negative radial gradient in the angular velocity of the disc is
MRI-unstable, and seed perturbations can grow exponentially on time
scales close to the rotational period.  During this phase,
\emph{channel modes} develop. Channel modes are pairs of coherent
radial up- and downflows stacked vertically and threaded by layers of
magnetic field of alternating radial and azimuthal polarity.  In these
modes, the magnetic tension (Maxwell stress tensor) transports angular
momentum along the field lines from the inner parts of the disc
outwards.

The criterion for the MRI onset can be formulated in a rather simple
manner, even if the thermal stratification (gradients of entropy or
molecular weight) and non-ideal effects (viscosity, resistivity) are
included
\citep{Balbus__1995__ApJ__stratified_MRI,Menou_etal__2004__ApJ__MRI-stability}.
This allows for its application beyond Keplerian discs, in particular
to proto-neutron stars (PNSs) resulting from the core-collapse of
rotating massive stars.  Simplified simulations by
\citet{Akiyama_etal__2003__ApJ__MRI_SN} showed that such PNSs possess
regions in which the MRI can grow on shorter time-scales than the time
between the bounce and the successful explosion.  This finding, later
confirmed in multi-dimensional models
\citep[e.g.][]{Obergaulinger_et_al__2006__AA__MR_collapse_TOV,
  Cerda-Duran_et_al__2007__AA__passive-MHD-collapse,
  Sawai_et_al__2013__apjl__GlobalSimulationsofMagnetorotationalInstabilityintheCollapsedCoreofaMassiveStar,Sawai_Yamada_2015},
presents the possibility of generating strong magnetic fields that can
tap the rotational energy of the core, power magnetohydrodynamics (MHD) turbulence
\citep{Masada_et_al__2015}, and become a potentially important
ingredient in rapidly-rotating core-collapse supernovae (CCSNe).

How much these systems are affected by the MRI crucially depends on
both its growth rate and on the final amplitude of the seed
perturbations.  We can give an upper limit by assuming that the MRI
ceases to grow once the magnetic field comes close (within a factor of
$\alpha$) to equipartition with the energy of the differential
rotation.  In CCSNe, this would correspond to dynamically important
field strengths up to $10^{15} \, \mathrm{G}$.  Similar energetic
arguments can be used to express the MRI-generated stresses in the
framework of $\alpha-$disc models
\citep{Shakura_Sunyaev__1973__AA__alpha_visco}.

This estimate neglects possible effects quenching the MRI before it
reaches its maximally-allowed energy,
and the effects of buoyancy as shown by \citet{Guilet_Mueller_2015}, who
            performed MRI simulations in the presence of  
            buoyancy. They  showed that the
            termination amplitude (at the end of the exponential growth) is
            not necessarily correlated to the magnetic field strength in the
            turbulent state that followed.
The physics of the termination
of the MRI growth remains an active field of research with many
studies devoted to finding the value of the $\alpha$ parameter for the
stress tensor.  We refer, among others, to the works of
\citet{Sano_etal__2004__ApJ__3d-local-MRI-sim-P,Sano_Inutsuka__2001__ApJL__MRI-recurrent-channels,Brandenburg__2005__AN__Turbulence_and_its_parameterization_in_accretion_discs,Fromang_Papaloizou__2007__AA__3d-local-MRI-disc-zero-net-flux_1,Gardiner__2005__MagneticFieldsintheUniverse__Energetics_in_MRI_driven_Turbulence,Knobloch__2005__PhFl__Saturation_of_the_MRI}.

The model of parasitic instabilities by \citet[][GX94
hereafter]{Goodman_Xu}, further studied and developed by
\cite{Latter_et_al}, \citet{Pessah_Goodman} 
and \citet{Pessah} provides a clear physical picture of the
termination mechanism.  The MRI channel modes are characterized by a
shear layer and a current sheet in the vertical profiles of velocity
and magnetic field, respectively.  Hence, the (laminar) channel flows
can be unstable against \emph{secondary} (or \emph{parasitic})
\emph{instabilities} of Kelvin-Helmholtz (KH) or tearing-mode (TM)
type.%
\footnote{\cite{Latter_et_al} classifies the types of parasitic
    modes differently.}
Initially, the role of the parasites is negligible, as they grow much
more slowly than the MRI.  However, since the growth rate of the
secondary instabilities is proportional to the channel mode amplitude,
it is clear that at some stage the parasites will grow faster than the
MRI channels whose growth rate is constant, whereas the parasites
  grow exponentially with time. Roughly at this point, the parasitic
instabilities should disrupt the channel modes and terminate the MRI
growth, marking the transition to the turbulent saturation phase
\citep{Pessah}. A further discussion of the MRI saturated state is
beyond the scope of this paper.

\citet{Pessah} analytically studied the MRI termination in
resistive-viscous MHD by solving simplified model equations for the
evolution of the parasitic instabilities.  He identified different
parameter-space regimes where, depending on hydrodynamic and magnetic
Reynolds numbers, either the KH instability or the TM is the dominant
(\ie faster developing) secondary instability. 
\cite{Obergaulinger_2014} found that the magnetic 
field at the surface of the PNS can be enhanced w.r.t.\ the interior regions. 
If this is also the case for rotating cores,  MHD phenomena (like, e.g.\ the MRI)
should be most prominent at the PNS surface.
In this region,    the Reynolds numbers are large if the surface is located above the
  neutrinosphere, which however may not always be the case \citep[see
  Fig.\,10 in][]{Guilet_2015}.  
 In this paper, we only  investigate the regime of 
 very high Reynolds numbers in which, according to the parasitic model,  the
MRI should be terminated by the KH instability. 

The parasitic model has not been tested with direct global numerical
simulations of the MRI in CCSNe. \ober found in their semi-global 2D
ideal MHD simulations that, because of numerical resistivity, the MRI
was terminated by TMs.  In their 3D simulations the MRI was
terminated by non-axisymmetric parasitic instabilities, although a
clear identification of their nature was not possible.

To test the predictions of \cite{Pessah}, we performed a set of 2D and
3D resistive-viscous MHD simulations of the MRI.  Given initial
conditions, we studied in particular the importance of non-ideal
effects by varying the values of both the uniform viscosity and
uniform resistivity, which influence the growth of KH modes and of
TMs, respectively.  In the case of the KH instability,
which is present already in ideal hydrodynamics, viscosity and
resistivity alter the properties of the unstable modes merely
quantitatively.  On the other hand, TM are essentially driven by
resistivity, \ie they do not grow in ideal MHD, while viscosity
plays a minor role by changing the properties of the unstable modes
only quantitatively.

Since we are mainly concerned in this work with identifying the type
of the parasitic instability that limits the growth of the MRI in
particular models, we performed all our simulations (except for one
control model) with a non-zero resistivity, but simulated models with
both zero and non-zero viscosity.  We defer a more thorough
quantitative analysis of the influence of non-ideal effects to a
subsequent work.

In Sec.\ \ref{sec:mri_theory} we give the criterion for the onset of
the MRI, and we describe the initial stage of the instability during
which channel modes develop.  Next, we discuss possible scenarios of
its termination, in particular, termination via the parasitic
instabilities.  In Sec.\ \ref{sec:numerics} we describe the numerical
code and the initial setup used in our 2D and 3D simulations. We
present the results of these simulations in Sec.\ \ref{sec:results},
and summarize our findings in Sec.\ \ref{sec:summary}.

\section{MRI exponential  growth phase and termination}
\label{sec:mri_theory}

%%%%%%%%%%%%%%%%%%%%%%%%%%%%%%%%%%%%%%%%%%%%%%%%%%%%%%%%%%%%%%%%%%%%%%
\subsection{Physical model}
\label{sSec:Physical_Model}

We consider flows that can be described by the equations of
resistive-viscous (non-ideal) MHD. In the
presence of an external gravitational potential, $\Phi$, these
equations read
\begin{align}
\label{eq:cont}
\partial_t \rho + \nabla \cdot ( \rho {\vek v} ) &= 0, \\ 
\partial_t( \rho {\vek v }) + 
  \divv \left(\rho {\vek v} \otimes {\vek v} + {\vek T} \right) 
 &= - \rho \nabla \Phi, \\
\partial_{t} e_{\star} +   \divv \left[ 
e_{\star} {\vek v} + {\vek v  \cdot T}
   + \eta \left( {\vek b} \cdot \nabla {\vek b}  
        - \mathsmaller{\frac{1}{2}} \nabla {\vek b}^2\right) 
  \right] &= -\rho {\vek v} \cdot \nabla \Phi, \\ 
\partial_t {\vek b} + 
  \divv \left({\vek v}\otimes{\vek b} - {\vek b}\otimes{\vek v} \right) 
 &= \eta \nabla^2 {\vek b}, \\
\divv {\vek b} &= 0,  
\label{eq:divb}
\end{align}
where $\vek v$, $\rho$, $\eta$, and
${\vek b} \equiv {\vek B}/\sqrt{4 \pi}$ are the fluid velocity, the
density, a uniform resistivity, and the redefined magnetic field
${\vek B}$, respectively. The total energy density, $e_{\star}$, is
composed of fluid and magnetic contributions, \ie
$e_{\star} = \varepsilon + \frac{1}{2} \rho {\vek v}^2 + \frac{1}{2}
{\vek b} ^ 2$
with the internal energy density $\varepsilon$ and the gas pressure
$p = p(\rho, \varepsilon, \dots)$.  The stress tensor ${\vek T}$ is
given by
\begin{equation}
  {\vek T} = \left[ P + \mathsmaller{\half} {\vek b}^2 
                      + \rho \left( \mathsmaller{\frac{2}{3}} \nu - \xi\right)
                    \divv {\vek v} \right] {\vek I} 
             - {\vek b} \otimes {\vek b} 
             - \rho \nu \left[ \nabla {\vek v} + (\nabla {\vek v})^T \right], 
\end{equation}
where ${\vek I}$ is the unit tensor, and $\nu$ and $\xi$ are the
kinematic shear and bulk viscosity, respectively.

\subsection{Magnetorotational instability}
\label{sSec:intability_criterion}

We study the MRI in a small portion of the rotating star at a given
distance $r$ from the rotation axis, embedded in a magnetic field. For
convenience, we use cylindrical coordinates $(r,\phi, z)$, hereafter.
We restrict our analysis to locations close to the equatorial plane
($z=0$) and vertical perturbation wavevectors for which the MRI is
known to develop fastest \citep[see,
e.g.][]{Balbus_Hawley__1998__RMP__MRI}. In this case, we can consider
a differentially rotating fluid with angular velocity $\Omega$ and
linear velocity
\begin{equation}
  {\vek v}=  \Omega r {\bm{ \hat{\phi}}},  
\label{eq:v_init}
\end{equation}
threaded by a uniform vertical magnetic field
\begin{equation}
  {\vek b} = \bz {\hat{\vek{z}}} 
\label{eq:b_init}
\end{equation}
in the local perturbation analysis. Here, ${\bm{\hat{\phi}}}$ and
$\hat{\vek{z}}$ are the unit vectors in $\phi$ and $z$ direction,
respectively.

With these assumptions and ignoring dissipative effects, the MRI
instability criterion is
\citep[c.f.][]{Balbus__1995__ApJ__stratified_MRI}
\begin{equation}
    N^2 + r \partial_{r} \Omega ^2  = N^2 + \kappa^2 - 4 \Omega^2< 0,
\label{eq:mri_instability}
\end{equation}
where
\begin{eqnarray}
  N^2 &=& \frac{\partial_r P}{\rho} 
          \left (\frac{\partial_r \rho}{\rho} - 
                 \frac{\partial_r P}{\Gamma_1 P} \right), 
\nonumber \\
  \kappa^2 &=& \frac{1}{r^3} \partial_r (r^4 \Omega^2),
\end{eqnarray}
are the square of the Brunt-V\"ais\"al\"a frequency and the epicyclic
frequency, respectively, and $\Gamma_1$ is the adiabatic index.

We consider an angular velocity with a radial dependence of the form
\begin{equation}
  \Omega = \Omega_0 \left( \frac{r}{r_0} \right)^{-q},
\label{eq:omega}
\end{equation}
where $\Omega_0$ is the angular velocity at the characteristic radius
$r_0$, and $q$ is the local rotational shear given by
\begin{equation}
  q = - \frac{ \der \ln \Omega}{\der \ln r}. 
\label{eq:q}
\end{equation}
The rotation profile \eqref{eq:omega} is quite generic for
astrophysical systems, e.g.\ for $q = 3/2$, one recovers the Keplerian
profile and for differentially rotating stars typical values of $q$
are in the range $0< q < 3/2$. The corresponding epicyclic frequency
(\ie the radial oscillation frequency) is
\begin{equation}
  \kappa = \sqrt{2(2-q)} \Omega,
\label{eq:kappa}
\end{equation}
which vanishes at $q=2$ (Rayleigh stability criterion limit for
unmagnetised rotating fluids).  We also assume that the entropy and
composition are constant within the simulated volume, and hence
$N^2=0$. The influence of entropy gradients on the MRI has been
studied by \citet{Obergaulinger_et_al_2009} and will not be discussed
here.

If condition~(\ref{eq:mri_instability}), is fulfilled, \ie
$0< q < 2$, any perturbation of the form
$e^{i k_z z + \gammamrismall t}$ (WKB ansatz) is unstable for
wavenumbers
\begin{equation}
  k_z < k_{\mathrm{crit}} =  \sqrt{2q} \frac{\Omega}{\caz},  
\label{eq:mri_kcrit}
\end{equation}
where $\caz\equiv b_{0z}/\sqrt{\rho}$ is the Alfv\'en speed in the
vertical direction.

BH91 considered small perturbations of the velocity, $\bfvc$, and the
magnetic field, $\bfbc$, in a system whose velocity and magnetic field
are given by \eqsref{eq:v_init} and (\ref{eq:b_init}), respectively,
\ie
\begin{align}
  {\vek v} &=  \Omega r  {\bm{ \hat{\phi}}} +  \bfvc, \\
  {\vek b} &=  b_{0z} {\vek{\hat{z}}} +   \bfbc.
\end{align}
We note that one must also  introduce  appropriate perturbations of the
other thermodynamical quantities to fulfil the MHD \eqsref{eq:cont} -
(\ref{eq:divb}), but we do not mention them here explicitly.

For the linearized ideal MHD equations in the incompressible limit,
BH91 found unstable solutions, which are usually referred to as {\it
  MRI channels},
\begin{align}
  \bfvc(t;\kmrismall) & = \tildevc e^{\gammamrismall t} 
               ( {\bf \hat{r}} \cos \phi_v + {\bm{ \hat{\phi}}} \sin \phi_v )
               \sin (\kmrismall z) , 
\label{eq:channel_v} \\
  \bfbc(t;\kmrismall) & = \tildebc e^{\gammamrismall t} 
               ( {\bf \hat{r}}  \cos \phi_b  + {\bm{ \hat{\phi}}} \sin \phi_b )   
               \cos (\kmrismall z)  
\label{eq:channel_b},  
\end{align}
where the subscript \emph{c} stands for \emph{channel}, $\bm{\hat{r}}$
is the unit vector in $r$ direction, $\tildevc$ and $\tildebc$ are the
initial amplitudes, $\phi_v $ and $\phi_b$ are the angles between the
$r$-axis and the direction of the velocity and magnetic field
channels, respectively. The wavenumbers $\kmrismall$ correspond to
values of $k_z$ fulfilling Eq.~(\ref{eq:mri_kcrit}).  To simplify the
notation, we define
\begin{align}
  \vc(t)  &= \tildevc  e^{\gammamrismall t}, \\
  \bc(t)  &= \tildebc  e^{\gammamrismall t},
\end{align}
and for brevity, we will often drop the explicit time dependence,
\ie $\vc = \vc(t)$ and $\bc = \bc(t)$.  GX94 generalized the results
of BH91, and they showed that the MRI channels are an exact solution
of the ideal incompressible MHD equations in the \emph{shearing sheet}
(local) approximation.  This approximation consists in transforming
the equations to a frame corotating with a fiducial fluid element and
linearising the rotational profile around a radius $r_0$, \ie
$\Omega(r) \approx (r-r_0) \partial_r \Omega(r)|_{r_0}$.  In this
frame, the gravitational force and the centrifugal force balance each
other for initially Keplerian accretion discs, but the Coriolis
force has to be taken into account. In differentially rotating stars,
additional pressure gradients are necessary to provide an equilibrium.

In the ideal MHD limit, the MRI growth rate and the channel angles,
$\phi_v $ and $\phi_b$, are given by
\begin{equation}
  \gammamrismall = \Omega \sqrt{ \sqrt{ (2-q)^2  
                   +  8q 
                        \left( \frac{\kmrismall}{k_{\mathrm{crit}}} \right)^2 } 
                   - (2-q) 
                   -  2q 
                        \left( \frac{\kmrismall}{k_{\mathrm{crit}} }\right)^2 }
\label{eq:gr_ideal}
\end{equation}
and
\begin{align}
  \phi_v & = \arctan \left[ \frac{ \caz^2 \kmrismall^2 + \gammamrismall^2 
                                 }{2 \gammamrismall \Omega_0 } \right], \label{eq:v_b_pi1} \\
  \phi_b & = \phi_v + \frac{\pi}{2}, 
\label{eq:v_b_pi} 
\end{align}
respectively. The amplitude ratio $\tildevc / \tildebc$ is a function
of $\kmrismall$ and $q$ \citep[cf.][PC08 hereafter]{Pessah_Chan}.  The
mode with the wavenumber
\begin{equation}
  \kmri = \sqrt{1 - \frac{(2-q)^2}{4}} \frac{\Omega}{\caz}
\label{eq:kmri}
\end{equation}
grows fastest at a rate
\begin{equation}
  \gammamri = \frac{q}{2} \Omega.
\label{eq:gammamri}
\end{equation}
Note that we use capital letters in the subscripts to refer to the
fastest-growing mode ($\gammamri$ and $\kmri$) to distinguish it from
generic unstable modes with growth rates
$\gammamrismall \le \gammamri$.

For the fastest-growing mode, the magnetic field and the velocity
amplitudes are related by
\begin{equation}
  v_{\mathrm{c}} = \sqrt{\frac{q}{4-q}} \cac
\label{eq:amplitude_ratio},
\end{equation}
where $\cac \equiv \bc/\sqrt{\rho}$ is the Alfv\'en speed parallel to
the MRI channel, and the channel angles are $\phi_v = \pi/4$ and
$\phi_b= 3\pi/4$.

The properties of MRI modes change when viscosity or resistivity are
present in the system.  PC08 generalized the results of GX94 and
showed that MRI channels (Eqs.\ (\ref{eq:channel_v}) and
(\ref{eq:channel_b})) are also exact solutions of the
resistive-viscous incompressible MHD equations in the shearing sheet
approximation.  They derived expressions for the growth rate
$\gammamrismall$, the amplitude ratio $\vc / \bc$, and the channel
angles $\phi_v$ and $\phi_b$ of MRI-unstable modes for arbitrary
hydrodynamic and magnetic Reynolds numbers. Following PC08, we define
these two (dimensionless) numbers as
\begin{align}
  \Ree &= \frac{\caz^2}{\nu \Omega},  
\label{eq:Re}  \\
  \Rm &= \frac{\caz^2}{\eta \Omega}.
 \label{eq:Rm}
\end{align}
We note that in the case of a zero entropy gradient the (most general)
dispersion relation of \cite{Menou_etal__2004__ApJ__MRI-stability}
reduces to the one analysed by PC08.  Since we limit our studies to
isentropic models, we can directly apply their results.

In resistive-viscous MHD, the channel angles, $\phi_v$ and $\phi_b$,
are given by
\begin{align}
  \phi_v & = \arctan \left[  
             \frac{\caz^2 \kmrismall^2 + 
                   (\gammamrismall + \kmrismall^2 \eta) 
                   (\gammamrismall + \kmrismall^2 \nu) 
                 }{2 \Omega_0 (\gammamrismall + \kmrismall^2 \eta)   } \right],
\label{eq:phi_v}  \\
\label{eq:phi_b}
  \phi_b & = \arctan \left[  
             \frac{- \Omega_0 \{  2 
                   (\gammamrismall + \kmrismall^2 \eta) + 
                   q (\nu - \eta) \kmrismall^2 \} 
                 }{\caz^2 \kmrismall^2 + 
                   (\gammamrismall + \kmrismall^2 \eta) 
                   (\gammamrismall + \kmrismall^2 \nu) } \right]
\end{align}
when transforming Eqs.~(50) and (51) of PC08 into dimensional units.
In general, velocity channels and magnetic channels are not
orthogonal. However, it is evident that if $\nu=\eta$ (\ie when the
magnetic Prandtl number $\Pm \equiv \Rm / \Ree=1$), \Eqref{eq:phi_b}
reduces to \Eqref{eq:v_b_pi}, so that the shift of $\pi/2$ between
$\phi_v$ and $\phi_b$ may also hold in the non ideal MHD limit.

As the expressions for the MRI growth rate $\gammamrismall$ and the
amplitude ratio $\vc / \bc$ are quite complex, we do not give them
here (the interested reader will find the details in PC08).  Instead,
we briefly discuss some of their key physical aspects.

Both resistivity and viscosity reduce the MRI growth rate. They also
shift both the critical wavenumber $k_{\mathrm{crit}}$ and the most
unstable wavenumber $\kmri$ to lower values than those given for ideal
MHD by Eq.~(\ref{eq:mri_kcrit}) and (\ref{eq:kmri}), respectively.
This behaviour can be readily understood, as dissipative effects are
more pronounced in the resistive-viscous regime (as they scale
$\propto k_z^2$).  The smaller the Reynolds numbers, the larger the
growth rate reduction and the larger the shifts.  Moreover, the
magnetic Reynolds number has a bigger influence than the hydrodynamic
one, because the nature of the MRI is more ``magnetic'' than
``hydrodynamic''.  For any Reynolds number, the MRI growth rate
$\gammamrismall (k_z)$ tends to zero for sufficiently short
wavevectors, \ie $\gammamrismall \rrr 0$ for $k_z \rrr 0$ (see
Eq.\,\ref{eq:gr_ideal}).

PC08 found analytic expressions for the growth rate $\gammamri$, and
the wavevector $\kmri$ of the fastest-growing mode only in some
limiting cases, usually for very large or very small Reynolds numbers
(e.g.\ for $\Rm \gg 1$ and $\Ree \ll 1$).  We note that for
$\Ree, \Rm \geq 10$ (which holds for core collapse supernovae
outside the neutrinosphere) the ideal MHD expressions given in
Eqs.\ (\ref{eq:gammamri}) and (\ref{eq:kmri}) determine the values of
the wavenumber and the growth rate of the fastest-growing mode with
relative errors $\leq 10\%$ compared to the values obtained by the
expressions of PC08.

The high conductivity of the degenerate matter in a supernova core
implies very high magnetic Reynolds numbers.  Using an
order-of-magnitude estimate of
$\eta \sim 10^{-4} \, \cm^2 \mathrm{s}^{-1}$
\citep{Thompson_Duncan__1993__ApJ__NS-dynamo}, Eq.~(\ref{eq:Rm})
yields
\begin{equation}
\label{eq:Rm-PNS}
  \Rm = 10^{13}
        \left( \frac{b_{0z}}{10^{13} \, \mathrm{G}} \right)^2
        \left( \frac{10^{14} \, \gccm}{\rho} \right)
        \left( \frac{10^{3} \, \mathrm{s}^{-1}}{\Omega} \right)
        \left( \frac{10^{-4} \, \cm^2 \mathrm{s}^{-1}}{\eta} \right)
     .  
\end{equation}
Because the magnetic Reynolds number is so large, we can safely
neglect resistivity in the expressions for the growth rate and the
wavelength of the MRI.  

The molecular viscosity of supernova matter
($\nu \sim 0.4$\,cm\,$^2$\,s\,$^{-1}$) is only a few orders of
magnitude larger than the resistivity, \ie the respective
hydrodynamic Reynolds number $\Ree \sim  2.5 \times 10^9 \gg 1$
\citep{Thompson_Duncan__1993__ApJ__NS-dynamo}.  In the region below
the neutrinosphere, however, the tight coupling between neutrinos and
matter
changes the situation. According to \citet{Guilet_2015} the
  neutrino-matter interaction results in an effective viscosity that
  varies between a few $10^{9}\, \cm^2 \mathrm{s}^{-1}$ deep inside
  the PNS and $10^{12}\, \cm^2 \mathrm{s}^{-1}$ near the
  neutrinosphere, \ie one finds
\begin{equation}
\label{eq:Re-PNS}
  \Ree = 0.1
         \left( \frac{b_{0z}}{10^{13} \, \mathrm{G}} \right)^2
         \left( \frac{10^{14}\, \gccm}{\rho} \right)
         \left( \frac{10^{3} \, \mathrm{s}^{-1}}{\Omega} \right)
         \left( \frac{10^{10} \, \cm^2 \mathrm{s}^{-1}}{\nu_{\nu}} 
         \right).
\end{equation}
 The above equation implies that the expressions for the growth
  rate and for the angles $\phi_b$ and $\phi_v$
  (Eqs.~\ref{eq:v_b_pi1}-\ref{eq:gammamri}), which are valid in the
  limit of ideal MHD, are not satisfied inside the PNS.
 Outside the neutrinosphere, the interaction between matter and
  neutrinos changes its character from diffusion to free streaming.
  In the free streaming regime neutrino drag damps fluctuations not
  like an effective viscosity, but like a drag force that is
  independent of the wavelength. Its impact on the MRI is therefore
  very different from the impact of a viscosity
  \citep{Guilet_2015}. Neutrino drag is relevant not only outside but
  also significantly below the neutrinosphere as long as the dynamics
  of interest is happening at a wavelength shorter than the neutrino
  mean free path, which varies from a metre inside the PNS to several
  kilometre near its surface \citep[see Fig.\,5 in][]{Guilet_2015}.
Because of these complications we will not investigate this effect
here and refer the reader to the studies of
\cite{Masada_etal__2007__ApJ__MRI-PNS-neutrino,
  Masada_et_al__2012__apj__LocalSimulationsoftheMagnetorotationalInstabilityinCore-collapseSupernovae}
and \cite{Guilet_2015}.

\subsection{MRI termination}

\subsubsection{Termination scenarios}

MRI channels cannot grow indefinitely, because their energy would
constantly increase, whereas the energy of the system contained in
differential rotation, which provides the reservoir for the MRI, is
finite. Hence, there must be a physical mechanism terminating MRI
growth.

GX94 suggested that MRI channels being exact solutions of the MHD
equations (in the shearing sheet and incompressible flow
approximations) may be unstable against parasitic instabilities, which
could terminate the MRI growth.  They found in their analytic
calculations (under the assumptions described in
Sec.~\ref{sec:termi_para}) that in ideal MHD (shear driven) KH modes
can develop on top of MRI channels.  GX94 also suggested that in
resistive MHD, parasitic instabilities of the (current driven) TM type
could develop, too.  We note that the importance of magnetic
reconnection for MRI termination was already discussed by BH91.
Analytic calculations by \citet{Latter_et_al} 
in resistive  MHD
confirmed this hypothesis.  
Alternatively, if this scenario does not hold, the magnetic
field of the MRI channels could grow to a dynamically relevant
strength (when the \alf speed becomes comparable to the sound speed),
violating the approximation of incompressibility.  Consequently, the
magnetic field pressure would become important and it could push
matter towards magnetic null surfaces of the MRI channels, or MRI
channels could become buoyancy unstable (GX94).
Finally, small amplitude MRI channels emerging in an already
  (MRI-driven) turbulent state could be destroyed by non-linear
  mode-mode interactions or turbulent mixing \citep{Latter_et_al}.

In this paper, we only consider the first scenario, \ie MRI
termination via parasitic instabilities.  In
Sec.~\ref{sec:termi_para}, we will briefly discuss assumptions and
findings of the parasite model of \citet{Goodman_Xu} and of its
extension to resistive-viscous MHD by \citet{Pessah}.

\subsubsection{Termination via parasitic instabilities}
\label{sec:termi_para}

GX94 considered perturbations in a system with already well developed
MRI channels (of the fastest-growing mode with $\kmri$)  given by
\begin{align}
  {\vek v} &= -q \Omega_0 (r - r_0) {\bm{ \hat{\phi}}}
              + \bfvc(t;\kmri) + \bfvp(r,\phi,z,t)  ,  
\label{channel_full_p1}\\
  {\vek b} &=  b_{0z} {\vek{\hat{z}}} 
              + \bfbc(t;\kmri) +  \bfbp(r,\phi,z,t) ,
\label{channel_full_p2}
\end{align}
where $\bfvp$ and $\bfbp$ are the velocity and the magnetic field of
the parasitic instabilities, respectively.

Solving the equations governing the evolution of the secondary (parasitic)
instabilities is a very challenging task, because MRI channels, which
are treated as a background field for the perturbations, are
non-stationary. Hence, standard techniques like an WKB ansatz cannot
be used. To make this task more tractable for analytic studies, GX94
considered a stage of MRI growth when the amplitude of the MRI
channels is much larger than the initial weak magnetic field, \ie
$ \bc \gg b_{0z}$.  The growth rate of the secondary instabilities
$\gammap$ (which scales $\propto \bc$) is then much larger than the
MRI growth rate, \ie $\gammap \gg \gammamri$. Under these conditions
the time evolution of the MRI channels, the Coriolis force (which is
of the order of $\gammamri$), the background shear flow, and the
initial background magnetic field $b_{0z}$ can be neglected.  Hence,
instead of searching for solutions to perturbations according to Eqs.\
(\ref{channel_full_p1}) and (\ref{channel_full_p2}), GX94 considered a
more simplified system where the velocity and the magnetic field are
given by
\begin{align}
  {\vek v}(t) &=  \bfvc(t_0;\kmri)+ \bfvp(r,\phi,z,t) , 
\label{eq:channel_stationary_v} \\
  {\vek b}(t) &= \bfbc(t_0;\kmri)+ \bfbp(r,\phi,z,t),  
\label{eq:channel_stationary_b}
\end{align}
with $t_0= \mathrm{const.}$ being the time at which the secondary
perturbations are imposed. Similar assumptions were also made by
\citet{Latter_et_al} and \citet{Pessah}. 

\cite{Pessah} identified regions in parameter space, where depending
on the values of the hydrodynamic and magnetic Reynolds numbers either
KH or TM is the dominant (\ie faster developing) secondary instability
that terminates MRI growth.  In particular, for the conditions
prevailing in core collapse supernovae outside the PNS
($\Ree, \Rm \gg 1$) the exponential growth phase of the MRI should be
terminated by KH instabilities.  TM should be dominant only in very
resistive media, \ie if $\Rm \lesssim 1$.  \cite{Pessah} also found
that, in general, (shear driven) KH modes grow fastest along the MRI
velocity channels, $\phikh = \phi_v$, whereas (current driven) TM grow
fastest along the magnetic field channels, $\phitm = \phi_b$.  One of
the aims of this work is to test these predictions through numerical
simulations.

%%%%%%%%%%%%%%%%%%%%%%%%%%%%%%%%%%%%%%%%%%%%%%%%%%%%%%%%%%%%%%%%%%%%%
%%%%%%%%%%%%%%%%%%%%%%%%%%%%%%%%%%%%%%%%%%%%%%%%%%%%%%%%%%%%%%%%%%%%%
\section{Method}
\label{sec:numerics}
%
%%%%%%%%%%%%%%%%%%%%%%%%%%%%%%%%%%%%%%%%%%%%%%%%%%%%%%%%%%%%%%%%%%%%%%

\subsection{Code}
%\label{sSec:Code}

We use the three-dimensional Eulerian MHD code \textsc{Aenus}
\citep{Obergaulinger__2008__PhD__RMHD} to solve the MHD equations
(\ref{eq:cont})--(\ref{eq:divb}).  The code is based on a
flux-conservative, finite-volume formulation of the MHD equations and
the constrained-transport scheme to maintain a divergence-free
magnetic field \citep{Evans_Hawley__1998__ApJ__CTM}.  Using
high-resolution shock-capturing methods
\citep[e.g.\,][]{LeVeque_Book_1992__Conservation_Laws}, the code
employs various optional high-order reconstruction algorithms
including a total-variation-diminishing piecewise-linear (TVD-PL)
reconstruction of second-order accuracy, a third-, \mbox{fifth-,}
seventh- and ninth-order monotonicity-preserving (MP3, MP5, MP7 and
MP9, respectively) scheme \citep{Suresh_Huynh__1997__JCP__MP-schemes},
a fourth-order, weighted, essentially non-oscillatory (WENO4) scheme
\citep{Levy_etal__2002__SIAM_JSciC__WENO4}, and approximate Riemann
solvers based on the multi-stage (MUSTA) method
\citep{Toro_Titarev__2006__JCP__MUSTA}.  
We add terms including viscosity and resistivity to the flux terms in
the Euler equations and to the electric field in the MHD induction
equation.  We treat these terms similarly to the fluxes and electric
fields of ideal MHD, except for using an arithmetic average instead of
an approximate Riemann solver to compute the interface fluxes. The
explicit time integration can be done with Runge-Kutta schemes of
first, second, third, and fourth order (RK1, RK2, RK3, and RK4),
respectively.

Choosing appropriate numerical schemes for our 3D simulations is an
important issue, because we want to keep the numerical viscosity and
resistivity as low as possible without unnecessarily increasing the
computational cost.  Therefore, before applying the code to the MRI,
we assessed its numerical viscosity and resistivity in an extended set
of auxiliary simulations (Rembiasz et al., in preparation).  We performed
test calculations of linear-wave propagation and the TM instability,
comparing the numerical solutions with (semi-) analytic solutions.
Our tests show that the very low numerical dissipation of the MP9
scheme well justifies its larger stencil (requiring more ghost zones).
In the TM simulations, the main contribution to the numerical
dissipation comes from the spatial rather than the temporal
discretisation errors. We do not observe any gain in accuracy when
using the RK4 instead of the RK3 scheme. Because we expect these
findings to hold also for MRI simulations, we performed the
simulations reported here with the MP9 scheme, a MUSTA solver based on
the HLLD Riemann solver, and an RK3 time integrator
\citep[]{Harten_JCP_1983__HR_schemes,HLLD}.

%%%%%%%%%%%%%%%%%%%%%%%%%%%%%%%%%%%%%%%%%%%%%%%%%%%%%%%%%%%%%%%%%%%%%%
\subsection{Equation of state}

We use the hybrid equation of state (EOS) of
\citet{Keil_Janka_Mueller__1996__ApJL__NS-Convection}, in which the
gas pressure $P$ results from the addition of two contributions,
namely a baryonic one $P_{\mathrm{b}}$ and a thermal one
$P_{\mathrm{th}}$. These pressure contributions are given by
\begin{align}
  P_{\mathrm{b}}  &= K \rho^{\Gamma_{\mathrm{b}}}, \\
  P_{\mathrm{th}} &= (\Gamma_{\mathrm{th}}-1) e_{\mathrm{th}}, 
\end{align}
where $K = 4.897 \times 10^{14}$ is the polytropic constant, and
$\Gamma_{\mathrm{b}}=1.31$ and $\Gamma_{\mathrm{th}}=1.5$ are the
barotropic index and the thermal adiabatic index, respectively.  The
quantity $e_{\mathrm{th}}$ is the thermal part of the internal energy
$e$, \ie
$e_{\mathrm{th}} = e - P_{\mathrm{b}}/(\Gamma_{\mathrm{b}}-1)$.

%%%%%%%%%%%%%%%%%%%%%%%%%%%%%%%%%%%%%%%%%%%%%%%%%%%%%%%%%%%%%%%%%%%%%%
\subsection{Computational grid and boundary conditions}
\label{sSec:BC}

Our study comprises a set of two-dimensional axisymmetric simulations
and a set of three-dimensional simulations. For both kinds of
simulations we employed cylindrical coordinates $(r, \phi, z)$ and a
computational domain centred around the equatorial plane at a radius
$r_0 = 15.5\,\km$. 
For this value of $r_0$ our computational box is located in the
  middle of a nascent PNS of radius
  $r_{\mathrm{PNS}} \approx 30\,\km$. 
However, \cite{Guilet_2015} have recently shown that the viscosity
  due to neutrinos can be much higher in collapsing cores than
  previously thought.  Close to the neutrinosphere, \ie whenever
  the neutrino mean free path is shorter than the length scale of
  interest, very low Reynolds numbers are expected. The most
  favuorable place for MRI amplification is located close to the PNS
  star surface, where differential rotation is stronger, and
  the Reynolds numbers are larger if the surface is located above the
  neutrinosphere, which however may not always be the case \citep[see
  Fig.\,10 in][]{Guilet_2015}. 
However this recent finding does not invalidate our studies as
 our models can be
  easily scaled for different initial conditions.  We may choose
  different values of the radius, $r$, rotational velocity, $\Omega$,
  or density, $\rho$, and scale the key physical quantities in
  following way:
\begin{align}
  \label{eq:1}
  \MMM &= \bar{\mathcal{M}}_{r\phi} 
          \left( \frac{r}{r_0} \right)^2 
          \left( \frac{\Omega}{\Omega_0} \right)^2  
          \left( \frac{\rho}{\rho_0} \right),
\\
  b   &= \bar{ b } \left( \frac{r}{r_0} \right)  
                   \left( \frac{\Omega}{\Omega_0} \right) 
                   \left( \frac{\rho}{\rho_0} \right)^{1/2},
\\
  t   &= \bar{ t }  \left(  \frac{\Omega}{\Omega_0} \right)^{-1},
\\
  \gammamri  &= \bar{ \gamma}_{\mathsmaller{\mathrm{MRI}}}  
                \left(  \frac{\Omega}{\Omega_0} \right),
\\
  \lambdamri &= \bar{\lambda}_{\mathsmaller{\mathrm{MRI}}} 
                \left( \frac{r}{r_0} \right),
\\
L_i &= \bar{L}_i\left( \frac{r}{r_0} \right), 
\end{align}
where $\Omega_0 = 1824\,\s^{-1}$, and
$\rho_0 = 2.47 \times 10^{13}\,\gccm$, 
$L_i$ is the box length in the direction $i$ (where $i = r,\phi,z$), and the barred quantities are
the values set up in or computed from our simulations.  
We stress
again that the most favorable place for the development of the MRI is
the region close to the surface of the PNS, where the differential
rotation gradient and the Reynolds numbers are larger than deep inside
the PNS.
Since the choice of $\Omega_0$ is ad hoc, it is evident that the
  ratio $\Omega/\Omega_0$ can be made as close as desired to one. This
  means that neither the MRI growth rate nor the typical growth time
  will change under a translation of the box to the surface of the
  PNS. Such a translation will also imply that $r/r_0\sim 2$, while
  $\rho/\rho_0\sim 0.1$, hence, the typical magnetic fields and
  Maxwell stresses would be 0.6 and 0.4 times smaller than in our
  models. However, as we will show, none of these variations would
  change the foremost qualitative prediction of our models, namely,
  that the termination of the MRI growth in collapsing stellar cores
  happens by the action of parasitic KH modes.

In the 3D simulations, the typical box size is
$L_r \times L_\phi \times L_z = 1\, \km \times 4\, \km \times 1\, \km$
in $r, \phi$, and $z$ direction, while the typical box size is
$1\, \km \times 1\, \km$ in the $(r, z)$ plane in our 2D
simulations. We performed simulations involving up to
$200 \times 800 \times 200$ zones (see Table\,\ref{tab:main_results}).

We assume periodic boundary conditions in both $\phi$ and
$z$-direction.  This choice is natural for the $\phi$ direction,
whereas in the vertical direction, $z$, the core is obviously not
periodic. However, for the simulated region near the equatorial plane
the vertical component of the gravitational force, $F_{\mathrm{g}z}$,
can be neglected, because it is much smaller than the radial one,
\ie $F_{\mathrm{g}z}/F_{\mathrm{g}r} \la 0.03$.

Unlike in local simulations of accretion discs, we cannot use the
shearing sheet boundary condition
\citep[see][]{Balbus_Hawley__1992__ApJ__MRI_3} in the radial
direction, because it does not allow for global gradients of
thermodynamic variables (which are present in core collapse
supernovae) in the simulation domain.  Therefore, like
\citet{Obergaulinger_et_al_2009} \citep[who
followed][]{Klahr_Bodenheimer__2003__ApJ__Global-baroclinic-inst-disc},
we use the shearing disc boundary condition, which allows one to take
these gradients into account.

We solve the full compressible MHD equations and do not perform a
transformation to the frame corotating with the fluid, which
discriminates our simulations from the common shearing-box approach.
In radial direction, we apply periodic boundary conditions to the
deviation of a variable (here, density) from its initial background
state, e.g.\
\begin{equation}
  \delta \rho(r,t) \equiv \rho(r,t) - \rho(r,0),
\label{eq:deltarho}
\end{equation}
\ie we enforce periodicity of the perturbations.  We apply
  these boundary conditions to angular velocity, density, momentum,
and entropy.  Because the initial magnetic field is homogeneous
  in all our simulations, we use periodic boundary conditions for this
  quantity too.

%%%%%%%%%%%%%%%%%%%%%%%%%%%%%%%%%%%%%%%%%%%%%%%%%%%%%%%%%%%%%%%%%%%%%%
\subsection{Initial conditions}
\label{sSec:Init}

Like \citet{Obergaulinger_et_al_2009}, we use equilibrium initial
models based on the final stages of post-bounce cores from
\cite{Obergaulinger_Aloy_Mueller__2006__AA__MR_collapse}, in which
(several tens of milliseconds after core bounce) the shock wave has
reached distances of a few hundred kilometres and the post-shock
region exhibits a series of damped oscillations as the PNS relaxes
into a \emph{nearly} hydrostatic configuration.

\begin{figure}%[t]
%mnrs  \sidecaption
\includegraphics[width=1.0\linewidth]{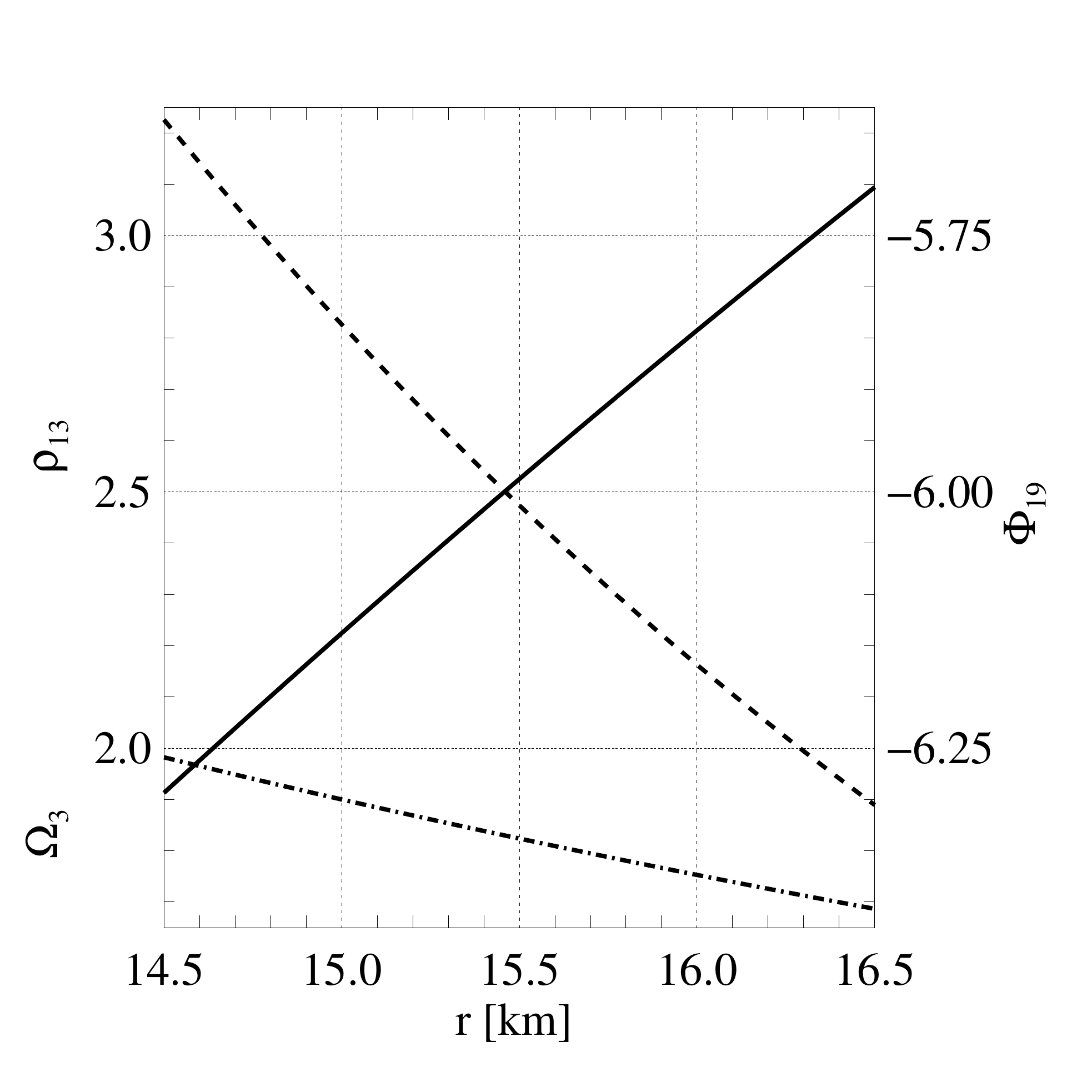}
\caption{Hydrostatic structure of the initial models.  The diagram
  shows the gravitational potential
  $\Phi_{19} = \Phi / (10^{19}~\mathrm{erg}~\mathrm{cm}^{-3})$ (solid
  line, right ordinate), the density
  $\rho_{13} = \rho / (10^{13} ~ \mathrm{g}~\mathrm{cm}^{-3})$ (dashed
  line, left ordinate), and the angular velocity
  $\Omega_3 = \Omega / ( 10^3 ~ \mathrm{s}^{-1})$ (dash-dotted line,
  left ordinate).  The entropy profile of this specific model is
  assumed to be flat. }
\label{fig:mri_init}
\end{figure}

The rotational profile (given by Eq.\,\ref{eq:omega} with $q = 1.25$) that we used in
our simulations, is similar to the one employed in the global MRI
simulations of
\citet{Obergaulinger_Aloy_Mueller__2006__AA__MR_collapse}.  Because
the resulting centrifugal force is insufficient to balance gravity,
the gas is kept in (an initial hydrostatic) equilibrium by an
additional pressure gradient, so that
\begin{equation}
  \rho \partial_r \Phi  -\partial_r P + r \rho \Omega^2 = 0.
\label{eq:equilibrium}
\end{equation}
The initial distributions of angular velocity, 
density, and gravitational potential are depicted in Fig.\
\ref{fig:mri_init}.

%%%%%%%%%%%%%%%%%%%%%%%%%%%%%%%%%%%%%%%%%%%%%%%%%%%%%%%%%%%%%%%%%%%%%%%%%
%%%%%%%%%%%%%%%%%%%%%%%%%%%%%%%%%%%%%%%%%%%%%%%%%%%%%%%%%%%%%%%%%%%%%%%%%

Unless otherwise stated, we set the initial magnetic field strength to
$b_{0z} = 4.6 \times 10^{13}\, \mathrm{G}$, the shear and bulk
viscosity to $\nu = \xi =0\, \mathrm{cm}^2\,\mathrm{s}^{-1}$,
and the resistivity to
$\eta = 4.45 \times 10^{8}\, \mathrm{cm}^2\,\mathrm{s^{-1}} $,
which implies a Reynolds number $\Ree=\infty$ and 
  $\Rm \approx 100$, respectively.  The Reynolds numbers slightly vary
  in the box from $\Rm = 89 \tto 125 $, because both the Alfv\'en
  speed and the rotational velocity are functions of radius (see
  Fig.\,\ref{fig:mri_init}, and Eqs.\,\ref{eq:Re} and \ref{eq:Rm}).
For these default parameters, the wavelength of the most unstable MRI
mode
 ranges from $\lambdamri = 0.314 \tto 0.385\, \mathrm{km}$, and
the corresponding velocity channels and the magnetic field channels
form (for a rotational shear $q = 1.25$) at an angle,
$\phi_v = 44.4 \tto 44.6\, ^{\circ}$ and
$\phi_b = 134.7 \tto 134.8^{\circ}$, respectively (see
\eqsref{eq:phi_v} and (\ref{eq:phi_b})). These angles differ from
those in ideal MHD ($\phi_v = 45^{\circ}$ and $\phi_b = 135^{\circ}$)
only very little.

In all simulations presented below the initial magnetic field has only
a uniform component in $z$ direction, as defined in
\Eqref{eq:b_init}. This field geometry is a popular choice in MRI
simulations, \citep[see, e.g.\
BH91;][]{Balbus_Hawley__1991__ApJ__MRI_2,
  Sano_Inutsuka__2001__ApJL__MRI-recurrent-channels,
  Obergaulinger_et_al_2009}, since the vertical component is the most
important one for the development of the instability
\citep[cf.][]{Balbus_Hawley__1998__RMP__MRI}.  Another common choice
is a so-called \emph{zero net flux} configuration \citep[see,
e.g.][]{Fromang_etal__2007__AA__3d-local-MRI-disc-zero-net-flux_2,
  Fromang_Papaloizou__2007__AA__3d-local-MRI-disc-zero-net-flux_1,
  Obergaulinger_et_al_2009}, in which the magnetic field has a
sinusoidal radial dependence, \ie
${\vek b} \propto {\vek{\hat{z}}} \sin(k_r r)$, where $k_r$ is chosen
in such a way that an integer number of wavelengths fits the
computational domain. Thus, $k_r = 2 \pi n/L_r$ with $n$ being a
natural number.

The value of the initial magnetic field amplitude, $b_{0z}$, requires
some further comments.  According to state-of-the-art stellar
evolution calculations the pre-collapse magnetic field for the most
strongly magnetised progenitors is less than about $10^9$\,G
\citep{Heger_2005}.  During the collapse phase the magnetic field can
be amplified by compression by two orders of magnitude to
$\approx 10^{11}$\,G \citep{Meier_etal__1976__ApJ__MHD_SN}.  From Eq.\
(\ref{eq:kmri}), which is valid in ideal MHD, we estimate that for
the PNS the wavelength of the fastest-growing MRI mode is
\begin{equation}
  \lambdamri \approx 70\ \mathrm{cm} 
                     \left( \frac{b_{0z}}{10^{11}\ \mathrm{G}} \right) 
                     \left( \frac{\rho}{2.5 \times 10^{13}\ 
                            \mathrm{g\ \cm}^{-3} } \right)^{-1/2} 
                     \left( \frac{\Omega}{1900 \ \mathrm{s}^{-1} } \right)^{-1}.
\end{equation}
For the typical Reynolds numbers used in our simulations
($\Ree = \Rm \approx 100$) the wavelength of the fastest-growing mode
is $\approx 1.5\%$ longer (and the growth rate
$\approx 1.6\%$ lower) than the one in ideal MHD.
\footnote{We obtained these values by plotting
  $\gammamrismall(\kmrismall)$ (for given $\Omega, q, \caz, \eta$, and
  $\nu$) using the expression from PC08 and determining the maximum
  $\gammamri$ and its location, $\kmri$, graphically.}
Assuming that $10$ zones per MRI channel are needed to resolve it
properly, a simulation with $b_{0z} =10^{11}\,$G would require a
resolution of the order of $10^5$ zones per dimension, which is
already unaffordable in 2D.  One could reduce the cost by using a
smaller computational domain, but the high rotational and sound speeds
($v_{\phi} \approx \cs \approx 3 \times 10^9\,\cm\ \s^{-1}$) would
still limit the timestep to $\deltt t \approx 3 \times 10^{-8}\,\ms$,
\ie almost $10^{7}$ iterations would be required to simulate the MRI
until its termination at $\approx 15\,\ms$. 

Therefore, following \citet{Obergaulinger_et_al_2009}, we use initial
magnetic fields which are two orders of magnitude higher
($b_{0z} \approx 10^{13}\,$G) to increase the wavelength of the
fastest-growing MRI mode to 
$\approx  100  \,\mathrm{m}$. This reduces the minimum
resolution to $\approx 100 \times 400 \times 100$ zones in a
$1\,\km \times 4\,\km \times 1\,\km$ box, and the number of iterations
to less than $10^5$ .

%%%%%%%%%%%%%%%%%%%%%%%%%%%%%%%%%%%%%%%%%%%%%%%%%%%%%%%%%%%%%%%%%%%%%%
\begin{figure*}
\centering
\includegraphics[width=0.42\textwidth]{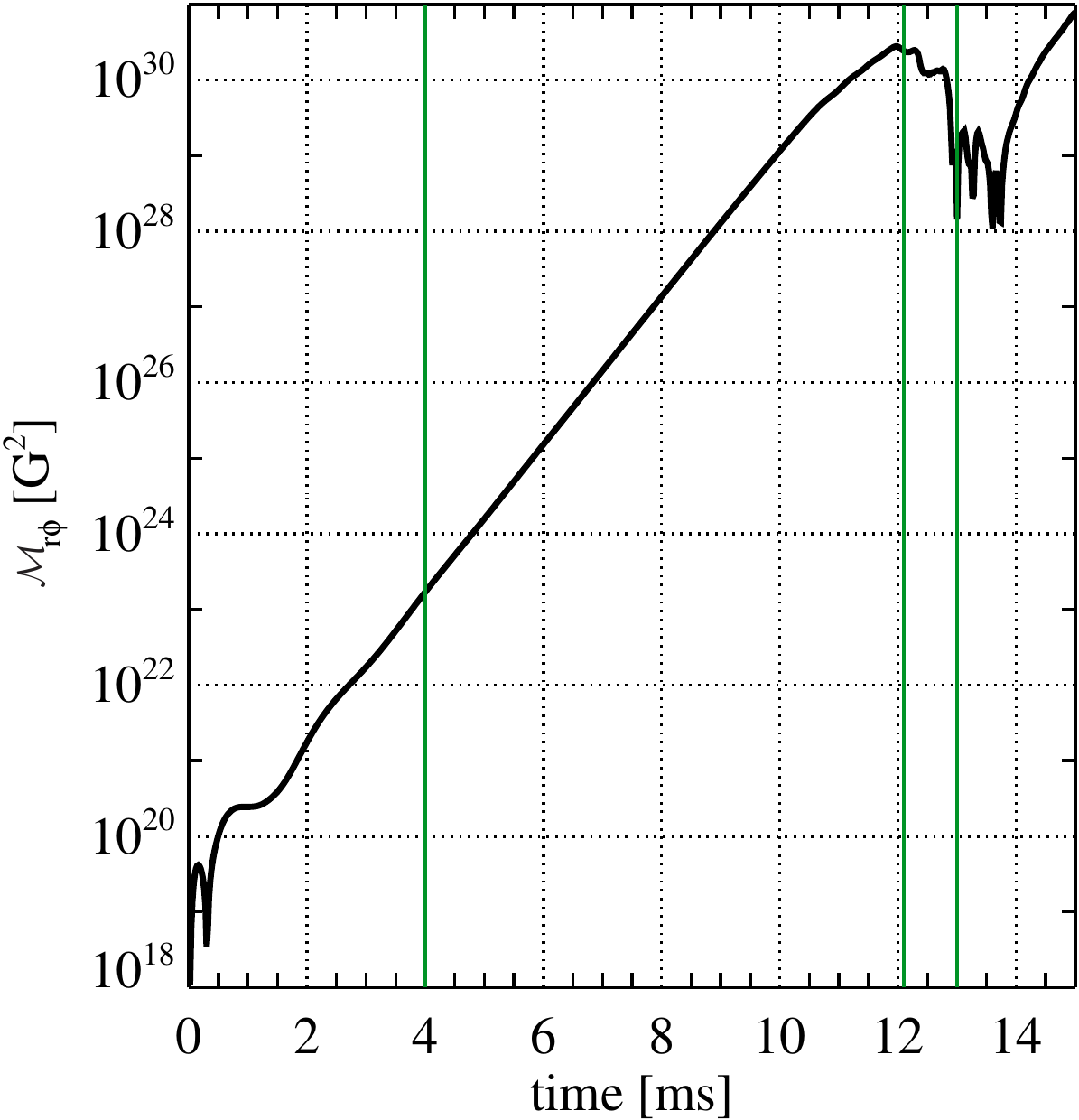}
\hspace{1.2cm}
\includegraphics[width=0.49\textwidth]{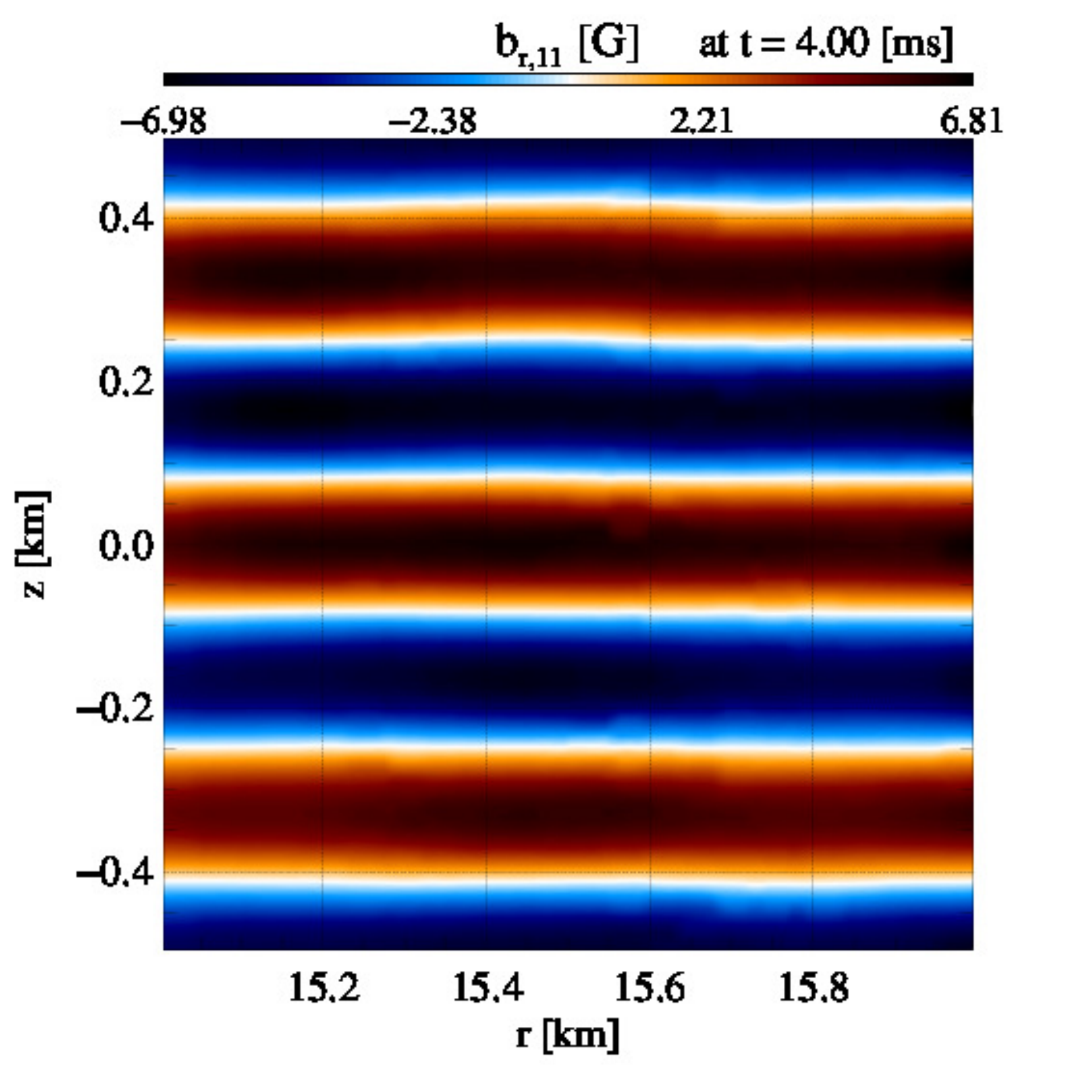}
\includegraphics[width=0.49\textwidth]{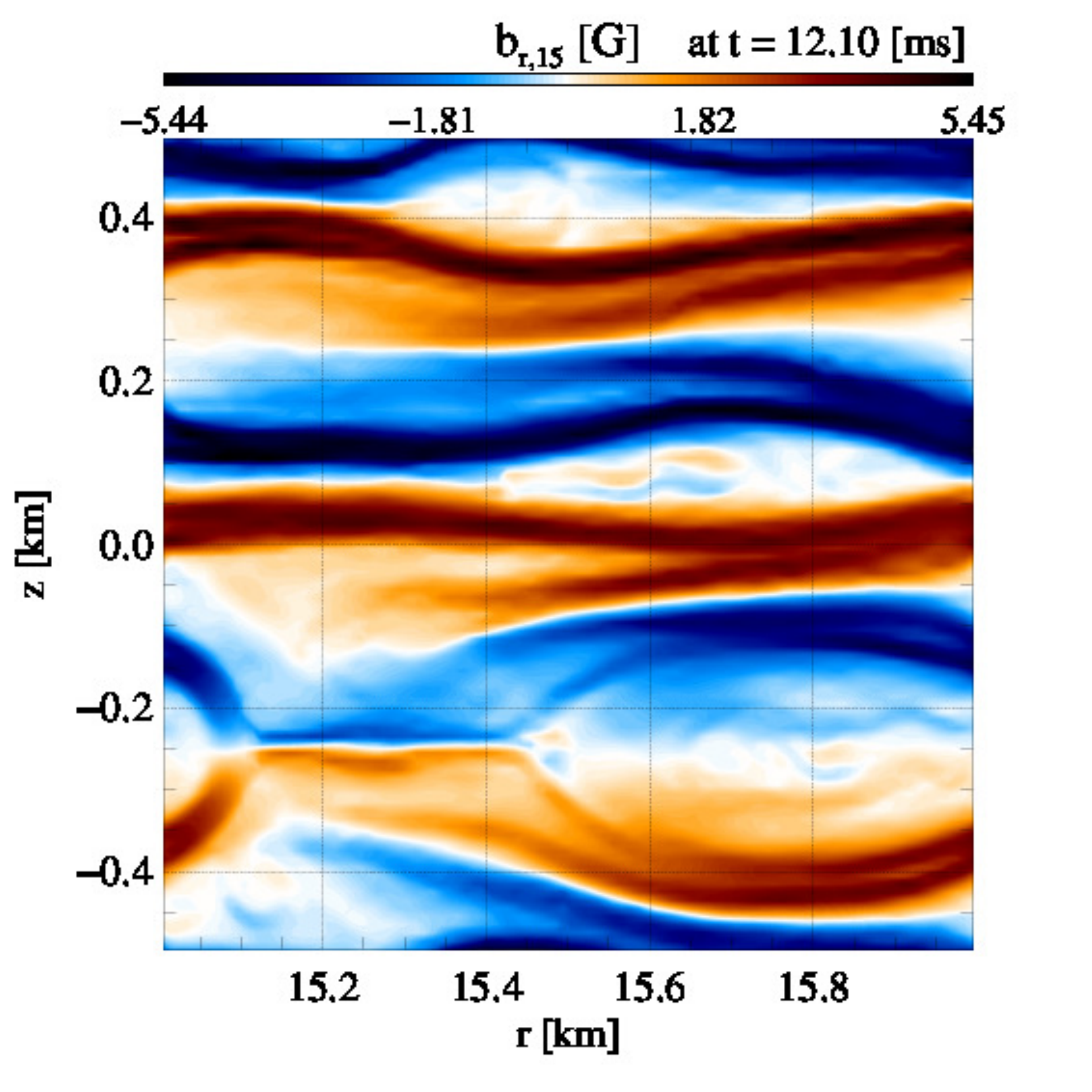}
\includegraphics[width=0.49\textwidth]{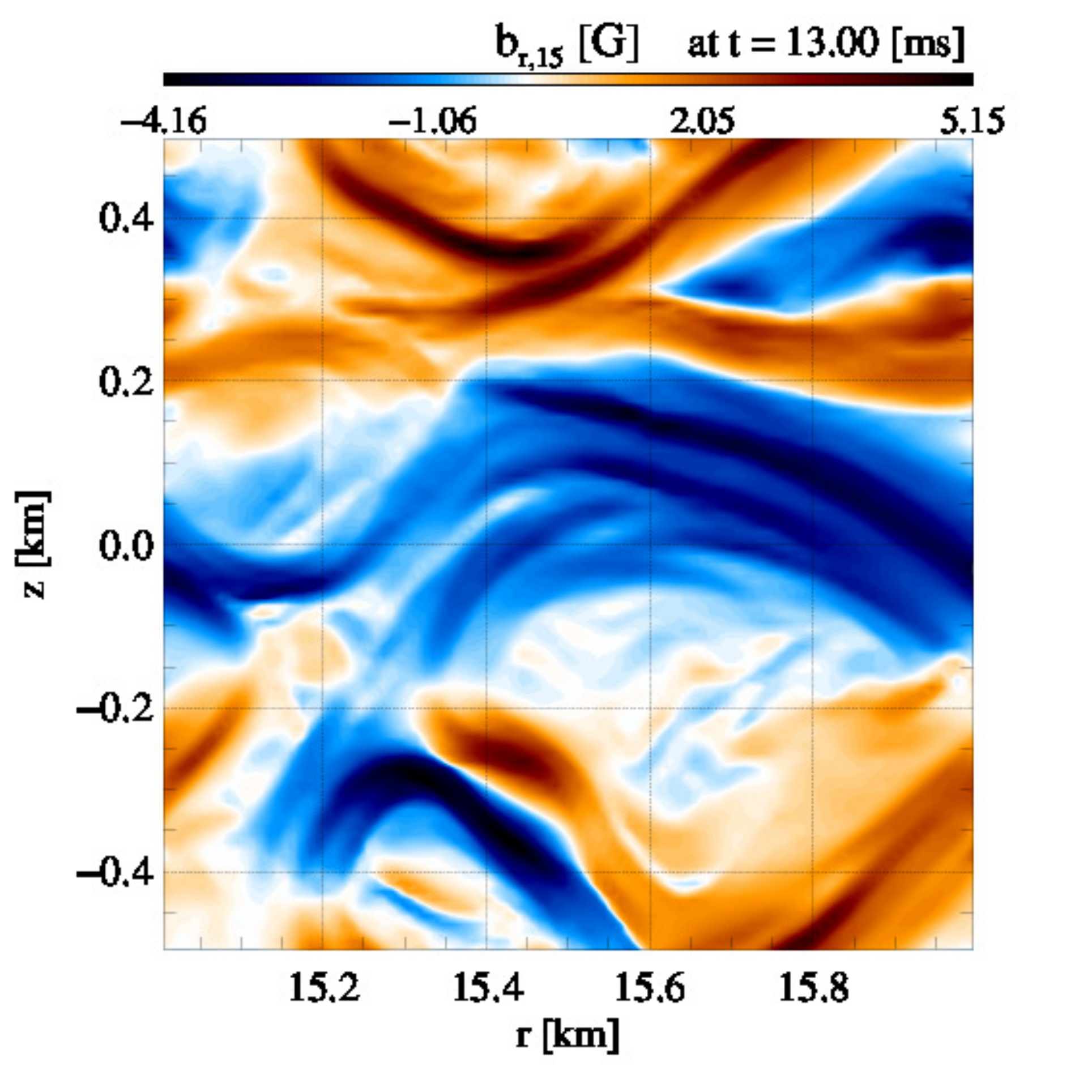}
\caption{Axisymmetric (2D) MRI simulation in which the instability is
  terminated by TMs.  Top left: Time evolution of the
  absolute value of the volume averaged Maxwell stress component
  $\MMM$ \Eqref{eq:MMM}.  The three green vertical lines mark the
  times corresponding to the snapshots shown in the other three
  panels, which display the structure of the radial magnetic field at
  $t=4\, \ms$ (top right), $t=12.1\ \ms$ (bottom left),
  and $t=13\, \ms$ (bottom right), respectively.}
\label{fig:2D_mri}
\end{figure*}
%%%%%%%%%%%%%%%%%%%%%%%%%%%%%%%%%%%%%%%%%%%%%%%%%%%%%%%%%%%%%%%%%%%%%%

We pay attention to choose the values of the initial magnetic field
strength such that an integer number of channels of the
fastest-growing MRI mode fits into the computational domain. In this
way this mode fulfils the periodic boundary conditions imposed in $z$
direction. Otherwise, the fastest-growing mode could be artificially
suppressed by an unfavourable box size.  \cite{Rembiasz} showed in
test simulations that if, e.g.\ a box of size $2.5 \lambdamri$ in $z$
direction is used, usually three MRI channels form. Either all three
of them have a wavelength smaller than $\lambdamri$ or he finds a
combination of larger and smaller channels. In any case, the MRI
developed at a rate which was lower than theoretically expected.

To trigger the MRI we impose an initial velocity perturbation on the
background velocity profile (defined by Eq.\,\ref{eq:v_init}) of the
form
\begin{align}
  {\vek v_1} = \Omega r \big[ & \{ \{ 
               {\delta}_{r} \mathfrak{R}_{r}(r,\phi,z) +
               \epsilon \sin(k_z z) \} {\vek{\hat{r}}} \} + 
\nonumber \\
             & \{ 
               \delta \mathfrak{R}_{\phi}(r,\phi,z) \} 
                      {\vek{\hat{\bm{\phi}}}} +  
               \delta \mathfrak{R}_{z}(r,\phi,z) {\vek{\hat{z}}} \big],  
\label{eq:v_with_sin}
\end{align}
where $\mathfrak{R}_{r}(r,\phi,z)$, $\mathfrak{R}_{\phi}(r,\phi,z)$,
and $\mathfrak{R}_{z}(r,\phi,z)$ are random numbers in the range
$[-1,1]$, $\delta$ and $\delta_r$ are the perturbation amplitudes,
$k_z$ is the radial perturbation wavenumber, and $\epsilon$ is the
amplitude of the sinusoidal perturbation.  If not otherwise written,
$k_z = \kmri$,\ $\delta_r = 10^{-6}$,\ $\delta = 10^{-5}$, and
$\epsilon = 2 \times 10^{-6}$. \citet{Obergaulinger_et_al_2009} used a
similar prescription for the initial perturbation, except that the
sinusoidal part was not present in their case, \ie $\epsilon=0$. We
find that the sinusoidal term is more robust in exciting MRI modes
from small perturbations, which are sometimes suppressed by numerical
effects if only random perturbations are imposed
\citep[see][]{Rembiasz}.

%%%%%%%%%%%%%%%%%%%%%%%%%%%%%%%%%%%%%%%%%%%%%%%%%%%%%%%%%%%%%%%%%%%%%%
%%%%%%%%%%%%%%%%%%%%%%%%%%%%%%%%%%%%%%%%%%%%%%%%%%%%%%%%%%%%%%%%%%%%%%
\section{Results}
\label{sec:results}
%
%%%%%%%%%%%%%%%%%%%%%%%%%%%%%%%%%%%%%%%%%%%%%%%%%%%%%%%%%%%%%%%%%%%%%%
\subsection{2D simulations}
\label{sSec:2D}

\subsubsection{Termination in 2D}

Imposing axisymmetry severely limits the number of modes that can grow
in 2D simulations.  While the fastest-growing MRI mode, which is an
axisymmetric one, can freely develop in such simulations, the dominant
parasitic instabilities, which for $\Rm > 1$ are non-axisymmetric KH
modes, are suppressed \citep{Pessah}.  Hence, among axisymmetric
secondary instabilities the fastest-growing modes are of TM rather
than KH type.  This even holds for simulations with a very low or even
a vanishing physical resistivity, as discussed by
\citep{Obergaulinger_et_al_2009}. These authors performed extensive
studies of the MRI by means of local 2D ideal MHD simulations.  Their
simulations confirmed the instability criteria and the growth rates of
the MRI for the flow regimes relevant to core collapse
supernovae. They also found that the growth of the MRI is terminated
by a tearing mode (TM) instability developing because of the
unavoidable presence of a numerical resistivity in (even ideal)
finite-volume MHD codes.

Figure\,\ref{fig:2D_mri} summarizes the evolution of an axisymmetric
model simulated with a resolution of $N_r = N_z = 100$ zones. We
performed appropriate convergence studies to ensure that the MRI is
properly resolved at this grid resolution \citep[see][for
details]{Rembiasz}.  The \emph{top left} panel displays the time
evolution of the absolute value of the volume-averaged Maxwell stress
component
\begin{equation}
  \MMM \equiv \frac{ \left| \int b_r b_{\phi} \ \mathrm{d} V \right|}{V} ,
\label{eq:MMM}
\end{equation}
where $V$ is the volume of the computational domain.  The other three
panels display the colour encoded value of the radial component of the
magnetic field in the $r-z$ plane at three different times.

From the initial velocity perturbations (see
  Eq.\,\ref{eq:v_with_sin}), three MRI channels have formed at
  $t = 4\, \ms$ (\emph{top right} panel) that grow exponentially with
  time at a constant rate (\emph{top left} panel).
A linear fit of $\log \MMM(t)$ in the time interval
  $t \in [6,8] \, \ms$ gives
  $\gamma_{\mathrm{MRI}} = 1127\, \mathrm{s}^{-1}$. 
This value is consistent with the local linear analysis,
Eq.\,(\ref{eq:gammamri}), which predicts values of
$\gamma_{\mathrm{MRI}}$ varying from $1087\,\mathrm{s}^{-1}$ to
  $1175\,\mathrm{s}^{-1}$ within the box boundaries because of its
dependence on $\Omega(r)$.

At $t = 12\,\mathrm{ms}$, the Maxwell stress reaches a value of
  $\MMM = 2.78 \times 10^{30} \,$G$^{2}$ and the MRI growth is
terminated by parasitic instabilities.  
We note that in 2D simulations, the stress at termination is
  highly sensitive to the initial random perturbation imposed in
  the simulation. Performing several realizations of the same
  simulation, with different seeds for the random number generator, we
  obtained a Maxwell stress at the termination varying from
  $\MMM = 2.19 \times 10^{30} \,$G$^{2}$ to
  $\MMM = 4.69 \times 10^{30} \,$G$^{2}$ \citep[see][for
  details]{Rembiasz}. As we discuss in the next section, we
    do not observe this large scatter in our 3D simulations.

The colour map of the radial component of the magnetic field
exhibits several X points at $t = 12.1\,\ms$ (\emph{bottom left}
panel), where field lines reconnect. These X points are located at
$(r,z) = (15.2, 0.4), \, (16,0.1)$, and $(15.2,-0.3)\, \km$,
respectively.  There are also three O points recognisable in the
centres of magnetic islands located at
$(r,z) = (15.5,0.4), \,  (15.7,0.1)$, and $(15.8,-0.3)\ \km$,
respectively.  These patterns indicate MRI termination by TM rather
than by KH instabilities.  Shortly afterwards, at $t = 13 \, \ms$
(\emph{bottom right} panel), the channel modes have been destroyed and
MHD turbulence sets in.

These results, which are qualitatively similar to those found by
\cite{Obergaulinger_et_al_2009}, confirm that in axisymmetric models
the MRI is artificially terminated by TMs, which would grow
slower than KH instabilities in 3D models.

%%%%%%%%%%%%%%%%%%%%%%%%%%%%%%%%%%%%%%%%%%%%%%%%%%%%%%%%%%%%%%%%%%%%%%
\begin{figure}%[t]
\includegraphics[width=0.9\linewidth]{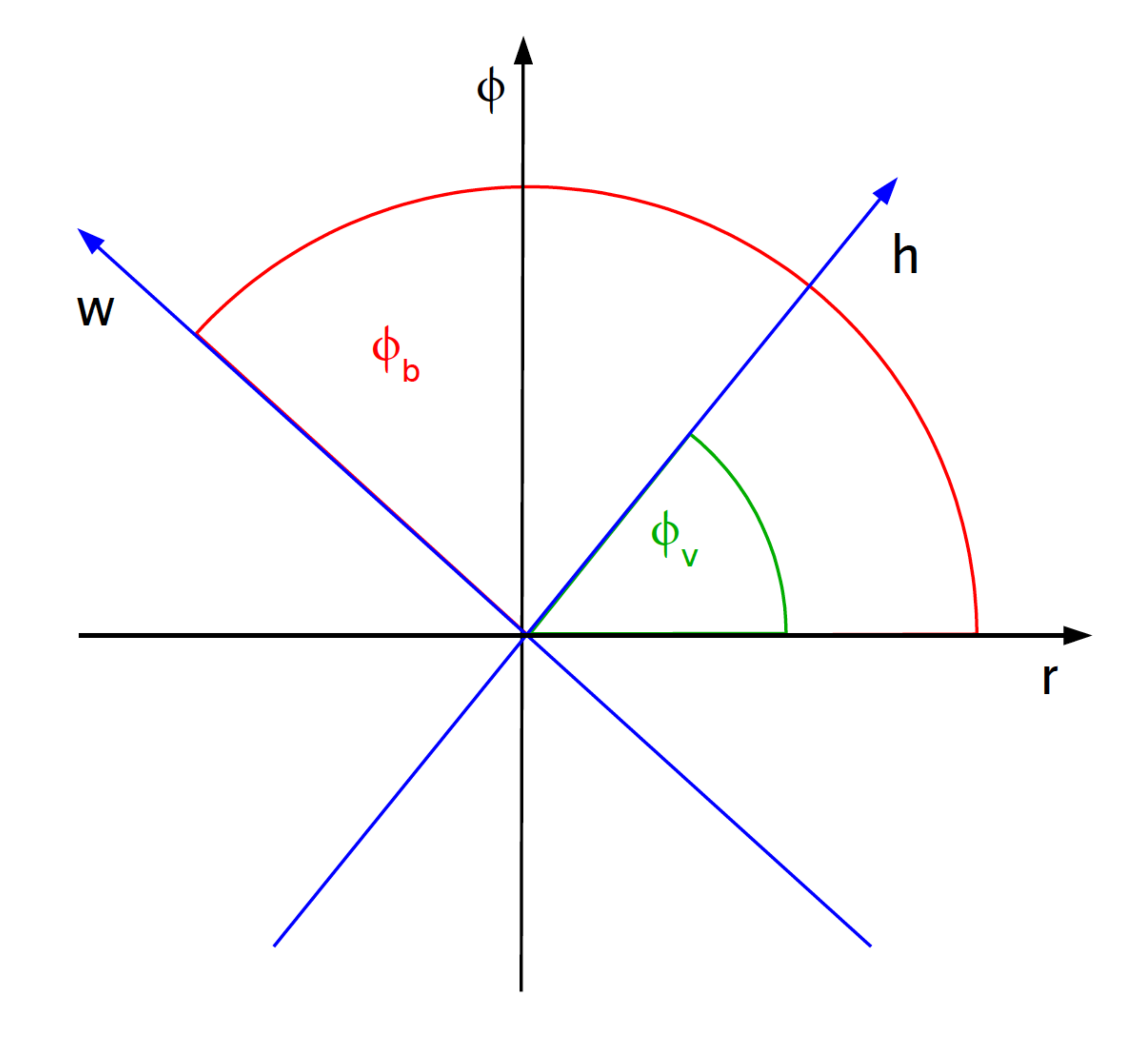}
\caption{Relation between the (not necessarily orthogonal) coordinates
  $(h,w)$ defined by \eqsref{eq:hhat} and (\ref{eq:what}) and the
  cylindrical coordinates $(r,z)$.  The angles $\phi_v$ and $\phi_b$
  are given by \eqsref{eq:phi_v} and (\ref{eq:phi_b}),
  respectively. Compare with Fig.\,2 of PC08. }
\label{fig:hw_coordinates}
\end{figure}
%%%%%%%%%%%%%%%%%%%%%%%%%%%%%%%%%%%%%%%%%%%%%%%%%%%%%%%%%%%%%%%%%%%%%%

%%%%%%%%%%%%%%%%%%%%%%%%%%%%%%%%%%%%%%%%%%%%%%%%%%%%%%%%%%%%%%%%%%%%%%
\subsection{3D simulations}
\label{sSec:3D}

\subsubsection{Termination in 3D}

According to predictions of \cite{Pessah}, in general, the
fastest-growing KH instabilities should develop along the velocity
channels, \ie $\phikh = \phi_v$, whereas the fastest TM should grow
along the magnetic field channels, \ie $\phitm = \phi_b$.  As these
channels define two important directions in the horizontal ($r, \phi$)
plane, we will sometimes use in the following discussion another
coordinate system $(h,w,z)$, the axes $h$ and $w$ being aligned with
the velocity and magnetic field channels, respectively.  The $(h,w)$
coordinates and the $(r,\phi)$ cylindrical coordinates are related
through the following coordinate transformation
(Fig.\,\ref{fig:hw_coordinates})
\begin{align}
  \label{eq:hhat}
  \bm{\hat{h}} & = \bm{\hat{r}}    \cos \phi_{v} + 
                   \bm{\hat{\phi}} \sin \phi_{v}, \\
\label{eq:what}
  \bm{\hat{w}} & = \bm{\hat{r}}    \cos {\phi}_b + 
                   \bm{\hat{\phi}} \sin {\phi_b},
\end{align}
where the
angles $\phi_v$ and $\phi_b$ are given by \eqsref{eq:phi_v} and
(\ref{eq:phi_b}), respectively.%
\footnote{This naming convention differs from that of
  \citet[][]{Latter_et_al} and \citet[][]{Pessah} who used $\vek{h}$
  to denote vectors (or their components) in the $(r,\phi)$ plane.  }
For $\Ree = \Rm$, the axes $h$ and $w$ are orthogonal, \ie during
the phase of exponential growth of the MRI $b_h=0$ and $v_w = 0$.

Figure\,\ref{fig:hw_coordinates}  depicts the horizontal plane in both
coordinate systems. The channel angles are set to
$\phi_v = 45^{\circ}$ and $\phi_b = 135^{\circ}$, which corresponds to
the ideal MHD limit.  In general, the vectors $\bm{\hat{h}}$ and
$\bm{\hat{w}}$ are not orthogonal to each other, and their exact
orientation depends on the Reynolds numbers. For example, in the limit
$\Ree \rrr \infty$ and $\Rm \rrr 0$, one has
$\bm{\hat{h}} || \bm{\hat{r}}$ and $\bm{\hat{w}} || \bm{\hat{\phi}}$,
\ie $\phi_v = 0^{\circ}$ and $\phi_b = 90^{\circ}$ 
(PC08).

For the values of the Reynolds numbers used in our simulations, \ie
$\Ree = \infty, \Rm \approx 100 $, and
$\Ree = \Rm \approx 100 $, we expect the channels to be
oriented along 
$\phi_v = 44.4 \tto 44.6\, ^{ \circ}$,
      $\phi_b = 134.7 \tto 134.8\, ^{\circ}$,
      $\phi_v = 44.3 \tto 44.5\, ^{\circ}$, and
$\phi_b = 134.5^{\circ}$, respectively. These angles differ only
little from those of the ideal MHD case.  Hence, in either case, the
fastest-growing KH mode should develop at an angle
$\phip = \phikh \approx 44.5^{\circ}$, where $\phip$ denotes
the angle at which the dominant parasitic instability
develops. Depending on the initial conditions, one either has
$\phip = \phikh$ or $\phip = \phitm$. In 2D axisymmetric simulations,
the only allowed angle is $\phip = 0^{\circ}$.

%%%%%%%%%%%%%%%%%%%%%%%%%%%%%%%%%%%%%%%%%%%%%%%%%%%%%%%%%%%%%%%%%%%%%%
\begin{figure}%[t]
  \centering
  \includegraphics[width=1.0\linewidth]{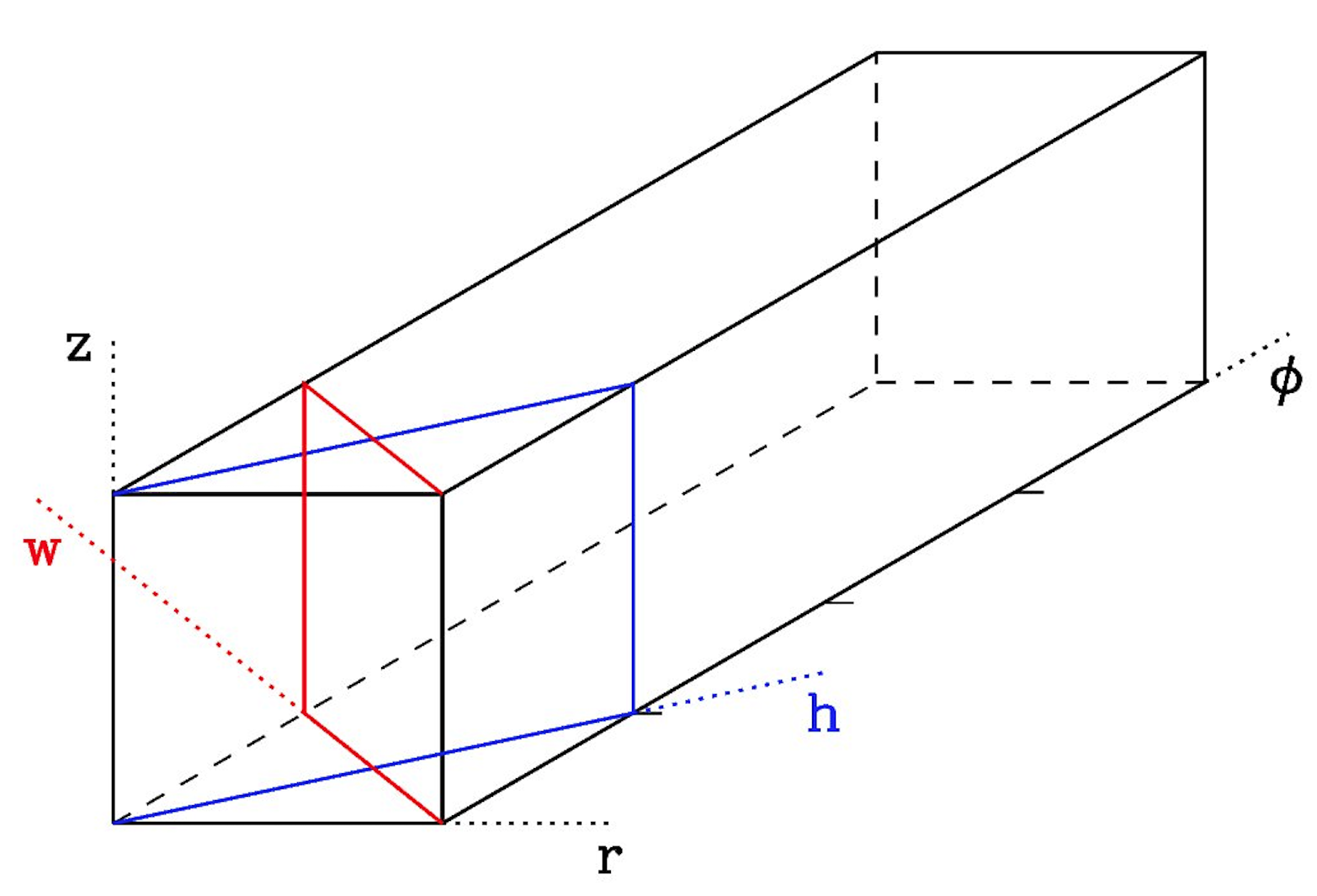}
  \caption{Sketch of the computational domain of size
    $1\times4\times1$ km, which was used to simulate models \#5 to \#8
    and \#10 (see Table\,\ref{tab:main_results}) together with two
    cuts corresponding to the $(h,z)$ plane (blue) and the $(w,z)$
    plane (red) used to display some results of model \#7 in
    \figref{fig:mri_cuts}. Note that the computational volume is
    actually bent in $\phi$ direction and the two cuts are no planes
    but curved surfaces in cylindrical coordinates.}
\label{fig:mri_cut_geometry}
\end{figure}
%%%%%%%%%%%%%%%%%%%%%%%%%%%%%%%%%%%%%%%%%%%%%%%%%%%%%%%%%%%%%%%%%%%%%%

We studied the termination process in a number of 3D simulations,
varying the size and aspect ratio of the computational domain, and the
grid resolution (see \tabref{tab:main_results} for the list of
models).  The default simulation box has a size of
$L_r \times L_{\phi} \times L_z = 1\,\km \times 4\,\km \times 1\,\km$,
as shown in \figref{fig:mri_cut_geometry}, which also depicts two cuts
corresponding to the $(h,z)$ and $(w,z)$ planes.  We note that the
computational volume is actually bent in $\phi$ direction and the two
cuts are not planes but curved surfaces in cylindrical coordinates.
However, we will call them planes for simplicity.

%%%%%%%%%%%%%%%%%%%%%%%%%%%%%%%%%%%%%%%%%%%%%%%%%%%%%%%%%%%%%%%%%%%%%%
%\afterpage{
\begin{table*}
  \caption[]{Overview of our 3D MRI simulations, which were all
    performed with the initial parameters chosen in such a way 
    that $\lambdamri \approx 0.333\,\km\,^{\ast}$.
    The columns give the model identifier, the magnetic field
    strength, the hydrodynamic and magnetic Reynolds numbers 
    ($\Ree$ and $\Rm$, respectively), $\,^{\ast}$
    the size of the computational domain, the resolution, the 
    number of grid cells per MRI wavelength, the measured MRI 
    growth rate, the volume-averaged Maxwell stress at termination,
    the azimuth angle at which the parasitic instability develops, and 
    its horizontal wavelength (\ie in the $(r,\phi)$ plane) measured at 
    $t = 11\,\ms$ and $t = 12.5\,\ms$ for KH instabilities and TMs,
    respectively. The azimuth angle and the horizontal wavelength
    are determined with 
    about a $9\,^{\circ}$ and $10\%$ accuracy, respectively
    (see also \figref{fig:parasitic_theta}). The final column gives
    the type of the instability. For models 19 and 20, we were unable 
    to identify the type of the parasitic instability terminating the 
    growth of the MRI.}
%
%\centering
\begin{center}
\begin{tabular}{|c|l|c|c|c|c|c|c|l|c|c|c|c|c| c|c|}
\hline
 \# & \pbox{5cm}{$b_{0z}$ \\ $[10^{13}$ G$]$ } & $\Ree$ & $\Rm$ &
      \pbox{5cm}{box size \\ $(r \times \phi \times z)$ [km]} & 
      \pbox{5cm}{resolution \\ $(r \times \phi \times z)$ \ } & 
      \pbox{5cm}{zones per \\ channel } & $\gammamri [\s^{-1}]$ & 
      \pbox{5cm}{$\MMM$ \\ $[10^{30}\ \mathrm{G}^2]$} & $\phip\ [^{\circ}]$ & 
      $\frac{\lambdap}{\lambdamri}$ & \pbox{5cm}{term. \\ instab.}  
\\  \hline

   1 & \hspace{0.3cm} 4.6 & $\infty$ & $\infty$ & $1 \times 4 \times 1$ 
     & $100 \times 400 \times 100$ & $ 33$ & 1137 & 0.96 & 45 & 0.94 & KH 
 \\
\hline
   2 & \hspace{0.3cm} 4.6 & $\infty$ & 100 & $1 \times 1 \times 0.333$    
     & $ 24 \times  24 \times   8$ & $  8$ & 
1104
 & 0.79 & 45 & 1.5  & KH 
 \\
   3 & \hspace{0.3cm} 4.6 & $\infty$ & 100 & $1 \times 1 \times 0.333$    
     & $ 30 \times  30 \times  10$ & $ 10$ & 
1122
& 1.3  & 44 & 1.2  & KH 
 \\
   4 & \hspace{0.3cm} 4.6 & $\infty$ & 100 & $1 \times 1 \times 0.333$    
     & $ 48 \times  48 \times  16$ & $ 16$ & 1130 & 1.1  & 47 & 1.1  & KH 
 \\
   5 & \hspace{0.3cm} 4.6 & $\infty$ & 100 & $1 \times 4 \times 1$     
     & $ 60 \times 240 \times  60$ & $ 20$ & 1126 & 1.1  & 44 & 0.92 & KH
 \\
   6 & \hspace{0.3cm} 4.6 & $\infty$ & 100 & $1 \times 4 \times 1$     
     & $ 76 \times 304 \times  76$ & $ 25$ & 1127 & 1.0  & 47 & 0.99 & KH 
 \\
   7 & \hspace{0.3cm} 4.6 & $\infty$ & 100 & $1 \times 4 \times 1$     
     & $100 \times 400 \times 100$ & $ 33$ & 1127 & 0.93 & 44 & 0.85 & KH  
 \\
   8 & \hspace{0.3cm} 4.6 & $\infty$ & 100 & $1 \times 1 \times 1$     
     & $100 \times 100 \times 100$ & $ 33$ & 1127 & 0.93 & 47 & 0.95 & KH 
 \\
   9 & \hspace{0.3cm} 4.6 & $\infty$ & 100 & $1 \times 1 \times 0.333$   
     & $100 \times 100 \times  34$ & $ 34$ & 1127 & 0.93 & 47 & 0.77 & KH 
 \\
  10 & \hspace{0.3cm} 4.6 & $\infty$ & 100 & $1 \times 4 \times 1$     
     & $200 \times 800 \times 200$ & $ 67$ & 1127 & 0.73 & 44 & 0.70 & KH 
 \\
  11 & \hspace{0.3cm} 4.6 & $\infty$ & 100 & $1 \times 1 \times 0.333$    
     & $400 \times 400 \times 134$ & $134$ & 1128 & 0.73 & 45 & 0.69 & KH 
 \\
\hline
  12 & \hspace{0.3cm} 4.6 & $   100$ & 100 & $1 \times 1 \times 0.333$    
     & $100 \times 100 \times  34$ & $ 34$ & 1120 & 0.90 & 44 & 0.85 & KH
 \\
  13 & \hspace{0.3cm} 4.6 & 100  & 100  & $1 \times 0.8 \times 0.333$ 
     & $100 \times  80 \times  34$ & $ 34$ & 1120 & 0.92 & 44 & 0.80 & KH 
 \\
  14 & \hspace{0.3cm} 4.6  & 100  & 100 & $1 \times 0.6 \times 0.333$ 
     & $100 \times  60 \times  34$ & $ 34$ & 1120 & 1.0  & 43 & 0.96 & KH 
 \\
  15 & \hspace{0.3cm} 4.6  & 100  & 100 & $1 \times 0.4 \times 0.333$ 
     & $100 \times  40 \times  34$ & $ 34$ & 1120 & 0.97 & 43 & 0.80 & KH
 \\
  16 & \hspace{0.3cm} 4.6  & 100  & 100 & $1 \times 0.3 \times 0.333$ 
     & $100 \times  30 \times  34$ & $ 34$ & 1120 & 1.1  & 52 & 0.70 & KH 
 \\
  17 & \hspace{0.3cm} 4.6  & 100  & 100 & $1 \times 0.2 \times 0.333$ 
     & $100 \times  20 \times  34$ & $ 34$ & 1120 & 4.5  &  - &    - & TM 
 \\
  18 & \hspace{0.3cm} 4.6  & 100  & 100 & $1 \times 0.1 \times 0.333$ 
     & $100 \times  10 \times  34$ & $ 34$ & 1120 & 4.9  &  - &    - & TM 
 \\
\hline
  19 & \hspace{0.3cm} 0.325 & $\infty$ & $0.1$ & $1 \times 1 \times 0.333$ 
     & $100 \times 100 \times  34$ & $ 34$ &   80 & 0.0020 & - & - & ?
 \\
  20 & \hspace{0.3cm} 0.163 & $\infty$ & $0.05$ & $1 \times 1 \times 0.333$ 
     & $100 \times 100 \times  34$ & $ 34$ &   34 & 0.0011 & - & - & ? 
 \\
\hline
  \end{tabular}
\label{tab:main_results}
\end{center}
\begin{flushleft}
  \footnotesize{$^{\ast}\,$Note that $\lambdamri$, $\Ree$, and
      $\Rm$ are not uniform throughout the computational domain, but
      vary by $\approx 20\%$ (see Sec.\,\ref{sSec:Init} for details).
    } 
\end{flushleft}
%}  
 \end{table*}
%}
%%%%%%%%%%%%%%%%%%%%%%%%%%%%%%%%%%%%%%%%%%%%%%%%%%%%%%%%%%%%%%%%%%%%%%

\begin{figure}%[t]
\centering
%mnras  \sidecaption
\includegraphics[width=1.0\linewidth]{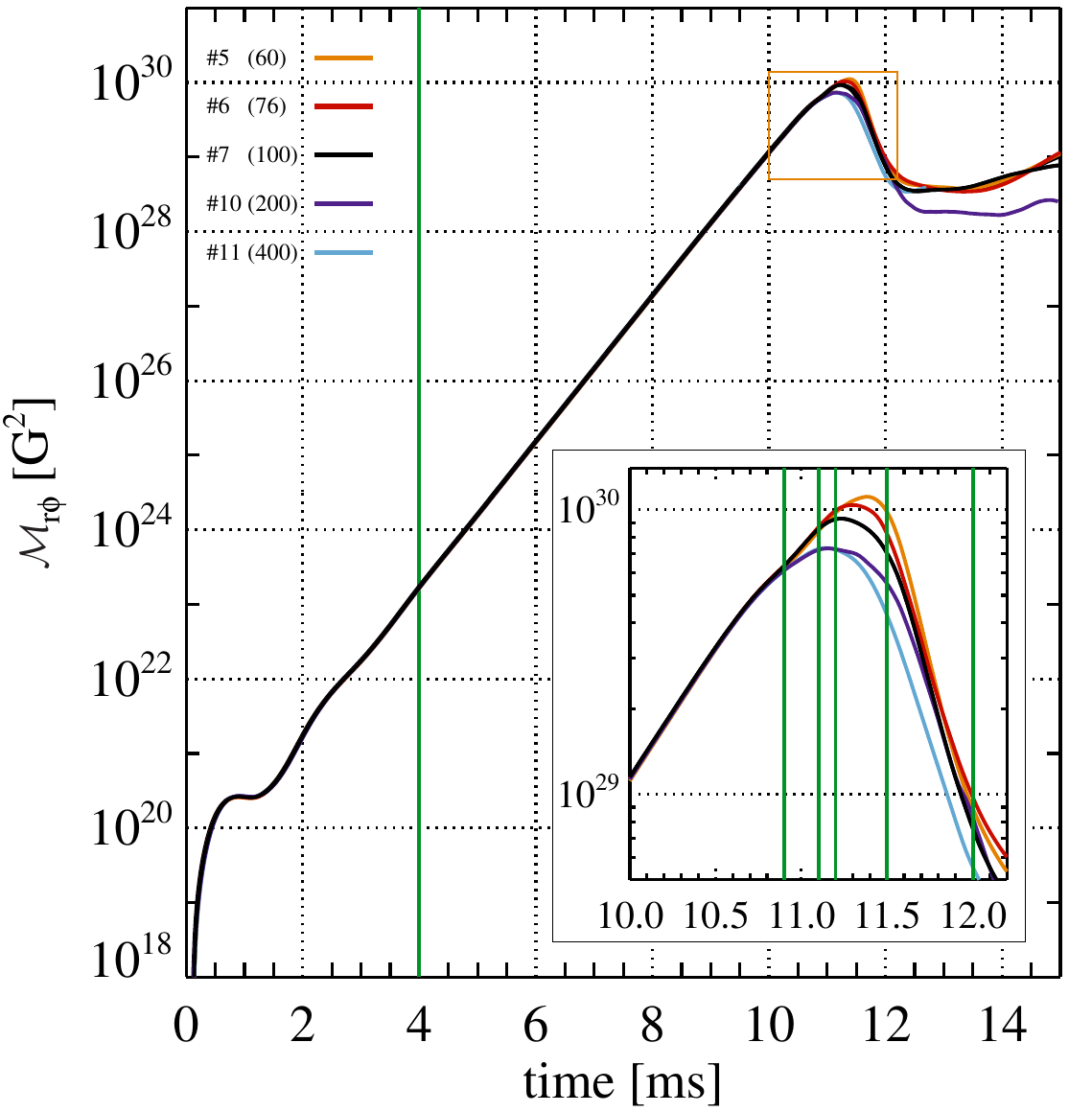}
\caption{Evolution of the volume-averaged Maxwell stress component
  $\MMM$ for the 3D
  models \#5 (orange), \#6 (red), \#7 (black), \#10 (violet), and
  \#11 (blue; run only until $t = 12 \, \ms$). Up to $t \approx
  10.5\,\ms$, the models give almost identical results, 
  differences being only visible around termination (see inset).
  Vertical lines indicate the times of the snapshots displayed for 
  model\,\#7 in \figref{fig:3D_mri}. 
 For a better readability of the figure, the vertical lines 
 around MRI termination are only displayed in the inset.}
\label{fig:mri_time_evolution}
\end{figure}

We begin the discussion of our results with model\,\#7 (see
\tabref{tab:main_results}), which we simulated in the default box
resolved with $100 \times 400 \times 100$ zones in $r, \phi$, and $z$
direction, respectively. Except for the third space dimension, the
initial conditions and the parameters of this 3D model are identical
to those of the axisymmetric model discussed in \secref{sSec:2D}.  As
illustrated by the time evolution of the Maxwell stress
(\figref{fig:mri_time_evolution}), the MRI grows exponentially with
time at the same rate as in the 2D model
($\gamma_{\mathrm{MRI}} = 1127\,\mathrm{s}^{-1}$) until saturation,
which occurs a bit earlier than in the 2D model at
$t = 11.2\,\mathrm{ms}$.

%%%%%%%%%%%%%%%%%%%%%%%%%%%%%%%%%%%%%%%%%%%%%%%%%%%%%%%%%%%%%%%%%%%%%%
%%%%%%%%%%%%%%%%%%%%%%%%%%%%%%%%%%%%%%%%%%%%%%%%%%%%%%%%%%%%%%%%%%%%%%

\begin{figure*}
\centering
%  \hspace{-0.4cm}
\includegraphics[width=0.33\linewidth]{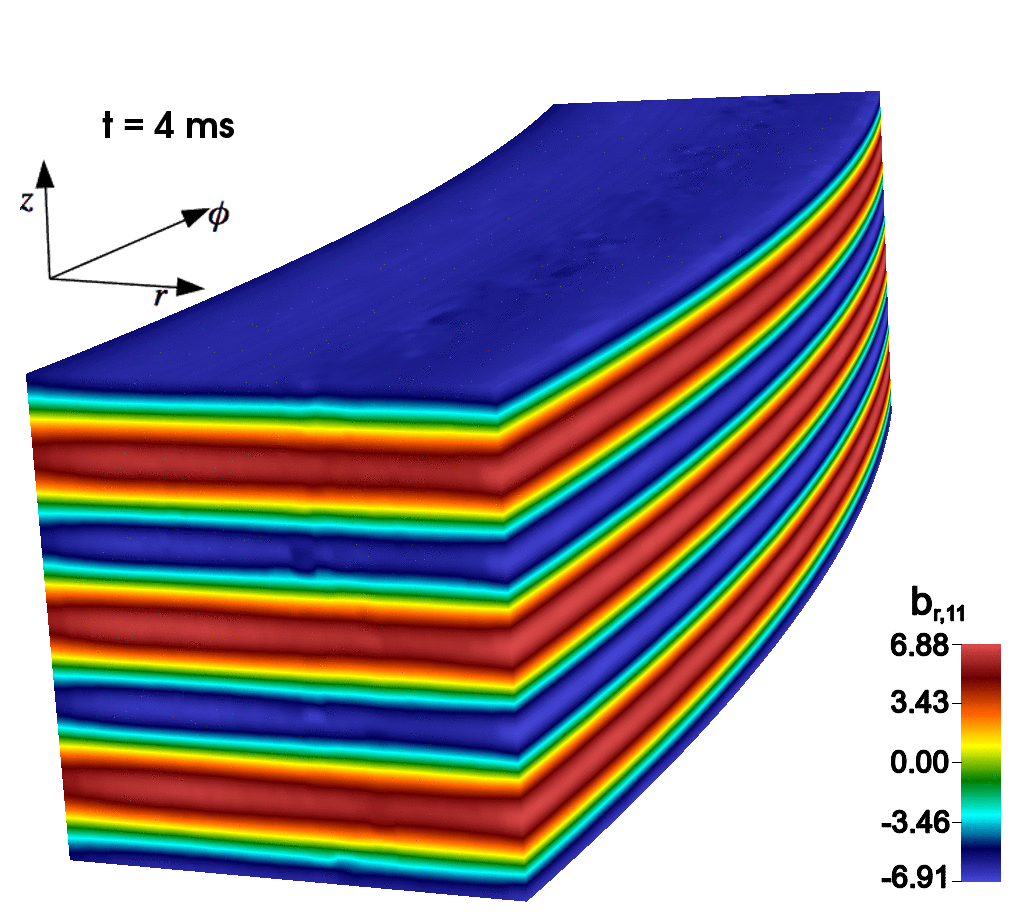}
\includegraphics[width=0.33\linewidth]{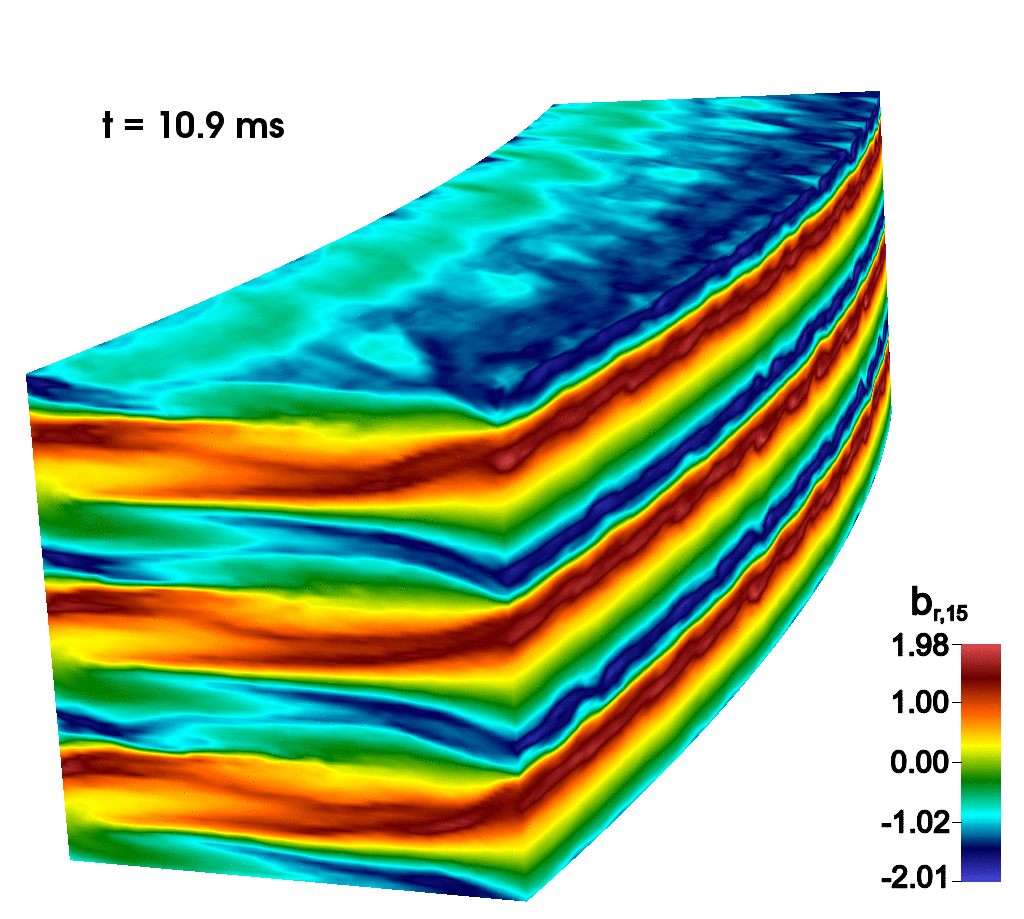}
\includegraphics[width=0.33\linewidth]{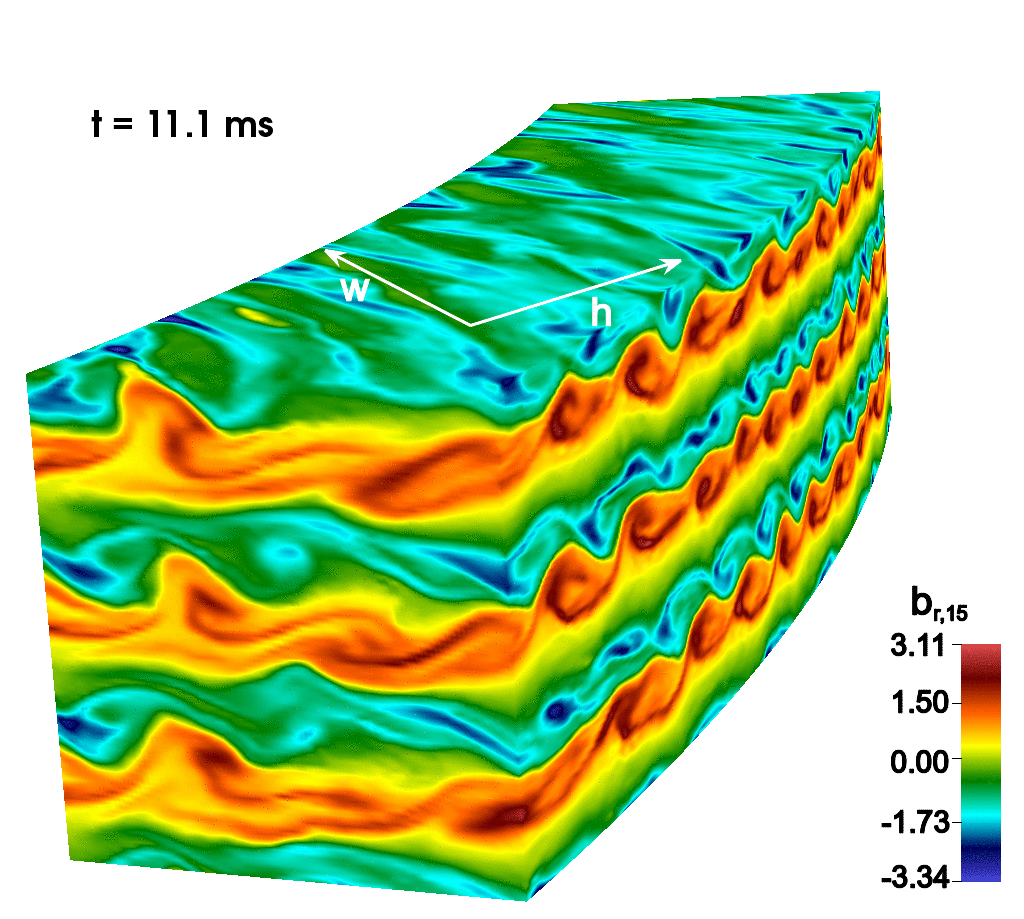}
\includegraphics[width=0.33\linewidth]{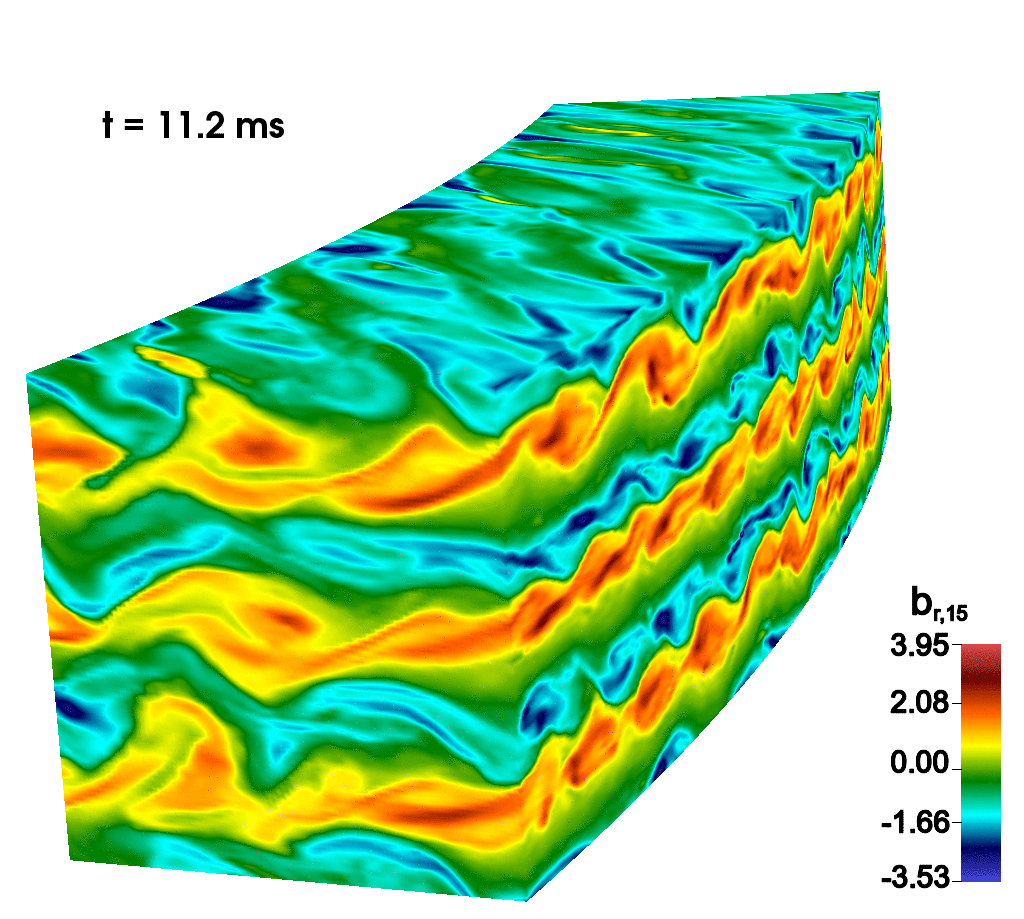}
\includegraphics[width=0.33\linewidth]{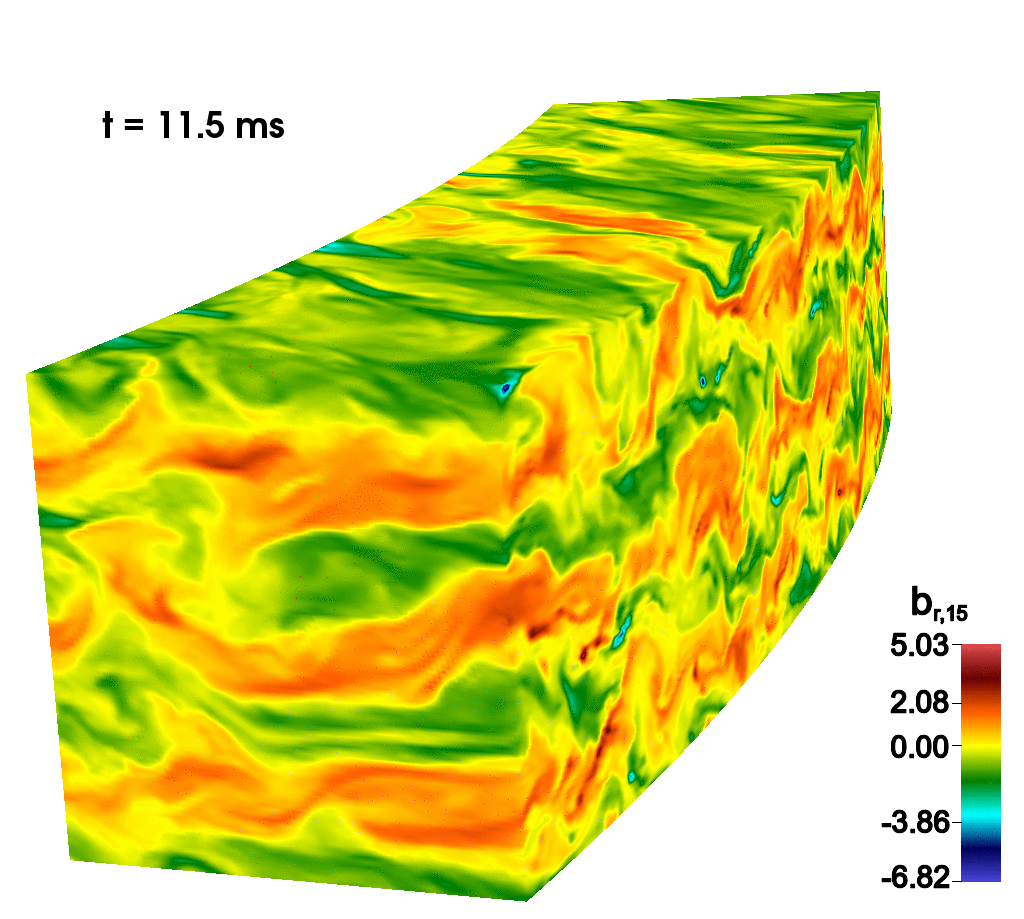}
\includegraphics[width=0.33\linewidth]{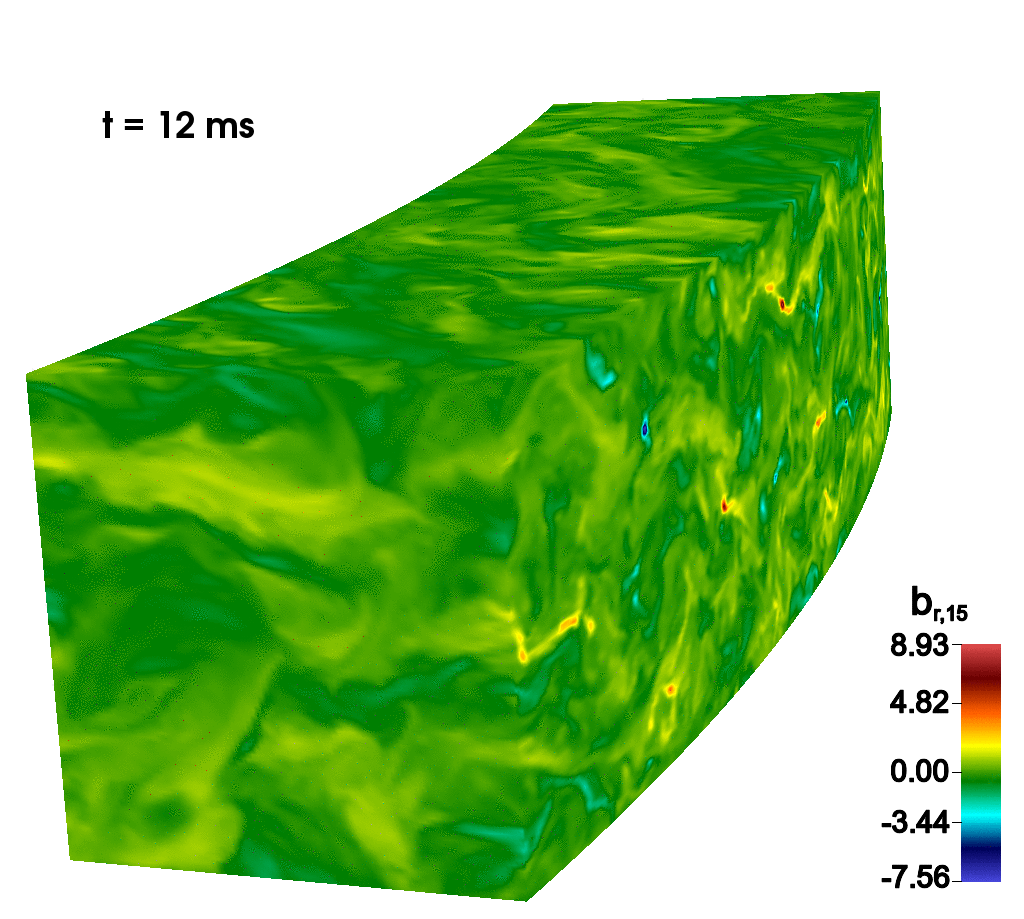}
\caption{Distribution of the radial component of the magnetic field
  $b_r$ across the surface of the computational domain for the 3D
  model\,\#7 at six different times. The times of the snapshot are
  marked by green vertical lines in \figref{fig:mri_time_evolution}. }
\label{fig:3D_mri}
\end{figure*}

\begin{figure*}%[h]
\centering
%  \hspace{-0.4cm}
\includegraphics[width=0.495\linewidth]{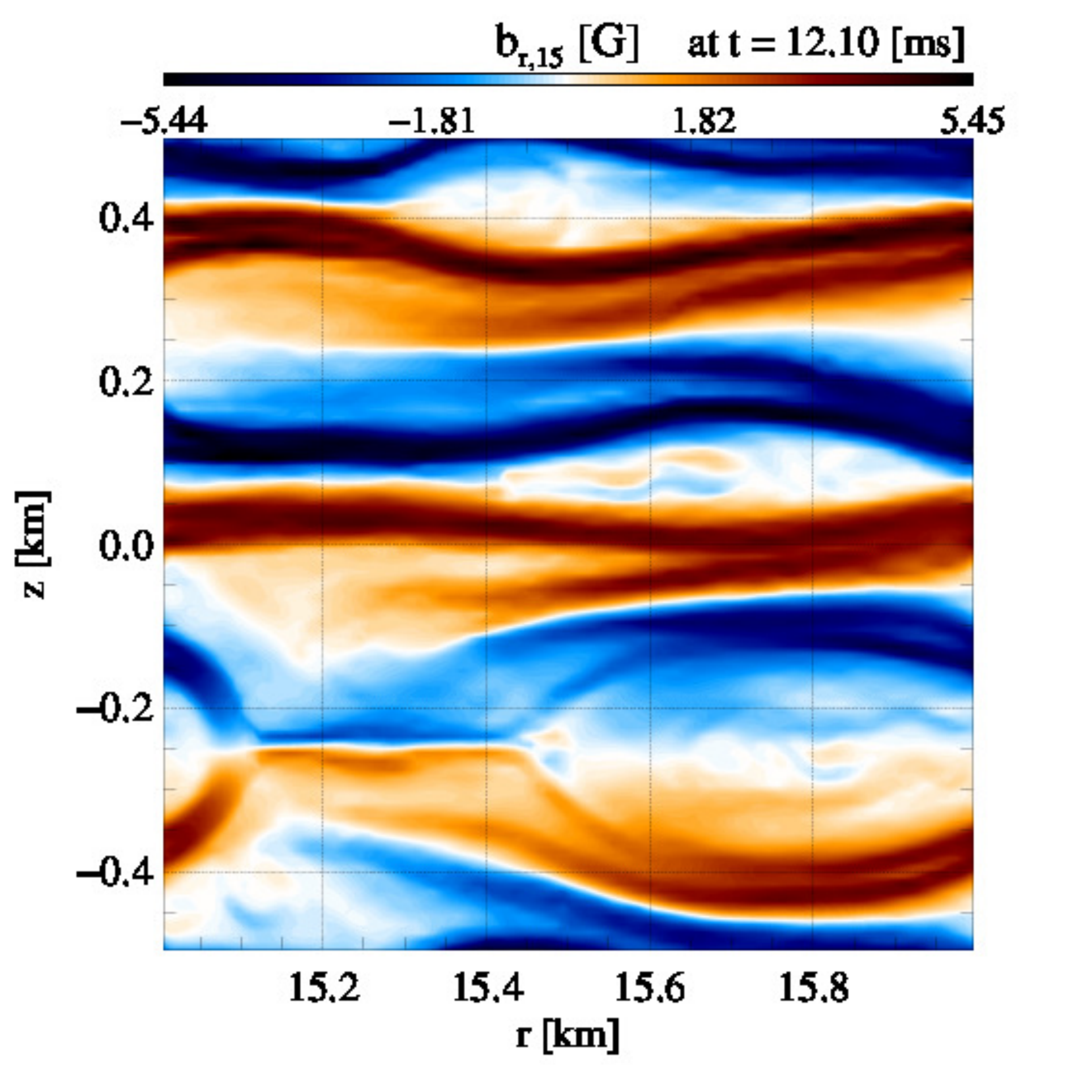}
\includegraphics[width=0.495\linewidth]{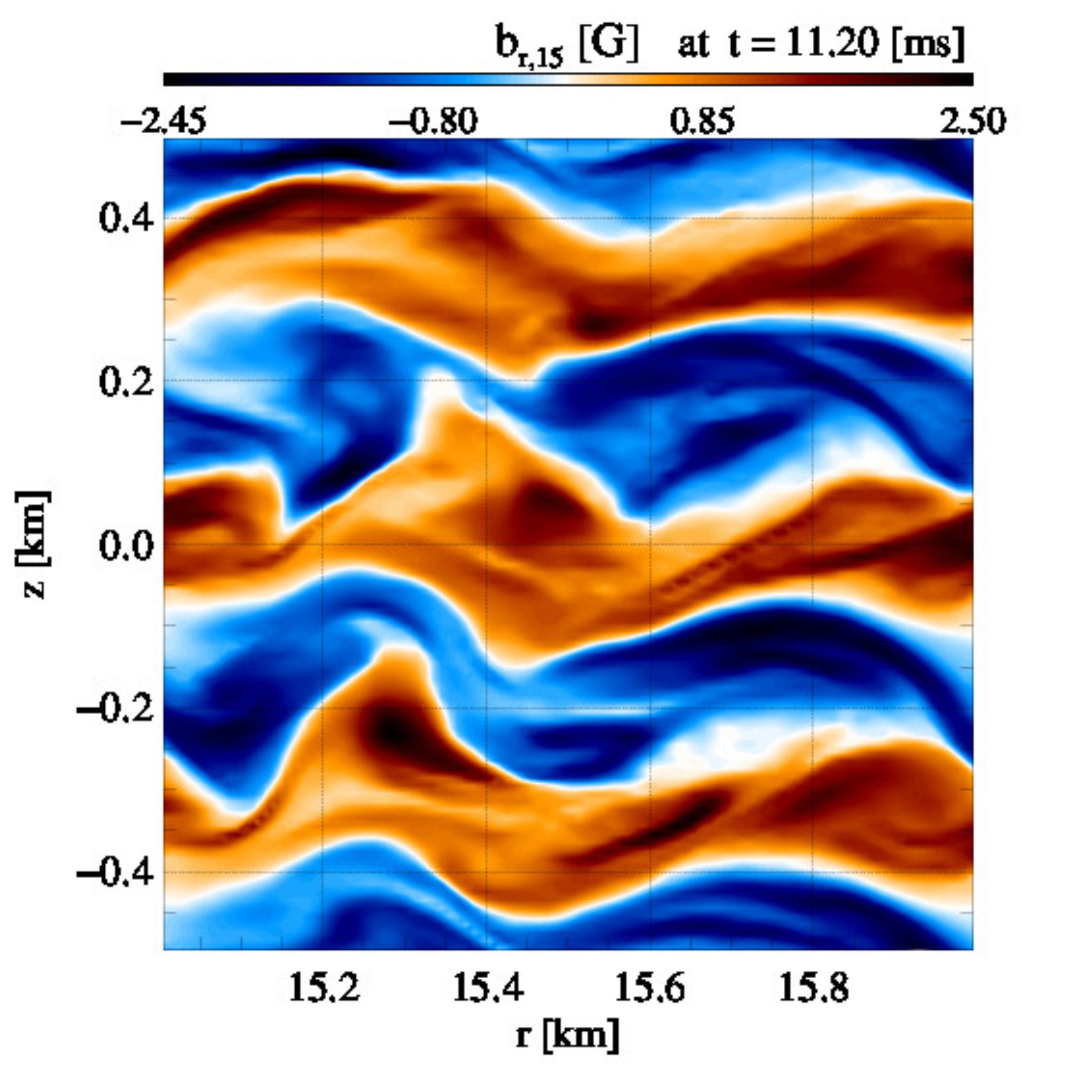}
\caption{Comparison of the distribution of the radial magnetic field
  $b_r$  around MRI termination in a 2D (left) and 3D (right)
  simulation (model\,\#7; cut at $\phi = -2\,\km$).  Note that the two
  snapshots are taken at different times.}
\label{fig:mri_2D_3D_term_comparison_local}
\end{figure*}

Figure\,\ref{fig:3D_mri} shows the 3D structure of the radial
component of the magnetic field at six distinct times (marked by
vertical lines in \figref{fig:mri_time_evolution}).  Magnetic field
perturbations grow in the form of three axisymmetric channels
(\emph{upper left} panel), which are perturbed in turn by secondary
instabilities that become recognisable only shortly before the MRI is
terminated at t = 11.2\,\ms.  The sequence of snapshots
 around this time (\emph{upper middle} to \emph{bottom middle})
demonstrates that non-axisymmetric parasitic modes are responsible for
MRI termination.  After exponential growth of the magnetic field has
ended, the simulation volume is dominated by a small-scale, turbulent
magnetic field (\emph{bottom right}).  
Comparing the structure of the radial magnetic field during MRI
termination (\figref{fig:mri_2D_3D_term_comparison_local}), we find
profound differences between the 2D model (\emph{left-hand} panel) and the
3D model (\emph{right-hand} panel).  The 2D model exhibits X and O points
that are characteristic of the TM instability, whereas in the 3D
simulation the channel modes bend strongly at the locations of vortex
rolls which are a typical feature of the KH instability.

We also studied some geometrical aspects of the parasitic
instabilities found in the 3D simulation. For this purpose, we use the
$(h, w, z)$ coordinates.Velocity-shear driven KH instabilities should
grow fastest in the $h$ direction and dominantly current driven TM
along the $w$ axis.  To simplify the expressions we performed the
coordinate transformation $(r,\phi,z) \rightarrow (h,w,z)$ with
$\phi_v = 45^{\circ}$ and $\phi_b = 135^{\circ}$ instead of the
theoretically expected angles $\phi_v = 44.5^{\circ}$ and
$\phi_b = 134.8^{\circ}$, which does not affect our qualitative
analysis, however.

\begin{figure*}%[htbp]
\centering
\includegraphics[width=0.498\linewidth]{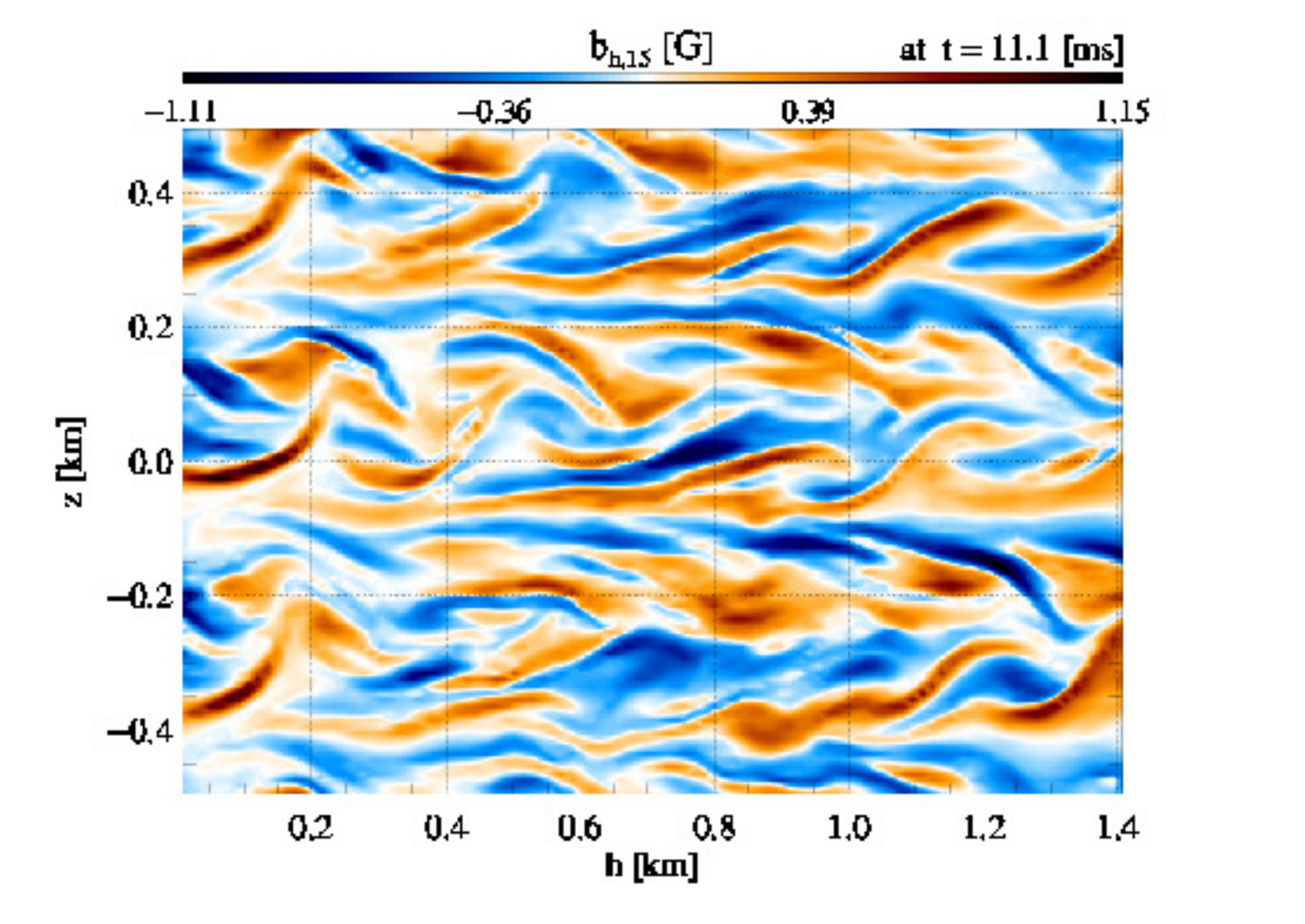}
\includegraphics[width=0.498\linewidth]{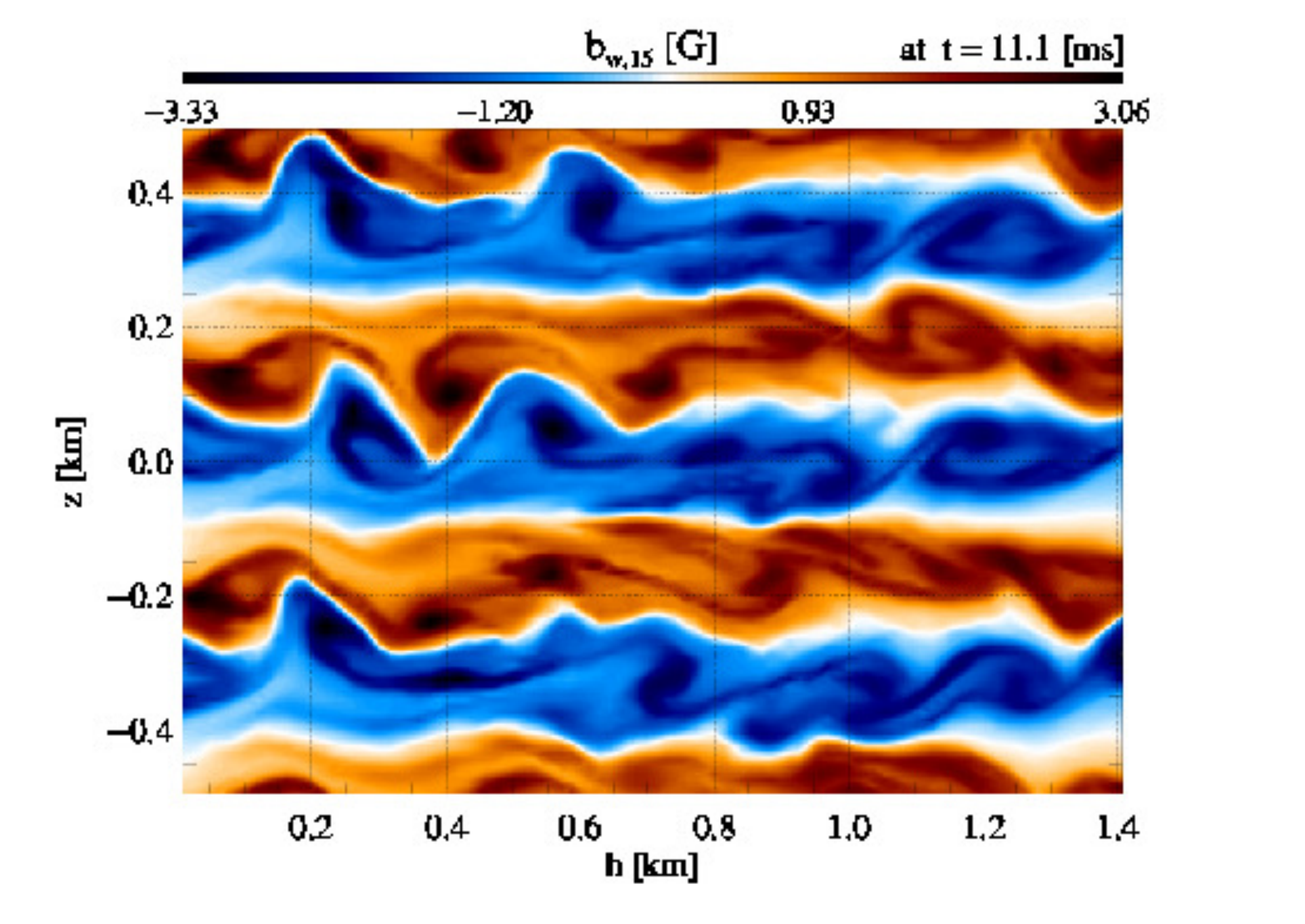}
\includegraphics[width=0.498\linewidth]{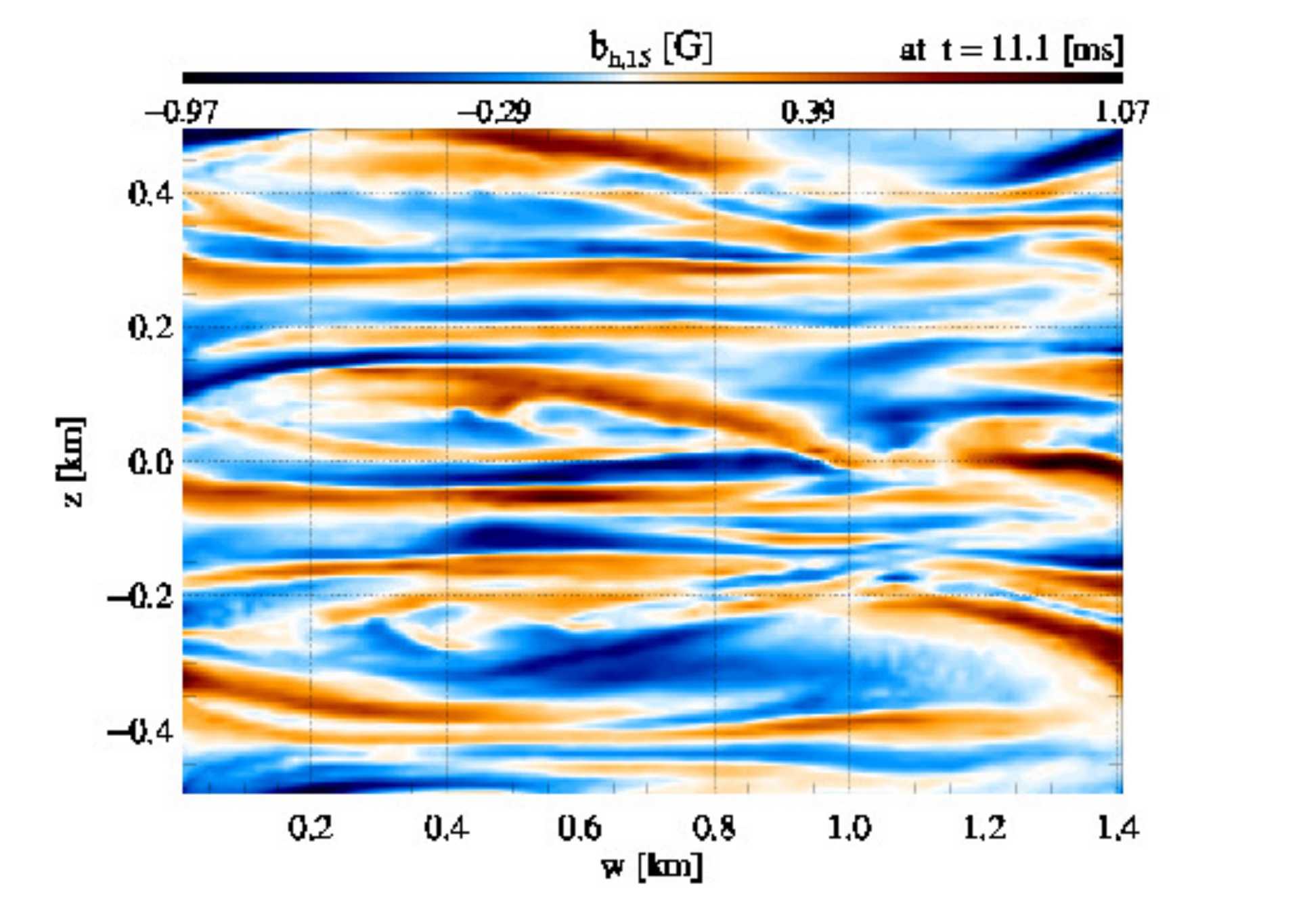}
\includegraphics[width=0.498\linewidth]{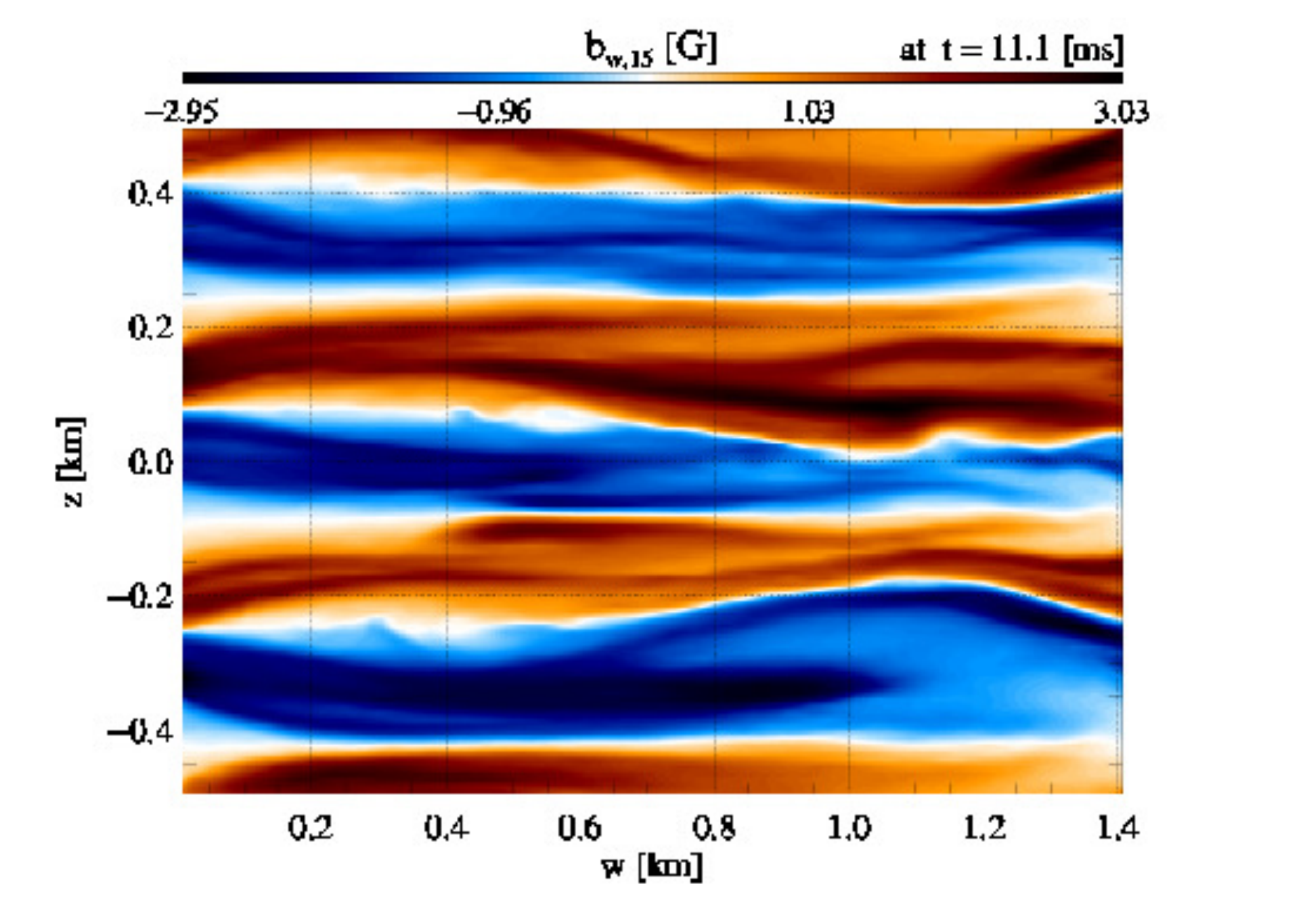}
\caption{The panels show the distribution of the magnetic field
  components $b_h$ (left) and $b_w$ (right) in two-dimensional cuts
  through the computational domain for the 3D model\,\#7 at time
  $t = 11.1\,$ms. The cut directions are perpendicular (top
  panels; $(h,z)$ cut) and parallel (bottom panels; $(w,z)$
  cut) to the direction of the magnetic MRI channels.  The locations
  of the cuts are marked in \figref{fig:mri_cut_geometry} in blue and
  red, respectively.}
\label{fig:mri_cuts}
\end{figure*}

Figure\,\ref{fig:mri_cuts} shows the distribution of the magnetic
  field components $b_h$ (\emph{left-hand panels}) and $b_w$ (\emph{right-hand
    panels}) in two-dimensional cuts through the computational domain
  of the 3D model\,\#7 
shortly before MRI termination. 
 The component $b_h$ is considerably smaller than the component
  $b_w$. This is consistent with the theoretical expectation for an
  MRI channel whose magnetic field should grow in the $w$ direction
  ($\phi_b = 135^{\circ}$) and vanish in the perpendicular $h$
  direction ($\phi = 45^{\circ}$).  Close to termination, the MRI
  channels are strongly perturbed and vortex rolls begin to form in
  the $h$ direction (\emph{upper panels}).  This indicates that KH
  instabilities developing along the velocity channels (\ie in the
  $h$ direction) are responsible for the disruption of the MRI
  channels.
  In the $w$ direction (\ie along the magnetic field channels
  separated by current sheets in which TM could develop), the channels
  remain almost unperturbed.  Only box-size structures appear which
  are most likely a consequence of the radial boundary conditions
  imposed in our simulations (see Sec.\,\ref{sSec:BC}). 
Although we cannot discard the presence of sub-dominant TM in this
projection, it is clear that TM do not play a dominant role in the
termination process, which can be understood completely in terms of
parasitic KH instabilities.

\begin{figure*}
\centering
%mnras  \sidecaption
\includegraphics[width=1\linewidth]{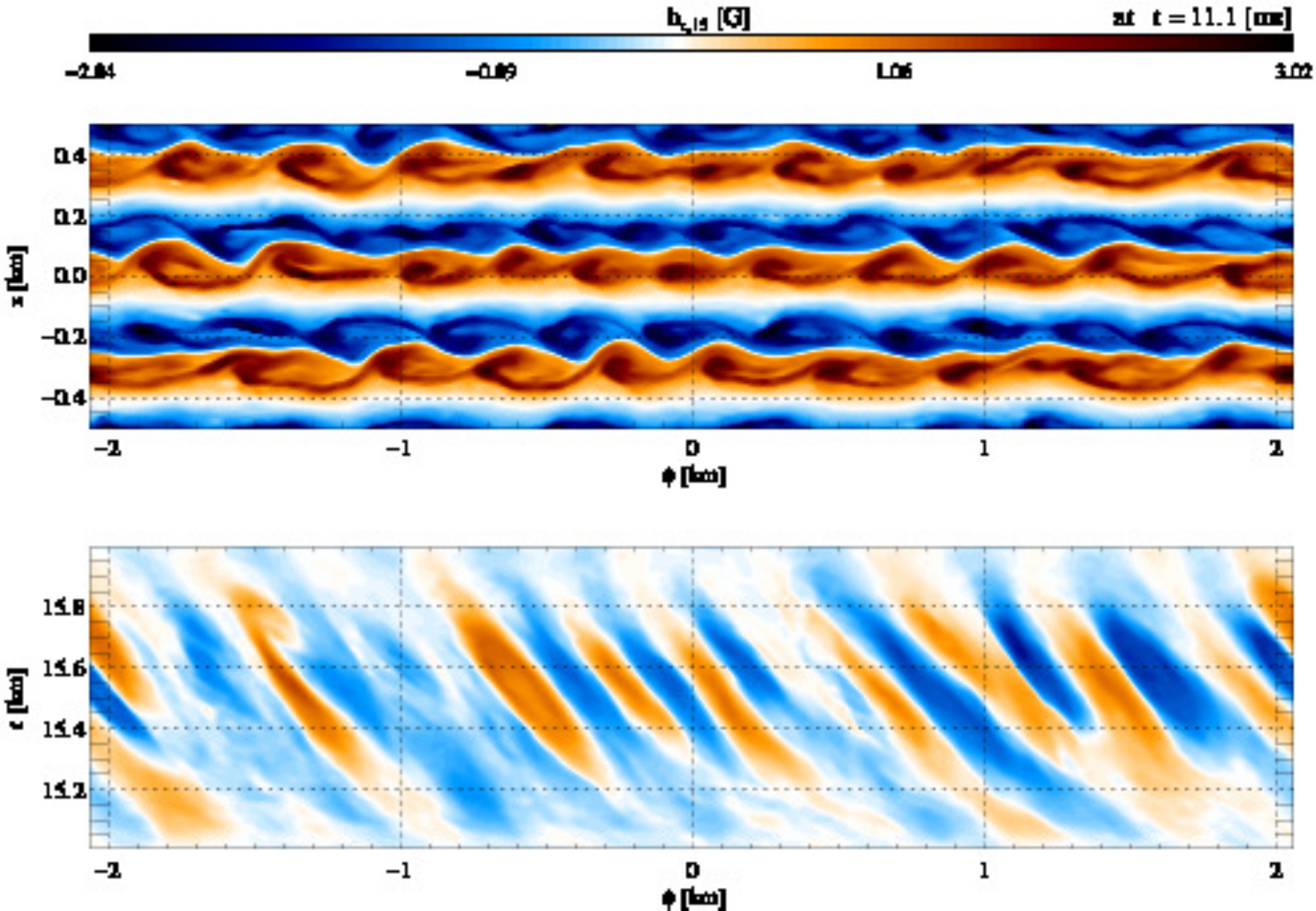}
\includegraphics[width=1\linewidth]{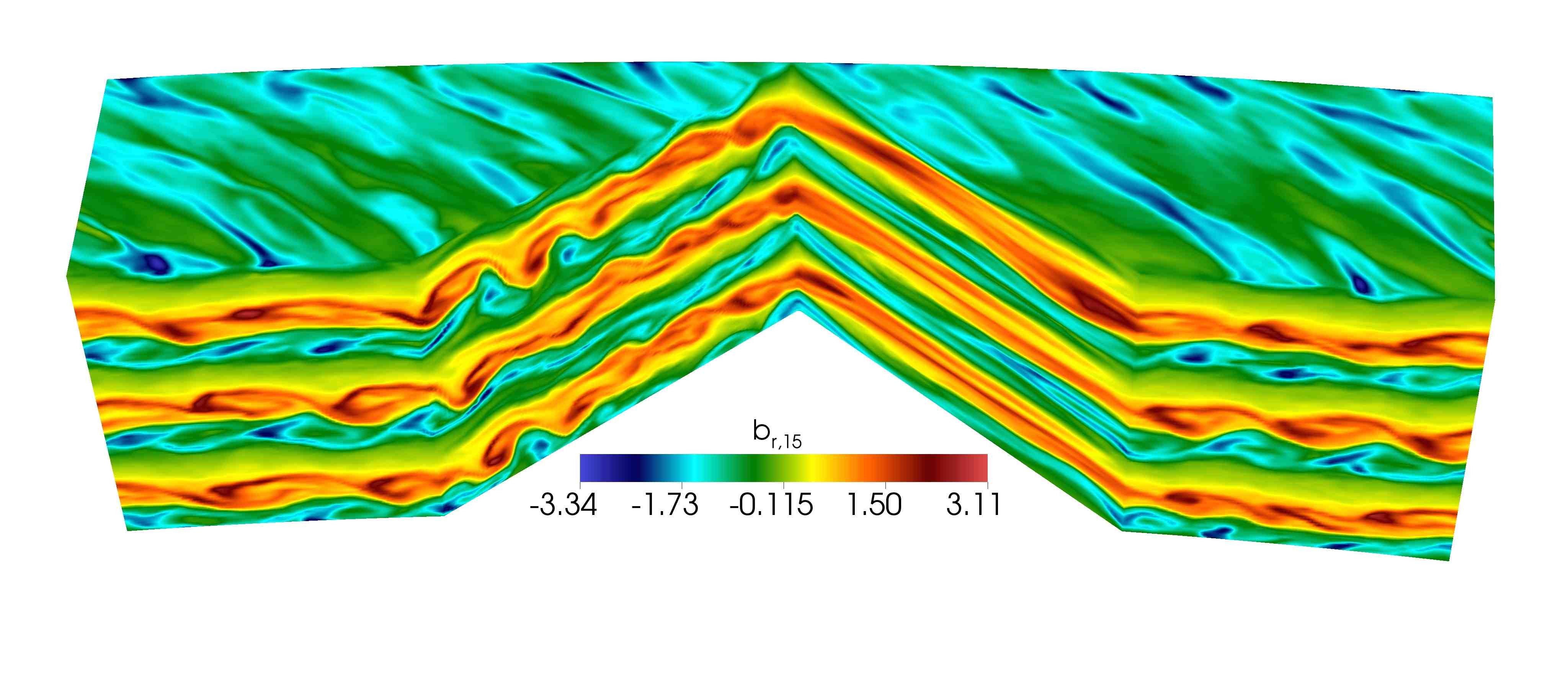}
\caption{Structure of the radial magnetic field component $b_r$ of the
  3D model\,\#7 at $t = 11.1\ \ms$.  \emph{Top:} $(z,\phi)$ cut at
  $r = 16\,$km.  Middle: $(r,\phi)$ cut at $z = -0.09\,$km.
  Bottom: 3D view of the surface of the computational domain
  where some part of it has been removed to show pieces of the $(h,z)$
  and $(w,z)$ cuts. Note that these cuts differ from those displayed
  in \figref{fig:mri_cuts}.}
\label{fig:deep_cuts}
%mnras \end{figure*}
\end{figure*}

Before we proceed further we take another look at the spatial
structure of $b_r$ during MRI termination at $t = 11.1\,$ms, which is
illustrated in \figref{fig:deep_cuts}.  In the upper panel depicting
$b_r$ in a $(\phi,z)$ cut at $r = 16\,$km (this cut corresponds to the
outer radial boundary of the computational domain shown in the upper
right panel of \figref{fig:3D_mri}), one can recognize vortex rolls
developing along the MRI channels. From the middle panel, which shows
the distribution of $b_r$ in the $(r-\phi)$ plane at $z =0.335\,$km,
one could determine the horizontal wavelength $\lambdap$ and the angle
at which parasitic instabilities develop ($\phip \approx 45^{\circ}$).
However, we can obtain these quantities much more accurately using
Fourier transforms (see next subsection).  The bottom panel provides
another view of the structure of $b_r$. By showing a part of both the
$(h,z)$ cut and the $(w,z)$ cut, it illustrates in which direction
parasitic instabilities develop. We note that the $(h,z)$ and $(w,z)$
cuts shown in Figs.~\ref{fig:deep_cuts} and \ref{fig:mri_cuts} are
different.

%%%%%%%%%%%%%%%%%%%%%%%%%%%%%%%%%%%%%%%%%%%%%%%%%%%%%%%%%%%%%%%%%%%%%%
%%%%%%%%%%%%%%%%%%%%%%%%%%%%%%%%%%%%%%%%%%%%%%%%%%%%%%%%%%%%%%%%%%%%%%

%%%%%%%%%%%%%%%%%%%%%%%%%%%%%%%%%%%%%%%%%%%%%%%%%%%%%%%%%%%%%%%%%%%%%%
%%%%%%%%%%%%%%%%%%%%%%%%%%%%%%%%%%%%%%%%%%%%%%%%%%%%%%%%%%%%%%%%%%%%%%

%%%%%%%%%%%%%%%%%%%%%%%%%%%%%%%%%%%%%%%%%%%%%%%%%%%%%%%%%%%%%%%%%%%%%%
%%%%%%%%%%%%%%%%%%%%%%%%%%%%%%%%%%%%%%%%%%%%%%%%%%%%%%%%%%%%%%%%%%%%%%

%%%%%%%%%%%%%%%%%%%%%%%%%%%%%%%%%%%%%%%%%%%%%%%%%%%%%%%%%%%%%%%%%%%%%%
%mnras \begin{figure*}[htbp]
%%%%%%%%%%%%%%%%%%%%%%%%%%%%%%%%%%%%%%%%%%%%%%%%%%%%%%%%%%%%%%%%%%%%%%

%%%%%%%%%%%%%%%%%%%%%%%%%%%%%%%%%%%%%%%%%%%%%%%%%%%%%%%%%%%%%%%%%%%%%%
\subsubsection{Fourier analysis}
\label{sec:FFT}

In order to confront the theoretical expectations with our numerical
results, we have calculated spatial discrete 3D Fourier transforms of
the magnetic field components $b_\alpha$ with
$\alpha \in{r, \phi}$ at a given time using a fast Fourier Transform
(FFT) algorithm.  We denote the complex FFT coefficients as
$a_\alpha$. The power spectral density is proportional to
$|a_\alpha|^2$, which is a measure of the average magnetic field
energy density of the component $b_\alpha$ in Fourier space.  We chose
the components $b_r$ and $b_\phi$ for the analysis, because they
contain the information about both the MRI channel flows and the
parasitic instabilities. We are particularly interested in determining
the wavevectors, ${\bf k}=(k_r,k_\phi,k_z)$, of the dominant modes,
which have the largest amplitudes in the Fourier space.  We expect
that MRI channel flows appear as structures with a wavevector
\begin{equation}
  {\bf k}_{\rm mri} = (0, 0, k_{\rm mri}),
\end{equation}
whose modulus is maximum for the fastest-growing mode, $k_{\rm MRI}$.

Parasitic instabilities develop in the whole Fourier space, whereas
the MRI does not contribute to modes with finite $k_{r}$ and
$k_{\phi}$, but $k_z = 0$. Hence, parasitic instabilities are expected
to produce a characteristic signature with wavevectors
\begin{equation}
  {\bf k}_{\rm p} = (k_r, k_\phi, 0),
\end{equation}
which should be distinguishable from the MRI signature.  The maximum
Fourier amplitude should be attained for the wavevector of the
fastest-growing parasitic mode.

\begin{figure}%[t]
\centering
\includegraphics[width=0.8\linewidth]{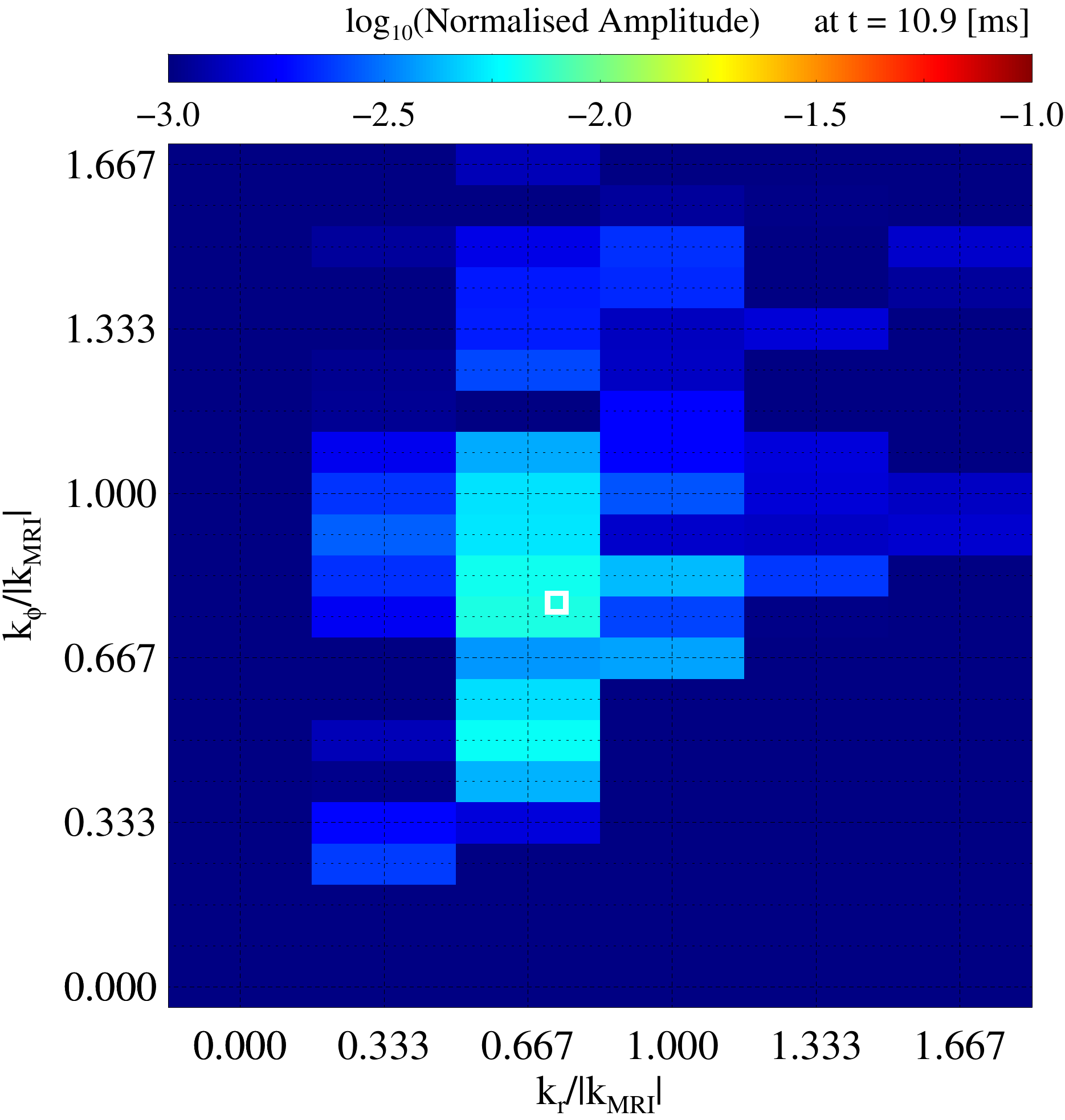}
\includegraphics[width=0.8\linewidth]{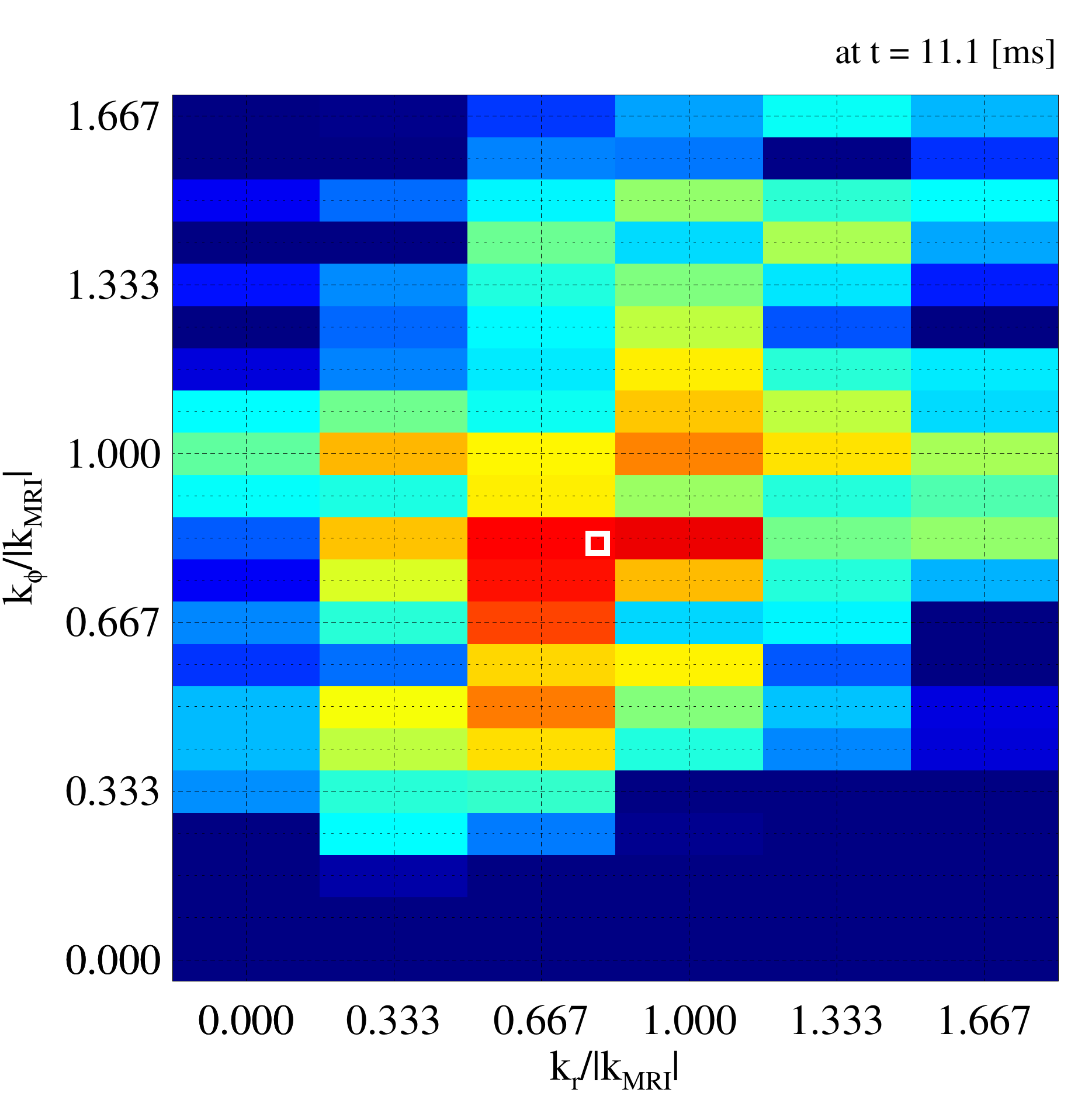}
\includegraphics[width=0.8\linewidth]{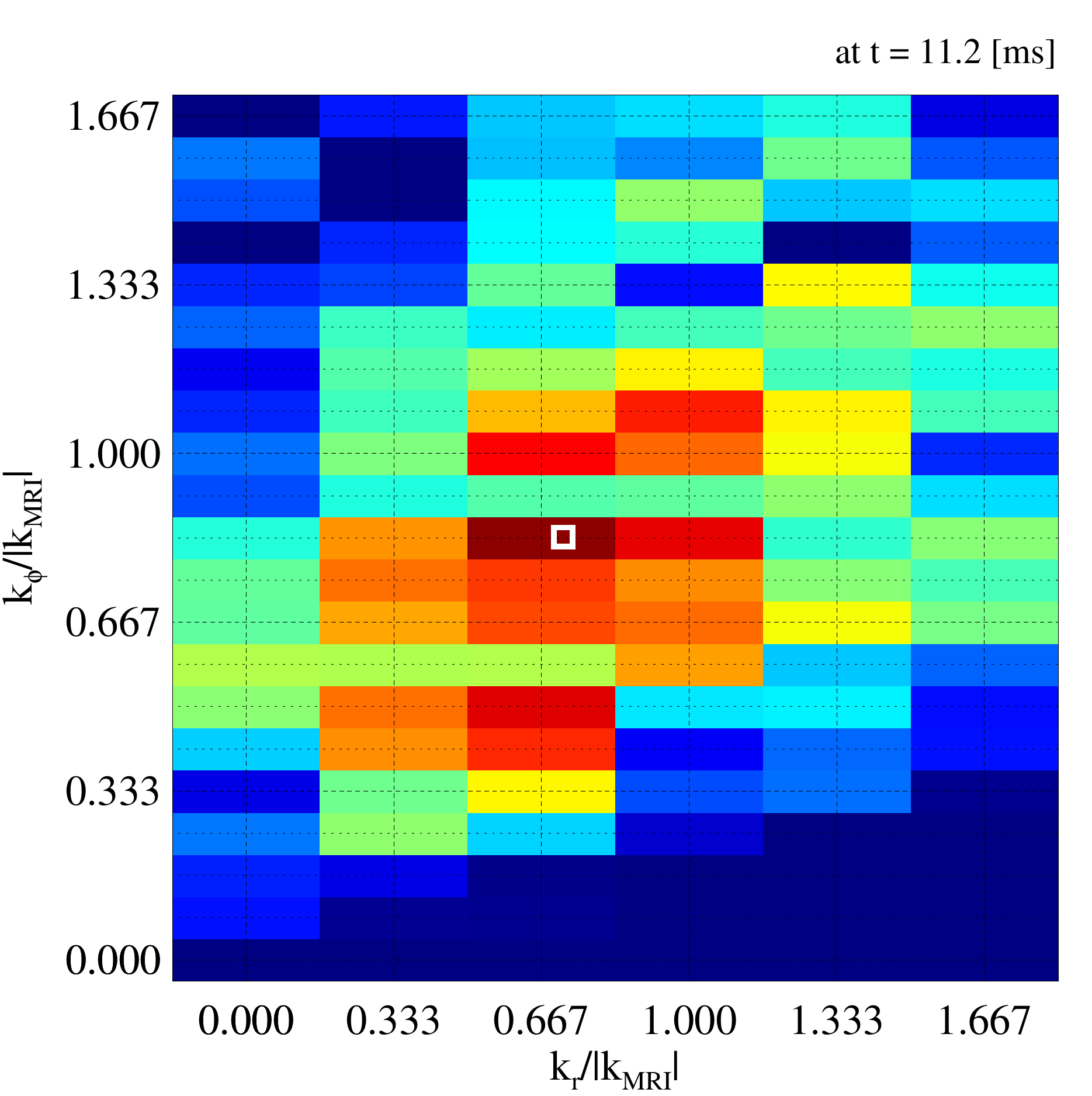}
\caption{Fourier amplitudes of the radial magnetic field $b_r$ for
  model\,\#7 at $10.9\,$ms (top), $11.1\,$ms (middle), and $11.2\,$ms
  (bottom), respectively. The colour coded quantity is the ratio
  $|a(k_r, k_\phi,0)| / |a(0,0 ,k_{\rm MRI})|$ of the Fourier
  amplitude representing parasitic instabilities and the one
  representing MRI channels. The white square is the location of the
  energy-weighted barycenter computed with Eq.\,(\ref{eq:barycenter}).}
\label{fig:fourier}
\end{figure}

The analysis of model\,\#7 reveals that the Fourier amplitude along
the line $k_r=k_\phi=0$ peaks at $k_z=18.8\,$km$^{-1}$, which
corresponds to the wavelength of the fastest-growing MRI mode,
 $\lambdamri \approx 0.333\,$km of this model. The latter result
holds for all our simulations during the phase of exponential growth
of the MRI.  The Fourier amplitudes in the plane $k_z=0$, displayed
for three different times for model\,\#7 in \figref{fig:fourier}, show
a power excess which peaks at
$(k_r, k_\phi) \approx (0.7,0.7)$~km~$^{-1}$.  The location of this
maximum barely changes with time, while the amplitudes of the Fourier
coefficients increase relative to those at $(0, 0, k_{\rm MRI})$. This
result is consistent with a super-exponential growth of parasitic
instabilities. In addition, as we will demonstrate below, the
behaviour of the Fourier amplitude in the $k_z=0$ plane is consistent
with the development of parasitic KH instabilities feeding off the MRI
channels.

%%%%%%%%%%%%%%%%%%%%%%%%%%%%%%%%%%%%%%%%%%%%%%%%%%%%%%%%%%%%%%%%%%%%%%
%%%%%%%%%%%%%%%%%%%%%%%%%%%%%%%%%%%%%%%%%%%%%%%%%%%%%%%%%%%%%%%%%%%%%%

The average magnetic energy density of the field component $b_\alpha$
can be computed from the Fourier amplitudes as
\begin{equation}
  e_{{\rm mag}, \alpha} =  \frac{1}{2} \sum_{l = -N_r/2}^{ N_r/2} \,  
                                     \sum_{m = -N_\phi/2}^{N_\phi/2}\,  
                                     \sum_{n = -N_z/2}^{N_z/2}
                         |a_\alpha(k_{l}, k_{m},k_{n})|^2,
\end{equation}
where
\begin{equation}
  (k_{l}, k_{m},k_{n}) = \left( \frac{2\pi \,l}{L_r}, 
                               \frac{2 \pi \,m}{L_\phi},
                               \frac{2 \pi \,n}{L_z} \right).
\end{equation}
Similarly, we can estimate the average magnetic energy density of the
MRI channels or parasitic instabilities restricting the summation to
locations in Fourier space relevant for each kind of
instability. Therefore, we define
\begin{align}
  e_{{\rm MRI}, \alpha} &= |a_\alpha(0, 0, k_{\rm MRI})|^2,
\\
  e_{{\rm p}, \alpha}   &= \frac{1}{2} \sum_{l = -N_r/2}^{N_r/2}\,  
                                     \sum_{m = -N_\phi/2}^{N_\phi/2}  
                          |a_\alpha(k_{l}, k_{m},0)|^2,
\end{align}
as proxies for the average magnetic field energy density associated
with these instabilities. In case of the MRI, the Fourier amplitudes
are distributed along the line $k_r = k_\phi = 0$, \ie
$e_{{\rm MRI}, \alpha}$ should be a good estimator. However, in the
case of the parasitic modes, $e_{{\rm p}, \alpha}$ contains
not all contributions to the energy density and thus provides only a
lower bound of the energy density associated with the parasitic
instabilities.

\begin{figure}%[t]
\centering
\includegraphics[width=1\linewidth]{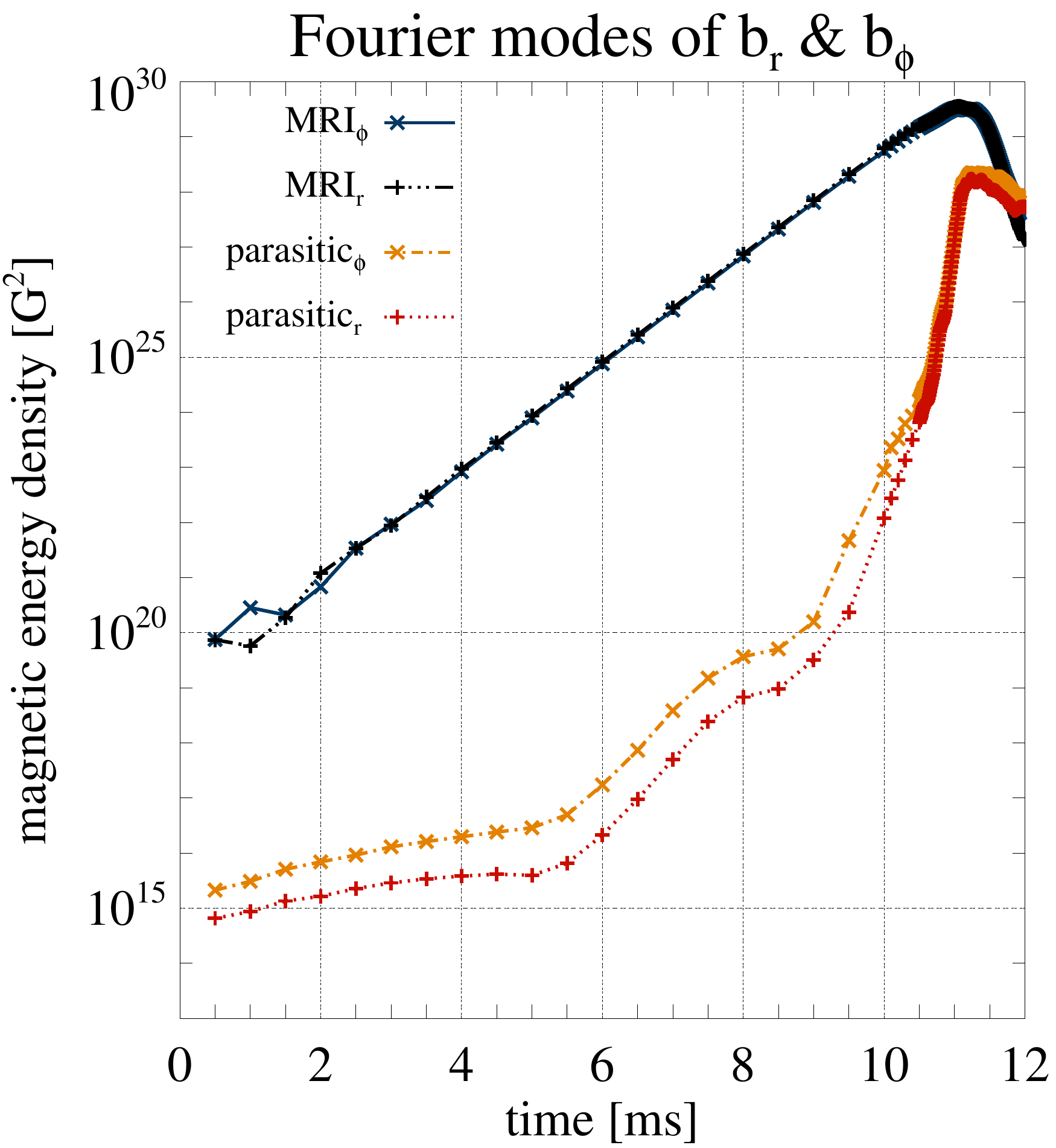}
\caption{Evolution of the average magnetic energy density associated
  with the MRI channels ($e_{\rm MRI, \alpha}$) and the parasitic
  instabilities ($e_{\rm p, \alpha}$) for different components
  $b_\alpha$ of the magnetic field for model\,\#7.}
\label{fig:parasites_time}
\end{figure}

\figref{fig:parasites_time} shows the time evolution of the magnetic
energy density for both the MRI and the parasitic instabilities. The
behaviour of $e_{{\rm MRI},\alpha}$ is very similar to the time
evolution of the Maxwell stress (see \figref{fig:mri_time_evolution}).
It is characterized by an exponential growth with the same growth rate
as for the Maxwell stress $\MMM$, and a termination point associated
with the disruption of the MRI channel flows.  The average magnetic
energy density $e_{{\rm p},\alpha}$ of the parasites 
 starts to grow
super-exponentially
 at $t \approx 9\, \ms$ 
from a value of about 
 $8$
 orders of magnitude smaller
than that of the MRI. At termination, the magnetic energy density of
the parasites amounts to, at least, 
 $6\%$
 of the magnetic energy
density of the MRI.

%%%%%%%%%%%%%%%%%%%%%%%%%%%%%%%%%%%%%%%%%%%%%%%%%%%%%%%%%%%%%%%%%%%%%%
%%%%%%%%%%%%%%%%%%%%%%%%%%%%%%%%%%%%%%%%%%%%%%%%%%%%%%%%%%%%%%%%%%%%%%

To determine the wavevector of the fastest-growing parasitic
instability, we search for the maximum Fourier amplitude with
$k_z=0$. For this purpose, it is sufficient to consider only positive
components of the wavevector ${\bf k}$. Numerically, we obtain the
wavevector by computing the energy-weighted barycenter in the relevant
part of Fourier space,
\begin{equation}
  k_{{\rm p}, \alpha} = \frac{\sum_{l= 1}   %^{N_r/2}
                            \sum_{m = 1}  %^{N_\phi/2}  
                            |a_r(k_{l}, k_{m},0)|^2 k_{\alpha}}{
                            \sum_{l = 1}  %^{N_r/2}
                            \sum_{m= 1}   %^{N_\phi/2}  
                            |a_r(k_{l}, k_{m},0)|^2}\, , 
\label{eq:barycenter}
\end{equation}
where we limited all summations to those Fourier modes displayed in
\figref{fig:fourier} to avoid high frequency contaminations.
Substituting $a_r$ for $a_{\phi}$ does not change the results
significantly.  

\begin{figure}%[t]
\centering
\includegraphics[width=1\linewidth]{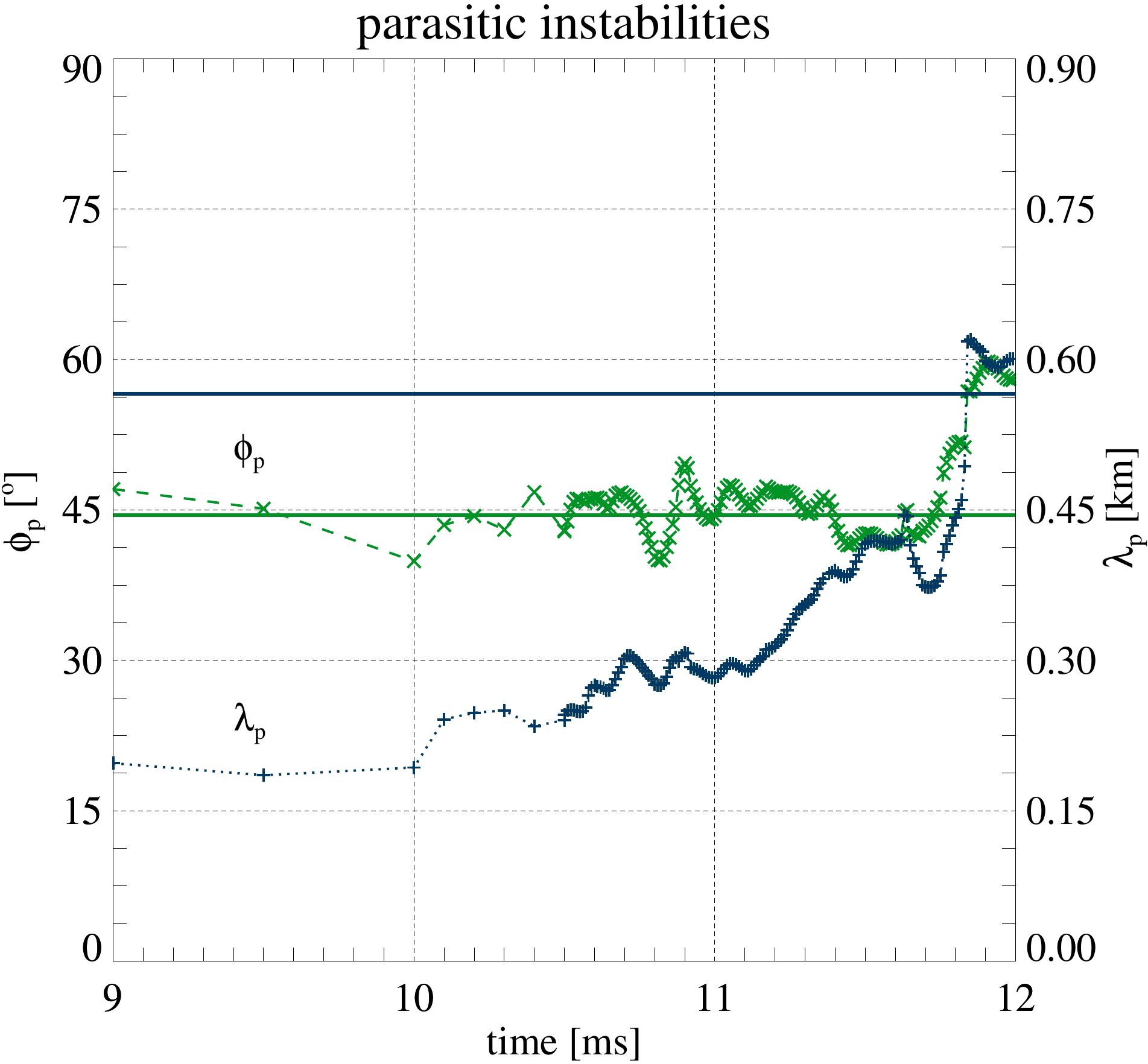}
\caption{Characteristics of the parasitic instabilities for model\,\#7
  during the late stage of MRI evolution.  Green crosses and blue
  diamonds depict the angle, $\phip$, and the wavelength, $\lambda$,
  of the parasites, respectively. Horizontal solid lines show the
  corresponding theoretical values for KH modes that one expects to
  develop in this simulation.}
\label{fig:parasitic_theta}
\end{figure}

The values of $k_{{\rm p},\alpha}$ obtained from 
\Eqref{eq:barycenter} (white squares in \figref{fig:fourier})
properly trace the location of the maximum Fourier amplitude. With
these values one can compute both the wavelength and the angle of the
parasitic instability according to
\begin{align}
  \lambdap      &= \frac{2 \pi}{\sqrt{k_{{\rm p},r}^2 + k_{{\rm p},\phi}^2} }, 
\\
  \phi_{{\rm p}} &= \arctan \left( \frac{k_{{\rm p},r}}{k_{{\rm p},\phi}} \right).
\end{align}
Figure\,\ref{fig:parasitic_theta} shows the time evolution of the
wavelength, $\lambdap$, and angle, $\phi_p$, of the parasitic
instabilities during the late stage of MRI evolution.  
Before MRI termination, the evolution of the angle is compatible
  with its theoretically expected behaviour for KH instabilities,
  \ie $ \phip \approx 45^{\circ}$ (horizontal green line), within
  the accuracy of the angle determination.%
  \footnote{The accuracy in the determination of the angle and the
    wavelength, $\approx 9^{\circ}$ and $\approx 10 \%$ respectively,
    depends on the accuracy in the determination of
    ${\bf k_{\rm p}}$, which is set by the size of the box. }
 The wavelength differs from its theoretically expected value
  $\lambdap \approx 0.56\ \km$ (horizontal blue line) by a factor of
  $\sim 2$. 
Whether the source of this disagreement is of a numerical origin
  or results from a limitation of the theoretical approach of
  \cite{Pessah} 
is   beyond the scope of this work.

%%%%%%%%%%%%%%%%%%%%%%%%%%%%%%%%%%%%%%%%%%%%%%%%%%%%%%%%%%%%%%%%%%%%%%
%%%%%%%%%%%%%%%%%%%%%%%%%%%%%%%%%%%%%%%%%%%%%%%%%%%%%%%%%%%%%%%%%%%%%%

We have performed a similar analysis for all our other simulations.
Table\,\ref{tab:main_results} gives the values of $\lambdap$ and $\phi_p$
at a representative time before termination (\ie at $11\,$ms for
models \#1 to \#16, and at $12.5\,$ms for models \#17 and \#18).  In
the following three subsections, we discuss the influence of different
numerical and physical parameters on the values of these two
quantities.

%%%%%%%%%%%%%%%%%%%%%%%%%%%%%%%%%%%%%%%%%%%%%%%%%%%%%%%%%%%%%%%%%%%%%%
\subsubsection{Box size}

Next, we address whether the size and aspect ratio of the
computational domain influences MRI termination {\cite[for a somewhat
  similar study of the post-termination phase, see][]{Bodo}}.
Following \citet{Obergaulinger_et_al_2009}, we simulated models \#5,
\#6, \#7, and \#10 in a box of (the default) size
($L_r \times L_{\phi} \times L_z = 1\ \km \times 4\ \km \times 1\
\km$).
We performed additional simulations reducing the size of the box in
$\phi$ direction ($1\ \km \times 1\ \km \times 1\ \km$, model \#8),
and both in $\phi$ and $z$ direction
($1\ \km \times 1\ \km \times 0.333\ \km$, models \#2, \#3, \#4, \#9,
and \#11).
Finally, we computed several models (\#12 to \#18) where we varied the
azimuthal size $L_{\phi}$ of the domain (see
Table\,\ref{tab:main_results}).

The main motivation for using a smaller box in some of our simulations
was computational cost reduction.  In accordance with theoretical
predictions of \citet{Pessah} for flows with $\Ree, \Rm \gg 1$, we
found that in model\,\#7 parasitic instabilities develop at an angle
$\phip \approx 45^{\circ}$ (see \figref{fig:parasitic_theta}, and
middle panel of \figref{fig:deep_cuts}).  This result suggests that it
is unnecessary to use a box elongated in azimuthal direction
($L_{\phi} > L_r$). Instead, one could rather study the MRI in a box
with a horizontal aspect ratio $L_r = L_{\phi}$ to minimize the volume
of the computational domain without affecting the development of
parasitic instabilities.  To test this hypothesis, we simulated
model\,\#8 in a box of size
$L_r \times L_{\phi} \times L_z = 1\ \km \times 1\ \km \times 1\ \km$
using the same spatial resolution as for reference model\,\#7.  In
both simulations, the Maxwell stress tensor at MRI termination
attained the same value, which confirms our expectations.

We can further reduce the computational domain in the vertical
direction based on the following observation.  In simulations
performed with the default box, we chose the magnetic field strength
in such a way that three of the fastest-growing MRI modes fit in the
computational domain.  Thus, it should be possible to reduce the
vertical extent of the box by a factor of $3$, \ie
$L_z = \lambdamri \approx 0.333\ \km$, without hindering the growth of
the dominant MRI mode.  However, in such a smaller domain parasitic
instabilities are restricted to modes having a vertical wavelength
equal to the width of an MRI channel (called Type-I modes by GX94),
and modes with a longer vertical wavelength (Type-II') are
suppressed.  Nonetheless, according to the predictions of GX94 and
\cite{Pessah}, in the ideal MHD limit or for $\Ree,\Rm > 1$, the
dominant parasitic modes should be (KH modes) of Type-I.  Hence,
we expect the MRI termination process to be unaltered by a box of
smaller vertical size.

To test this theoretical prediction, we simulated model\,\#9 in a box
of size
$L_r \times L_{\phi} \times L_z = 1\ \km \times 1\ \km \times 0.333\
\km$
using the same grid resolution as in models\,\#7 and \#8. We found
that the Maxwell stress tensor reached the same value at MRI
termination in all three models.  This result not only confirms (given
our initial conditions) the termination of the MRI by ``Type-I''
parasitic modes, which is an important result, but it also justifies
the use of a computational box with a $12$ times smaller volume for
studying the MRI termination process.  We made use of this fact in our
resolution studies (see next subsection).

In Section \ref{sSec:2D}, we showed that the MRI is terminated by
TMs in axisymmetric 2D simulations.  One can also impose
axisymmetry in 3D simulations by choosing a box of vanishing azimuthal
length, \ie $L_{\phi} \rightarrow 0$.  Therefore, we expect that for
a certain small, but non-zero $L_{\phi}$, KH instabilities should be
suppressed even in 3D simulations, and the MRI should be terminated by
TM instead.  To determine this critical azimuthal box length, we
performed a series of simulations varying $L_{\phi}$ from 1.0 to
0.1\,km (models \#12 to \#18).  Moreover, to investigate the influence
of viscosity, we ran these simulations with a non-zero shear viscosity
corresponding to a hydrodynamic Reynolds number $\Ree = 100$. The
shear viscosity should change the wavelength of the fastest growing
MRI mode, but only by a negligible amount from the numerical point of
view (\ie less than $1\%$).

As expected, the MRI is terminated by KH instabilities in model\,\#12,
which only differs from model\,\#9 by the value of $\Ree$.  In both
simulations, the parasitic instabilities have the same horizontal
wavelength, $\lambdap$, and they develop at the same angle $\phip$
within the measurement error. In models\,\#13--\#16, the MRI is also
terminated by KH instabilities, but the value of the Maxwell stress
tensor at termination is somewhat larger.  This result can be easily
understood: parasitic modes that would grow fastest in a box with
$L_{\phi} = 1\,\km$ are affected (or even suppressed) in narrower
boxes, \ie the MRI can operate a bit longer before it is finally
terminated. 
The smaller the box, the stronger  the
 suppression of the KH modes and the larger  the deviation of
such quantities as the Maxwell stress at termination from model \#12.
Finally, in models\,\#17 ($L_{\phi} = 0.2\,\km$)
and \#18 ($L_{\phi} = 0.1\,\km$), KH modes are suppressed and the MRI
is terminated by TMs, showing similar features as in the 2D
simulations described in section~\ref{sSec:2D}.  We note that the
Maxwell stress reaches considerably higher values at termination in
these models, because TM are growing more slowly.
A similar behaviour was observed by \cite{Lesaffre_et_al} in
  their mean-field shearing-box simulations of accretion discs. In
  simulations done in a box of size
  $L_{r} = L_{\phi} = L_z = \lambdamri$, KH modes were suppressed and
  the MRI channels were disrupted by TM, whereas in simulations done
  in larger boxes in the horizontal directions (\ie
  $ L_{r} = L_z = \lambdamri$ and $ L_{\phi} = 4 \lambdamri$, or
  $ Lz = \lambdamri$ and $ L_r  = L_{\phi} = 4 \lambdamri$), the MRI was
  terminated by KH instabilities.

To summarize, MRI termination in models\,\#12--\#18 is well
explained within the parasite model, and the minimum azimuthal box
length for which KH modes are not severely affected by boundary
conditions is $L_{\phi} \approx  0.3  \,\km \approx  1  \lambdamri$.

%%%%%%%%%%%%%%%%%%%%%%%%%%%%%%%%%%%%%%%%%%%%%%%%%%%%%%%%%%%%%%%%%%%%%%
\subsubsection{Influence of grid resolution}

To properly resolve MRI termination, one needs to use a resolution
that is high enough not only to resolve the MRI channels, but also
parasitic instabilities.  The latter criterion is more stringent, as
parasitic instabilities develop finer structures in 
vertical
direction than MRI channels do \citep[GX94,][]{Pessah}.

We explore this issue in a series of simulations performed in the
small
($L_r \times L_{\phi} \times L_z = 1\,\km \times 1\,\km \times 0.333\,\km$;
models \#2, \#3, \#4, \#9, and \#11) and in the default
($L_r \times L_{\phi} \times L_z = 1\,\km \times 4\,\km \times 1\,\km$;
models \#5, \#6, \#7, and \#10) computational box.  As we have
demonstrated in the previous subsection, parasitic instabilities can
develop equally well in both computational domains, \ie the boxes
are equally well suited for resolution studies.

In models \#2--\#4, which were simulated with 8--16 zones per MRI
channel in the vertical direction, we observe differences already
during the phase of exponential growth. First, channel modes emerge
$\approx 2\,\ms$ later than in the other better resolved simulations.
Secondly, the channel modes somewhat differ from the analytic
solution, \ie some imperfections develop on channels which are of
numerical and not physical origin.  Thirdly, the MRI growth rate
measured with the help of $\MMM$ slightly fluctuates during the phase
of exponential growth.  These fluctuations of the order of a few
percent can be explained by the fact that the above mentioned
imperfections also contribute to $\MMM$.

In models \#5--\#11, which were simulated with at least 20 zones per
MRI channel, the channel modes appear roughly at the same time and
grow at the same constant rate during the phase of exponential growth.
Based on these observations, we conclude that 20 zones per MRI channel
are sufficient to resolve the phase of exponential growth with the MP9
scheme (but more zones will be required for lower order reconstruction
schemes).
 This result is consistent with previous estimations of the
  numerical viscosity and resistivity of the code done by
  \cite{Rembiasz}.  With the help of tests involving Alf\'en wave
  propagation and TM instabilities, he determined scaling laws for the
  numerical viscosity and resistivity of the code as a function of
  resolution and initial conditions.  Given our numerical setup, the
  numerical viscosity and resistivity of the code are
  $<10^7\, \cm^2 \s^{-1}$, if the relevant length scale is covered by
  at least 20 zones (see also section\,4.2.5.1).

To investigate the nature of the parasitic instabilities that are
responsible for quenching the MRI, we used both Fourier analysis and
the local magnetic field structure during MRI termination.  In
models\,\#2 to \#11, we find parasitic instabilities of KH type
developing at an angle consistent with $\phip = 45^{\circ}$
independent of resolution. The horizontal wavelength of the parasitic
modes ranges from $\lambdap \approx 0.50\,\km$ in the coarsest resolved
model\,\#2 to $\lambdap \approx 0.23\,\km$ in the best resolved
models\,\#10 and \#11.

Remarkably KH instabilities develop even in model\,\#2, although it
was simulated with the coarsest grid which noticeably affects the
structure of the channel modes. Their identification was not
straightforward, however.  From the patterns recognizable in the
magnetic field structure during MRI termination, one can exclude TM
and other obvious numerical artefacts, e.g., arising from imperfect
boundary conditions.  We also find patterns resembling underresolved
vortex rolls which we attribute to KH modes, although without the help
of Fourier analysis, our experience gained from observing similar
patterns in better resolved simulations, and most importantly a
theoretical model \citep[GX94;][]{Pessah}, this identification would
have been impossible.

\begin{figure}
\vspace{0.05cm}
\hspace{0.75cm}
\includegraphics[width=0.9\linewidth]{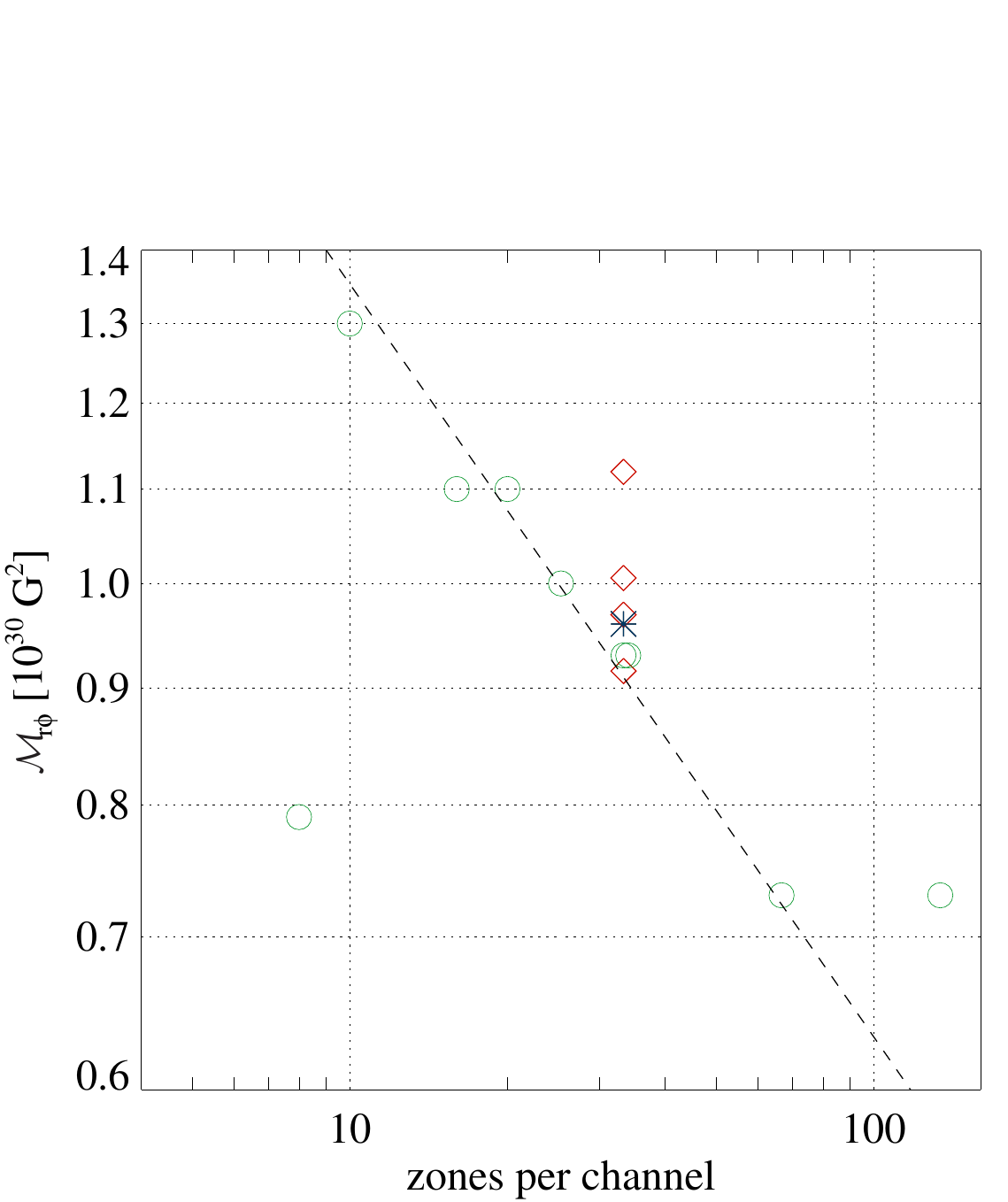}
\caption{MRI termination amplitudes as a function of grid resolution
  for models \#1 (blue asterisk), \#2--\#11 (green circles), and
  \#12--\#18 (red diamonds), respectively.  The dashed line
  represents a power law in the grid resolution with an index of
  $-1/3$. }
\label{Fig:reso}
\end{figure}

The value of the Maxwell stress at MRI termination depends on the grid
resolution (see \tabref{tab:main_results} and green circle symbols in
\figref{Fig:reso}). It is larger for coarse grids, the only exception
being model\,\#2.  Using the MP9 reconstruction scheme, the Maxwell
stress converges to a value of
$\MMM = 0.73 \times 10^{30}\, \mathrm{G}^2$ in models \#10 and \#11
simulated with 67 and 134 grid cells per MRI channel, respectively.
The dashed line in \figref{Fig:reso} shows that $\MMM$ follows roughly
a power law with index $-1/3$ as a function of the number of zones per
MRI channel, except for the under-resolved model\,\#2 with only eight
zones per channel, in which MRI growth terminates at a much smaller
value of $\MMM$.

This behaviour of $\MMM$ can be explained within the parasite model.
 The KH modes responsible for MRI termination grow from the
  shear layers created by the MRI channels.  During their growth,
  these secondary instabilities develop structures that are even
  smaller than the MRI channels.  Therefore, to resolve these
  structures 
a finer grid is necessary than for the MRI channels themselves.  As a
result, simulations may not be converged even if the numerically
obtained MRI growth rate is close to the theoretical value.  Because
badly resolved KH modes grow at a rate below the theoretical
expectation, parasitic instabilities reach the necessary amplitude to
disrupt the channels slightly later during the evolution, and the MRI
will terminate at a somewhat higher amplitude than in well-resolved
simulations.

The appearance of the outlier (model\,\#2) in \figref{Fig:reso} is not
surprising.  We have already seen that MRI channels are
underresolved in this model. Therefore, one would not expect to find
any reasonable termination amplitude for this model, \ie it is
probably a chance coincidence that this model gives roughly the same
results as the other better resolved models.

For completeness, we included also models\,\#12--\#18 in
\figref{Fig:reso} (red diamond symbols), for which the parasitic modes
are restricted not only by the grid resolution but also by the box
size and some physical viscosity.  The data of these models lie all
above the dashed line indicating the influence of resolution alone.
Finally, the asterisk symbol marks the result of the ideal MHD
model\,\#1, which we discuss in the next subsection.

%%%%%%%%%%%%%%%%%%%%%%%%%%%%%%%%%%%%%%%%%%%%%%%%%%%%%%%%%%%%%%%%%%%%%%
%%%%%%%%%%%%%%%%%%%%%%%%%%%%%%%%%%%%%%%%%%%%%%%%%%%%%%%%%%%%%%%%%%%%%%

%%%%%%%%%%%%%%%%%%%%%%%%%%%%%%%%%%%%%%%%%%%%%%%%%%%%%%%%%%%%%%%%%%%%%%
\subsubsection{Influence of viscosity and resistivity}

\paragraph{Viscosity}

We have performed the reference simulation (model \#7; see
\tabref{tab:main_results}), and the set of models described above
without any explicit physical viscosity.  To study whether a finite
physical viscosity corresponding to $\Ree \approx 100$ (for
which the equations of ideal hydrodynamics should still approximately
hold) leads to more than mere quantitative changes, we compare the
results of models \#9 and \#12. As expected, in both of these two
models, which differ only in the value of the hydrodynamic Reynolds
number ($\Ree = \infty$ for model \#9, and $\Ree \approx 100$
for model \#12) the growth of the MRI is terminated by KH modes.

 Even though no physical viscosity was used for model \#9, it is
  affected by a non-zero numerical viscosity.  To infer the
  quantitative importance of the numerical viscosity, we first compare
  models which differ only in grid resolution in the inviscid limit,
  e.g., model \#9, where the evolution is still notably affected by
  numerical viscosity, and models \#10 and \#11, which are already
  converged, as we have argued in the previous section. We note that
  the numerical viscosity accounts for the $\simeq 22\%$ difference in
  the volume-averaged Maxwell stress at termination (\ie the
  parasitic instabilities are somewhat under-resolved), and has a
  negligible impact on the MRI growth rate.

We now turn to the comparison of an inviscid model (\#9) with
  one where a physical viscosity has been included (\#12). Quantifying
  the exact contribution of the numerical viscosity in a model that
  includes also physical viscosity is much more involved. The reasons
  for this difficulty are multiple. On the one hand, the numerical
  viscosity is a scale dependent effect, affecting more the smallest
  scales. Hence, it is relatively more important for the development
  of parasitic modes than for the MRI growth.  On the other hand, the
  volume-averaged Maxwell stress at termination depends chiefly on a
  delicate balance between two competing factors. First, models
  computed with a finite physical viscosity have a smaller
  $\gamma_{\mathrm{MRI}}$. Thus, for a similar growth time, ${\cal
    M}_{r\phi}$ should be smaller than for models computed without a
  physical viscosity. Secondly, this effect is offset, however, by the
  also smaller growth rate of the KH parasitic modes, which defers the
  saturation process and allows for a further growth of ${\cal
    M}_{r\phi}$. Thereby, our comparison is of qualitative nature
  only. In view of the former reasoning, we tentatively attribute the
  $\sim 4\%$ difference in ${\cal M}_{r\phi}$ between the inviscid
  model \#9 and the viscous model \#12 to the effect of a finite
  viscosity in the latter one, though we cannot unambiguously say
  which part of this difference is caused by numerical viscosity. To
  support the argument that a finite physical viscosity and not a
  numerical viscosity is driving the dynamics, we note that the MRI
  growth rate decreases significantly in model \#12 (as also in all
  other viscous models from \#13 to \#18). This expected behaviour (see
  Section \ref{sSec:intability_criterion}) implies that including a
  physical viscosity causes measurable differences already in the MRI
  growth phase.

\paragraph{Resistivity}
In a final set of simulations, we investigated the influence of a
finite (physical) resistivity on MRI termination.  Model\,\#1,
simulated in ideal MHD, represents the typical conditions prevailing
in core collapse supernovae close to the surface of PNS, whereas
in models\,\#19 and \#20 we explored the (opposite) limit of a very
high resistivity.  With the latter simulations we also address a
prediction of \citet{Pessah}, namely that for $\Rm < 1$, the MRI
should be terminated by TM.  We performed the ideal MHD simulation of
model\,\#1 in the default size box, the results being very similar to
those obtained for model\,\#7, which we simulated with the identical
setup, but with $\Rm  \approx 100$.

As expected, the MRI grew in the ideal MHD simulations at a somewhat
higher rate ($\gammamri = 1137\,\s^{-1}$), because its growth was not
damped by resistivity.  We found that the MRI is terminated by
parasitic KH instabilities in this model, too. The parasites develop
at an angle $\phip \approx 45^{\circ}$ in accordance with the parasite
model of GX94.  The Maxwell stress at MRI termination is
$\MMM = 0.96 \times 10^{30}\,\mathrm{G}^2$, a value which is similar
to those obtained for models\,\#7 and \#12 (the latter one being
simulated with $\Ree  \approx   \Rm =100$).  From these
findings we conclude that for Reynolds numbers $\Ree,\Rm  
  \gtrsim   100$ viscosity and resistivity have no strong influence
on the process of MRI termination and the saturation value of the
Maxwell stress.  However, one should not forget that in all three
simulations (\ie models\,\#1, \#19, and \#20), numerical resistivity
and numerical viscosity do contribute to the 'total' hydrodynamic and
magnetic Reynolds numbers (by lowering them).

In the limit of very high resistivity, \ie for $\Rm \ll 1$, the MRI
modes that grow fastest in ideal MHD are completely suppressed by
magnetic dissipation (PC08).  Only modes with
sufficiently long wavelengths survive the interplay between MRI
amplification and resistive damping in this limiting case, because
magnetic dissipation is weaker for these modes. The wavelength of the
fastest-growing mode is $\lambdamri \propto b_{0z} \Rm^{-1}$
(PC08).  To keep $\lambdamri \approx 0.333\,\km$ in
models\,\#19 ($\Rm = 0.1$) and \#20 ($\Rm = 0.05$), we had to reduce
the initial magnetic field strength to values of
$b_{0z} = 3.25 \times 10^{12} \,\mathrm{G}$ and
$b_{0z} = 1.63 \times 10^{12} \,\mathrm{G}$ , respectively.

\begin{figure}%[t]
\centering
\includegraphics[width=1\linewidth]{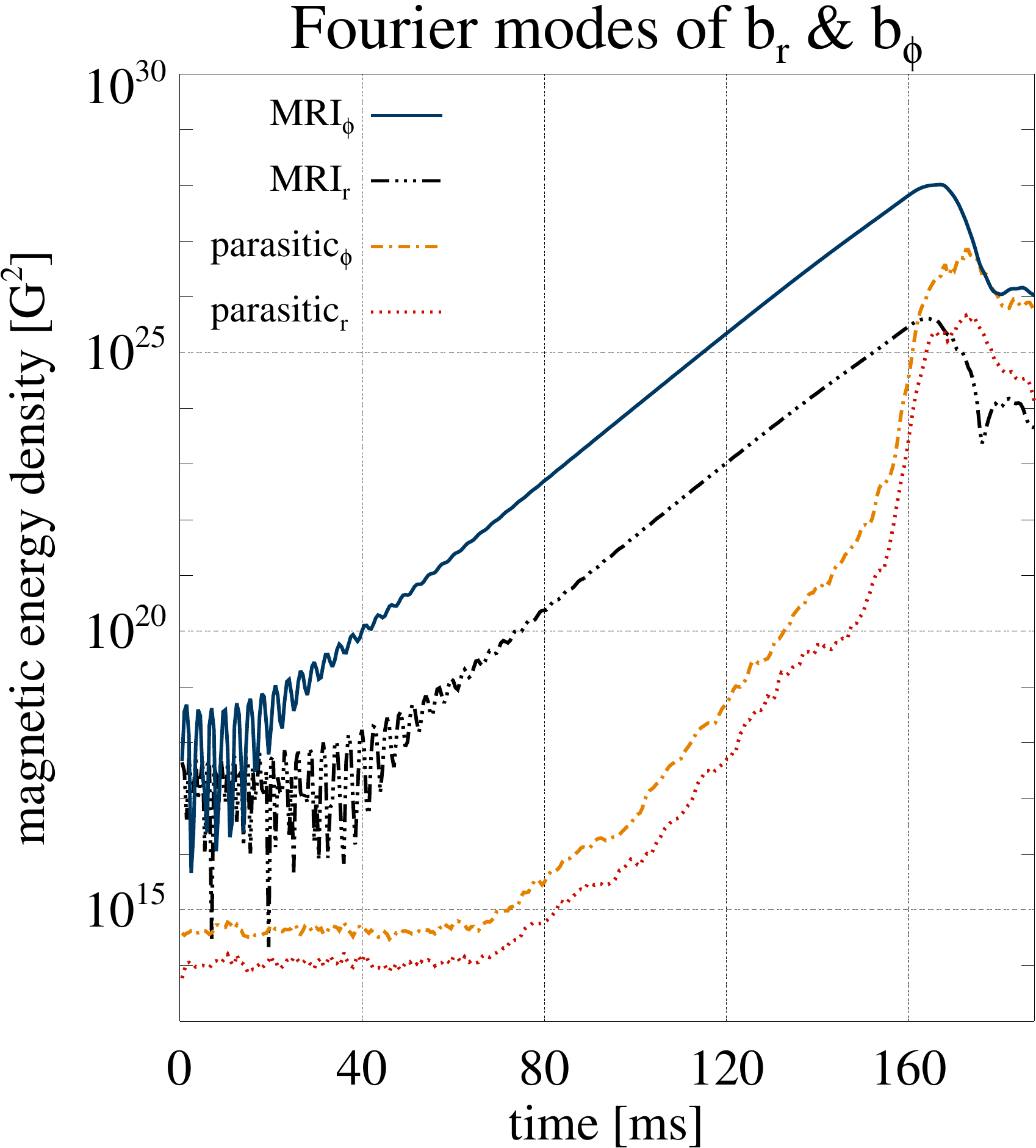}
\caption{Same as \figref{fig:parasites_time} but for model\,\#19
  simulated with a high resistivity ($\Rm = 0.1$). Note the very
  different time-scale at which the MRI evolves in this case.}
\label{fig:parasites_TM_time}
\end{figure}

We find that the fastest MRI mode grows at a considerably slower rate
than in the ideal MHD limit in both models, the measured growth rates
being $\gammamri = 80\ \s^{-1}$ (model\,\#19) and
$\gammamri = 34\ \s^{-1}$ (model\,\#20). These values agree very well
with those of  PC08, who predicted
$\gammamri = 79\ \s^{-1}$ and $\gammamri = 40\ \s^{-1}$, respectively.
PC08  also predicted that for $\Ree \gg 1$ and
$\Rm = 0.1$, we should find $\phi_v = 2^{\circ}$ and
$\phi_b = 94^{\circ}$, \ie the velocity channels and the magnetic
field channels should almost be aligned with the $r$ axis and the
$\phi$ axis, respectively. Indeed, our simulations show that
 $|b_{\phi}| \gg |b_{r}|$, \ie $\phi_b \approx 90 ^\circ$,
during the exponential growth phase (see
\figref{fig:parasites_TM_time}).

Finally, according  to \citet{Pessah}, the MRI should be
terminated by TM and not KH instabilities for $\Rm < 1$.  For
$\Rm = 0.1$, the dominant parasitic mode should have a wavelength
\begin{equation}
\lambdap =   \lambdatm = 2.1 \lambdamri
\end{equation}
developing at an angle $\phip = \phi_b = 94^{\circ}$.  A Fourier
analysis of the MRI modes and their parasites (see
\figref{fig:parasites_TM_time}) shows that shortly before MRI
termination ($t \approx 160\ \ms$) the energy stored in horizontal
Fourier modes increases significantly growing super-exponentially with
time.  However, because of numerical noise introduced by our imperfect
boundary conditions, we could neither identify parasitic instabilities
nor determine their angle with the help of the Fourier analysis.
Therefore, the type of the parasitic instability terminating the MRI
in model\,\#19 (as well as in model\,\#20) remains unknown. It is
probably neither KH nor TM, because we could not recognise the
characteristic features of none of these parasites.  The bottom line
is that based on models\,\#19 and \#20, we cannot confirm the
predictions of \citet{Pessah} that for $\Rm < 1$ the MRI is terminated
by TM.  As the highly resistive limit is of no direct relevance for
core collapse supernovae, we did not investigate this regime in more
detail.

%%%%%%%%%%%%%%%%%%%%%%%%%%%%%%%%%%%%%%%%%%%%%%%%%%%%%%%%%%%%%%%%%%%%%%
%%%%%%%%%%%%%%%%%%%%%%%%%%%%%%%%%%%%%%%%%%%%%%%%%%%%%%%%%%%%%%%%%%%%%%

%%%%%%%%%%%%%%%%%%%%%%%%%%%%%%%%%%%%%%%%%%%%%%%%%%%%%%%%%%%%%%%%%%%%%%
%%%%%%%%%%%%%%%%%%%%%%%%%%%%%%%%%%%%%%%%%%%%%%%%%%%%%%%%%%%%%%%%%%%%%%
\section{Summary and discussion}
\label{sec:summary}

\citet{Akiyama_etal__2003__ApJ__MRI_SN} first pointed out that the MRI
can develop in CCSNe and could lead to a strong amplification of the
magnetic field during the proto-neutron star phase. The presence of a
dynamically relevant magnetic field is of great interest because it
may have implications for the supernova explosion mechanism and for
explaining the origin of neutron stars magnetic fields. However, for
the typical magnetic field strength present in supernova progenitors,
it is necessary to resolve numerically MRI channels with a length
scale $\sim 1$\,m, which makes global simulations unfeasible even with
present day supercomputers. To overcome the limitation of global
simulations \ober performed semi-global simulations in a kilometre
size box located near the surface of the proto-neutron star. They
confirmed the hypothesis of \citet{Akiyama_etal__2003__ApJ__MRI_SN}
showing that the MRI can strongly amplify the magnetic field and
parasitic instabilities are able to destroy the MRI channels, thereby
quenching the amplification of the magnetic field. However, they were
unable to identify the agent terminating the MRI growth. Hence, \ober
could not give an upper limit on the MRI-driven magnetic field
amplification.

This  work tries to shed some light on the termination process.
We numerically studied the evolution of the MRI in the context of
a CCSN.  To this end, we performed a set of local 3D resistive-viscous
MHD simulations in a box representing typical conditions in a
proto-neutron star.  In all of our simulations we observed the growth
of the MRI and its subsequent termination. We identified that
secondary KH instabilities are responsible for MRI termination by
acting as parasites on the MRI channels. Our results are consistent
with the predictions of the parasite model proposed by GX94 and
further developed by \cite{Latter_et_al} and \cite{Pessah}.  Hence,
the parasite model, which is based on a local stability analysis of
the system, allows some insight into the termination process and
provides a valuable guidance for interpreting numerical results.  This
conclusion was obtained by an analysis of our numerical simulations
based on MRI theory and the parasite model in a number of tests:
\begin{enumerate}
\item We observe an exponential growth of MRI channels of a size
  compatible with the predicted value, $\lambdamri$, for the
  fastest-growing mode (BH91). The growth rates measured in the
  numerical simulations agree with those obtained from the local
  analysis within the simulation box. The presence of viscosity or
  resistivity lowers the values of the growth rates as expected from
  theory. In the limit of large hydrodynamic and magnetic Reynolds
  numbers, $\Ree, \Rm \gg 1$, the angle between velocity (magnetic
  field) in the MRI channels and the radial direction is close to the
  theoretically expected one in the ideal MHD limit by GX94, \ie
  $\phi_v = 45^\circ$ ($\phi_b = 135^\circ$).
  
\item For simulations with large hydrodynamic and magnetic Reynolds
  numbers, $\Ree, \Rm \gg 1$, the MRI is terminated by KH
  instabilities, as predicted by GX94 and \citet{Pessah} for this
  regime. To identify the KH instabilities we had to reduce the time
  interval between outputs of our simulations around termination to a
  value below the typical growth time of the parasites, in our case
  $0.1\, \ms$. This explains why \ober did not observe the development
  of KH instabilities, since they employed insufficiently frequent
  output. To help the analysis we project the magnetic field
  components into the directions $\phi_v$ and $\phi_b$. Shortly before
  termination we observe the development of vortex rolls at the
  position of the shear layer located along an angle
  $\phi_v = 45^\circ$. We identify the vortices as a consequence of KH
  instabilities, which eventually disrupt the channels causing
  turbulence. This behaviour is in agreement with the predictions of
  GX94 and \citet{Pessah}.

\item Analysing the numerical simulations by means of Fourier
  transforms, we have been able to determine the properties of the
  parasitic instabilities, \ie their horizontal wavelengths and the
  angles at which they develop. Our results confirm the theoretical
  predictions of GX94 and \cite{Pessah} that for $\Ree, \Rm \gg 1$ the
  dominant parasitic instabilities are KH modes developing parallel to
  the MRI velocity channels (\ie at an angle
  $\phip \approx 45^{\circ}$ in the ($r, \phi$) plane).  The
  horizontal length of these KH modes is in a reasonable agreement
  (within a factor of $2$) with the predictions of \citet{Pessah}.
  
\item Motivated by the good agreement with the parasite model, we
  explored the regime of very high resistivity (\ie $\Rm < 1$),
  although it is not of direct relevance for CCSNe. In this regime
  \citet{Pessah} predicts that the MRI should be terminated by TMs.
  To this end we performed two simulations with $\Rm = 0.1$ and
  $ 0.05$, but we were unable to identify the agent terminating
  the MRI growth, because no clear signature of TM or KH instability
  could be identified in our simulations. We conclude that possibly
  even higher resistivities are required for TM to become the dominant
  secondary instability.  
  A possible explanation of this discrepancy between
    theory and our simulations is the fact
    that the calculations of \citet{Pessah} are based
    on somewhat 
simplified assumptions.  Therefore, his results
    should be treated more like a guideline rather than exact
    predictions.

\end{enumerate}

To confirm the above conclusions, we have studied systematically the
effect of resistivity, viscosity, box size, and grid resolution. There
are a number of numerical artefacts that can affect the simulations:

\begin{enumerate}

\item In axisymmetric (2D) simulations, the MRI growth is terminated by
  TM. This was already reported by \citet{Obergaulinger_et_al_2009},
  who observed TM in their 2D ideal MHD simulations, in accordance
  with theoretical results obtained by \citet{Pessah}.
  \citet{Obergaulinger_et_al_2009} argued that this resistive MHD
  instability must have been triggered by numerical resistivity.  In
  2D simulations only axisymmetric parasitic modes can develop.  As
  axisymmetric KH modes are strongly suppressed by the magnetic field
  tension of the MRI channels, TM become the dominant secondary
  instability.  They suffer less strongly from this constraint,
  develop faster than axisymmetric KH modes (but slower than KH modes
  would grow in full 3D), and terminate the MRI growth. As a result
  the MRI in 2D is terminated at unrealistically large magnetic
  stresses and continues to grow after termination, a behaviour not
  observed in our 3D simulations.

\item We performed a set of simulations to study the dependence of the
  properties of the parasite modes. Using a 9th-order spatial
  reconstruction scheme (MP9), we reached convergent results only in
  simulations with at least 60 zones per MRI channel, \ie the value
  of the magnetic stress at termination differed by less than $10\%$
  between the two highest resolution runs. This grid resolution is
  significantly higher than the one necessary to obtain convergence in
  the growth rate of the MRI, for which 8 zones are sufficient to
  obtain a $10\%$ accuracy.  We note that the required number of zones
  will be significantly higher, if lower order reconstruction methods
  are used. Our result is not surprising when viewed in the light of
  the parasite model.  The KH instability is triggered by the shear
  layer between MRI channels. At this layer structures develop that
  are much finer than the width of the channel itself. Failing to
  capture the KH instabilities properly because of a lack of
  resolution leads to artificially large magnetic stresses at MRI
  termination. However, the qualitative behaviour of the flow seems
  not to be affected by a lack of resolution. Even for our lowest
  resolution simulation (8 zones per channel) we are able to identify
  KH instabilities developing at $\phip \approx 45^{\circ}$ as the
  main termination agent.

\item We studied how the box size can affect the development of
  parasitic instabilities. In 3D simulations with an azimuthal box
  length of at least $ \approx 1 \lambdamri$ the MRI is
  terminated by KH modes, whereas simulations with an azimuthal box
  length $ \lesssim 1 \lambdamri$ gave very similar results
  as 2D simulations and the MRI was terminated by TM (because KH modes
  are suppressed; compare models \#16 and \#17 in
    \tabref{tab:main_results}).
Taking into account that the parasitic KH instabilities  develop at a $45^\circ$
  angle, one should use  a box size of at least $1 \approx \lambdamri$ is in the radial  direction.
  However, we recommend using larger
  boxes in the horizontal directions  to reduce the influence of the boundary conditions on the
  development of the parasitic instabilities. In the vertical
  direction it is sufficient to use a box size of $\lambdamri$, \ie
  it is sufficient to consider the evolution of one single MRI channel
  to capture the termination process correctly. 
Determining the minimum box size has been critical to be able
  to perform the simulations with the highest resolution presented in
  this work (model \#11 with 134 zones per channel).
    
\end{enumerate}

Our results have some important implications for the community
modelling magnetorotational collapse of stellar cores. Studying the
termination of the exponential growth phase of the MRI, we learned
that it involves parasitic KH instabilities, but does neither depend
on the physical resistivity nor viscosity present in CCSNe. 
The effects of the interaction of neutrinos and matter, which at
  different locations in the PNS can be described either as an
  effective viscosity or by a drag term, affect the properties of the
  MRI, in particular its growth rate and wavelength \citep{Guilet_2015}.
Consequently, this interaction should induce modifications
  of the termination amplitude.  On the other hand, we do not expect
  these effects to change the most important qualitative result of our
  study, namely that the MRI is not terminated by TM, but by the KH instability. 

   The presence of background flows in CCSNe (like convection
      or turbulence) can affect the growth and termination of the
    MRI, too. 
In 2D global simulations,
    \cite{Cerda-Duran_2008} observed the growth of coherent channel
    flows, albeit deformed, while
    \citet{Sawai_et_al__2013__apjl__GlobalSimulationsofMagnetorotationalInstabilityintheCollapsedCoreofaMassiveStar}
    observed the disruption of channel flows on timescales of
    milliseconds due to the presence of a dynamical background.  Both,
    the parasitic instabilities and the effect of the background flow,
    seem to disrupt channel flows on similar timescales. Therefore, it
    is likely that both effects will be of relevance in studying the
    MRI termination in global simulations.

Furthermore, the presence of parasitic instabilities implies the
existence of a maximum magnetic stress at termination. This limits the
ability of the MRI to amplify the magnetic field to dynamically
relevant values in CCSNe. The turbulence triggered by the MRI may
amplify the global magnetic field further, if conditions for dynamo
action are encountered. A study of the latter process is beyond the
scope of this work, and will be addressed in a future
  publication. The MRI has been traditionally invoked in the CCSN
community as a justification to use supernova progenitors with an
artificially enhanced initial magnetic field strength, $10^3$--$10^4$ 
times larger than the values expected in stellar evolution
models.  This argumentation should be revisited if we want to move
towards a more realistic modeling of magnetorotational core collapse.

Unfortunately, given the minimum numerical requirements that we
provide in this work, 3D simulations seem nowadays unfeasible. To
resolve MRI channels of $1\,$m length scale in a proto-neutron star of
about $30$\,km radius, with a resolution of $\sim 60\,$zones per
channel, it would require about $5\times 10^{19}\,$grid zones.
Evolving this system for a typical dynamical timescale of the
proto-neutron star evolution, $\sim 100\,$ms, would require about
$10^{8}$ time steps, several zettabytes of memory and would consume
$\sim 10^{14}$ CPU hours ($10^{8}\,$iterations times
$5\times 10^{19}\,$zones times $100\,$operations/zone at
$500\,$Gflops).  With current petascale supercomputers this would
  amount to $\sim 3\times10^5\,$yr of
uninterrupted computation with $100\,000\,$cores.  A simulation which
 includes the  whole iron core with a radius of
$\sim 1000$\,km would be even more demanding. Reducing the
dimensionality of the system by imposing axisymmetry reduces the
computational time by a factor $\sim 4\times 10^6$.  The use of
adaptive mesh refinement (AMR) techniques would alleviate, but not
solve the problem.

The CCSN community has been performing 2D and 3D simulations with
resolutions of up to $12.5\,$m
\citep[2D;][]{Sawai_et_al__2013__apjl__GlobalSimulationsofMagnetorotationalInstabilityintheCollapsedCoreofaMassiveStar}
and $15.6\,$m \citep[3D;][]{Masada_et_al__2015}.  We note that these
authors could afford such high resolutions only by accepting other
limitations in terms of geometry and input physics, while other
simulations use resolutions of several 100\,m
\citep{Obergaulinger_Aloy_Mueller__2006__AA__MR_collapse,
  Obergaulinger_et_al__2006__AA__MR_collapse_TOV, Shibata_2006,
  Burrows_2007, Cerda-Duran_2008, Scheidegger_2008, Mikami_2008,
  Kuroda_2010, Takiwaki_2011, Winteler_2012, Mosta_2014,
  Obergaulinger_2014}, which even for the artificially enhanced
initial magnetic field, fail to resolve the termination process of the
MRI. Given the above computational requirements, the community should
rethink the way to model this scenario.  The agreement that we have
found here between local simulations and the parasitic instability
analysis of GX94 and \cite{Pessah} make us hope that there could be a
way out. If we can understand how MRI and turbulence work at sub-meter
scales, this information could be incorporated in global simulations
using appropriate subgrid models. As a first step along this path we
plan to gain a deeper understanding in the maximum magnetic stress
achievable by the MRI and in the properties of its turbulent saturated
state.

\section*{Acknowledgements}
TR acknowledges support from The International Max Planck Research
School on Astrophysics at the Ludwig Maximilians University Munich, EM
\& TR acknowledge support from the Max-Planck-Princeton Center for
Plasma Physics, and MA, PCD, TR and MO acknowledge support from the
European Research Council (grant CAMAP-259276). We also acknowledge
support from grants AYA2013-40979-P and PROMETEOII/2014-069. The
authors thank Rob Yates for proof reading the manuscript, and
M.~Pessah, C.~McNally, and H.~Latter and J.~Guilet for helpful
discussions. 
 The authors thank the anonymous referee whose useful remarks
greatly improved the quality of this manuscript.
 The computations have been performed at the Leibniz
Supercomputing Center of the Bavarian Academy of Sciences and
Humanities (LRZ), the Rechenzentrum Garching of the
Max-Planck-Gesellschaft (RZG), and at the Servei d'Inform\`atica of
the University of Valencia.

%%%%%%%%%%%%%%%%%%%%%%%%%%%%%%%%%%%%%%%%%%%%%%%%%%%%%%%%%%%%%
%% Bibliography
%%%%%%%%%%%%%%%%%%%%%%%%%%%%%%%%%%%%%%%%%%%%%%%%%%%%%%%%%%%%%

\bibliographystyle{mn2e}

\label{lastpage}

\end{document}